\documentclass[showpacs,twocolumn,amssymb]{aastex631}
\usepackage{float}
\usepackage{amssymb}
\usepackage{amsmath}
\usepackage{appendix}

\newcommand{\cm}{\,\text{cm}}
\newcommand{\g}{\,\text{g}}
\newcommand{\km}{\text{ km}}
\newcommand{\s}{\text{ s}}
\newcommand{\E}{\text{ e.s.u}}

\shorttitle{Accretion of CO in AGN disk : Outflow feedback}
\shortauthors{Chen et al.}

\begin{document}

\title{The Role of Outflow Feedback on Accretion of Compact Objects 
	in Accretion Disk of Active Galactic Nuclei}

\author[0000-0001-8955-0452]{Ken Chen}
\author[0000-0002-9037-8642]{Jia Ren}

\affiliation{School of Astronomy and Space Science, Nanjing University, 
	Nanjing 210023, China}

\author[0000-0002-7835-8585]{Zi-Gao Dai}

\affiliation{Department of Astronomy, School of Physical Sciences, 
	University of Science and Technology of China, Hefei 230026, China}
\affiliation{School of Astronomy and Space Science, Nanjing University, 
	Nanjing 210023, China}

\begin{abstract}
Compact objects (COs) can exist and evolve in an active 
galactic nuclei (AGN) disk, triggering a series of 
attractive CO-related multi-messenger events 
around a supermassive black hole. To better 
understand the nature of an embedded CO and its 
surroundings and to investigate CO-related 
events more accurately, in this paper, we study the
specific accretion process of a CO in an AGN disk 
and explore the role of outflow feedback. We show that 
the asymptotically isotropic outflow generated from 
the CO hyper-Eddington accretion would truncate 
the circum-CO disk and push out its surrounding gas, 
resulting in recurrent formation and refilling of an
outflow cavity to intermittently stop the 
accretion. Applying this universal cyclic process to 
black holes (BHs) and neutron stars (NSs), we find 
that, even if it is above the Eddington rate, the mass 
rate accreted onto a BH is dramatically reduced compared 
with the initial gas captured rate and thus consumes 
few mass of the AGN disk; outflow feedback on a NS is 
generally similar, but possesses complexities on the 
existence of a stellar magnetic field and hard surface. 
We demonstrate that although outflow feedback itself 
may be unobservable, it remarkably alters the CO 
evolution via reducing its mass growth rate, and the AGN 
disk can survive from the otherwise drastic 
CO accretion overlooking outflow. In addition, we discuss 
the potential influence of underdense cavity 
on CO-related events, which embodies the significant role 
of outflow feedback as well.
\end{abstract}


\section{Introduction} \label{sec:intro}

Accretion disks can exist around the galaxy central 
supermassive black holes (SMBHs), lighting up SMBHs to 
form the so-called active galactic nuclei (AGN).
Stars and compact objects (COs), e.g. black holes (BHs) and 
neutron stars (NSs), are predicted to widely reside in the
AGN disk via gas-capture of orbiters 
from the galaxy nuclei cluster to the disk, 
or in-situ star formation and evolution, or geometrical 
coincidence of a few nuclear stars and COs with the 
renascent disk
\citep[e.g.][]{Ostriker83, Syer91, Cheng99, Goodman03,TQM05, 
McKernan12, Fabj20}.

The AGN disk provides a venue for various CO-related 
astrophysical events. Especially since the 
BH-BH merger event GW190521 \citep{Abbott20} along with  
the plausible EM counterpart ZTF19abanrhr \citep{Graham20} 
indicates an AGN disk environment origin, the 
studies on CO-related events in the AGN disk are 
in the ascendant lately. The AGN disk channel can 
contribute to a large amount of unique BH-BH, 
BH-NS, NS-NS mergers \citep[e.g.][]{Tagawa20a, 
Tagawa21, McKernan20a, McKernan20b, Samsing22} 
and extreme mass ratio inspirals 
\citep[e.g.][]{Pan21a,Pan22} events for gravitational 
wave (GW) observations. Electromagnetic (EM) events, such 
as gamma-ray bursts and kilonovae \citep{Zhu21a, Perna21a, 
Yuan22, Ren22, Wang22}, 
supernovae \citep{Zhu21b, Grishin21, Moranchel21},
accretion-induced collapses of 
NSs \citep{Perna21b} and white dwarfs \citep{Zhu21b}, 
micro-tidal disruption events \citep{Yang22}, 
occurring in the AGN disk show 
characteristic transient features. 
All the GW and EM transients are distinguishable from 
the same events taking place in the classical 
interstellar medium on account of the COs 
lying within the extremely dense environment.
Moreover, accretion in the AGN disk gives 
opportunities for BHs to be observed 
\citep{Wang21a, Wang21b, Kimura21b}. On the other hand, 
mass and spin of the embedded stars and COs would
increase rapidly via accreting gas in the AGN disk
\citep[e.g.][]{Jermyn21, Dittmann21, Tagawa20b, Tagawa22}, 
markedly changing the features of stars and COs 
to potentially impact CO-related events and 
AGN disk structure. Therefore, as the prerequisite, 
a more comprehensive understanding of 
COs\textquoteright \ surroundings is very 
important to further explore the CO evolution and 
the CO-related events in the AGN disk.

Gas accreted onto an embedded CO from the AGN disk 
environment is inevitable. The CO mass capturing 
rate is generally hyper-Eddington, which may even 
exceed the mass inflow rate of the AGN disk. 
Previous works have artificially set 
upper limits for the accretion rate onto CO 
at the Eddington rate with various 
radiative efficiencies, or at the AGN disk mass 
inflow rate, or directly at a certain fraction of CO mass 
capturing rate \citep[e.g.][] {Tagawa20a, McKernan20a, 
Kaaz21, Kimura21b, Wang21a}, but none yet 
connected the rate to the actual accretion processes of CO.
Hyper-Eddington accretion is 
predicted both theoretically and by numerical 
simulations to be accompanied with powerful 
outflow of mass and energy in the form of 
circum-CO disk wind and collimated jet 
\citep[e.g.][]{Blandford99, Begelman12, Gu12, Gu15, 
Ohsuga05, Jiang14, Yang14, Hashizume15, 
Sadowski151, Sadowski162, Kitaki18, 
Hu22}, which would reduce the mass rate 
eventually accreted onto COs \citep{Pan21b}, and 
would conversely interact with 
surroundings, acting as feedback.
Several recent works have investigated the 
feedback of CO hyper-Eddington accretion on 
the AGN disk environment \citep{Wang21a, Kimura21b, 
Tagawa22}, where \cite{Wang21a} and \cite{Kimura21b} 
did not concretely discuss the properties 
of the disk wind, while \cite{Tagawa22} mainly 
focused on the jet-cocoon feedback 
from high-spin stellar-mass BHs, which may 
not be practicable for low-spin BHs and NSs. 
So, a detailed study for the formation, evolution, 
and feedback of outflow is needed, 
and then we can better understand the 
nature of CO accretion, evolution, 
and COs\textquoteright \ surroundings.

In this paper, we construct  
the circum-CO accretion disk and the induced 
outflow via the widely accepted description of 
hyper-Eddington accretion. The asymptotically 
isotropic outflow then interacts with the AGN disk 
environment, circularly opens and refills an outflow 
cavity, resulting in the CO undergoing periodic 
accretion processes. We focus on the role of 
outflow feedback on CO accretion and
its surroundings, manifesting as the 
prominent reduction of time-averaged mass 
accretion rate, and CO mainly staying 
in the underdense cavity rather the AGN disk 
environment, which significantly impact the 
evolution of CO and the features of CO-related 
events. 

This paper is organized as follows. 
In Section \ref{sec:model}, we show 
the construction of a circum-CO accretion disk and an
outflow. We describe the outflow evolution in the 
AGN disk, then show the features of outflow cavity, 
and reduction of the mass rate accreted onto a
BH or NS in Section \ref{sec:feedback}. 
Several discussions are presented in Section 
\ref{sec:Discussion}, involving the potential 
influence of shock radiation-cooling efficiency and 
CO-gravity-induced gap in Sections \ref{subs41} 
and \ref{subs42}, the depletion of AGN disk 
inflow mass by CO accretion process in Section
\ref{subs43}, the jet effect and the outflow EM 
radiation in Section \ref{subs44}, the role of 
outflow cavity on CO evolution and CO-related 
events in Section \ref{subs45}, and the probable 
accretion feedback of CO with large bulk velocity 
in Section \ref{subs46}.
We summarize our conclusions in Section 
\ref{sec:Summary}. We use the convention of 
$Q_x=Q/10^x$ in cgs units unless otherwise noted. 
The $c$ and $G$ are speed of light and 
gravitational constant.

\section{Hyper-Eddington Accretion in the AGN Disk} \label{sec:model}

The accretion disk of an AGN is typically modeled 
as a geometrically-thin, optically-thick 
disk with the $\alpha$-prescription representing 
the viscosity torque \citep{Shakura73}. We define the 
dimensionless accretion rate of SMBH by 
$\dot{\mathcal{M}}=\dot{M}/\dot{M}_{\rm{Edd}}$, where $\dot{M}$ 
is the accretion rate of SMBH, $\dot{M}_{\rm{Edd}}=L_{\rm{Edd}}/c^2$ is the 
Eddington limit accretion rate, 
$L_{\rm{Edd}}=4\pi G M m_pc/\sigma_T=1.26\times10^{46}M_8\rm{erg\,s}^{-1}$, 
$M$ is the mass of SMBH in units of $M_\odot$, $m_p$ 
is the proton mass, and $\sigma_T$
is the Thompson cross section. Because the evolution of 
most COs should take place in the region $\gtrsim O(10^3)R_g$ of
AGN disk during the AGN lifetime owing to 
the larger disk volume holding more COs and 
the potential existence of migration trap \citep{Bellovary16}, 
where $R_g=GM/c^2$ is the gravitational radius, we use the 
outer region solution of the standard disk model. 
The half-thickness, density, and mid-plane temperature of 
the AGN disk are \citep[e.g.][] {Kato08, Wang21a}
\begin{equation}\label{Eq:disk}
\left\{
\begin{array}{l}\vspace{1ex}
H=4.3\times 10^{14}\,\alpha_{-1}^{-1/10}M_{8}^{9/10}\dot{\mathcal{M}}^{3/20} R_{4}^{9/8}\,{\rm cm},\\ \vspace{1ex}
\rho_{\rm d} =6.9\times 10^{-11}\,(\alpha_{-1}M_{8} )^{-7/10}\dot{\mathcal{M}}^{11/20}R_{4}^{-15/8}\,{\rm g\,cm^{-3}},\\ \vspace{1ex}
T_{c}=4.6\times10^{3}\,(\alpha_{-1}M_{8} )^{-1/5}\dot{\mathcal{M}}^{3/10}R_{4}^{-3/4}\,{\rm K},
\end{array}\right.
\end{equation}
where $\alpha$ is the constant viscosity parameter \citep{Shakura73} and $R$ is 
the radius of disk in units of $R_g$. The outer disk can be 
self-gravity unstable and would fragment into gas clouds and stars if the 
Toomre parameter 
$Q=\tilde{\Omega}_{\rm K} \tilde{c}_{s}/2\pi G \rho_{\rm d} H<1$ \citep{Toomre64}, 
where the local sound speed $\tilde{c}_{s}\approx \sqrt{3kT_{c}/m_{p}}\approx 15.7\,T_{4}^{1/2}\km\s^{-1}$ and the angular velocity 
$\tilde{\Omega}_{\rm K}=\sqrt{GM/R^{3}}$ for 
Keplerian-rotating disk.

Equation (\ref{Eq:disk}) would be inaccurate to describe the outer disk region 
because of the gas self-gravity leading to accretion processes more 
complicated \citep[e.g.][]{ Goodman03}. 
Many plausible theoretical models have been proposed to describe
the structure of a quasi-stationary self-gravitating accretion disk 
\citep[e.g.][et al.]{SG03, TQM05, Mishra20, Gilbaum22}; the disk structure is
diverse in different models, and yet there is no general agreement. As 
discussed in \cite{Wang21a}, though the standard disk model 
may be inexact, the main characteristics of disk can 
be predicted qualitatively; the analytic solution can also help to better
understand the dependence of CO accretion and outflow feedback properties 
on AGN disk parameters. So we use Equation (\ref{Eq:disk}) as an 
environment to investigate the CO accretion and outflow feedback 
in the AGN disk. Future works can use the CO accretion-feedback processes proposed
in this work and employ specific AGN disk model to better study the evolution of
CO and the astrophysical events relevant to CO.

When $\dot{\mathcal{M}}$ is extremely low, the AGN disk is likely to be 
advection-cooling dominated \citep{Narayan94,Narayan95}, its properties  
differ from those of the radiation-cooling dominated thin disk, e.g. the much 
higher temperature and the much geometrically thicker inflow 
\citep[see][for a review]{Yuan14}. 
The evolution of CO in these low-accretion-rate AGN disks 
should be distinctive and thus would be studied separately in a future work.

\subsection{Disk Formation around CO} \label{subs21}
As embedded in the AGN disk 
environment, COs (e.g. BH or NS) will capture and accrete gas 
within its sphere of gravity. 
AGN disk rotates differentially, the accreted gas hence holds shear velocity 
w.r.t the CO, and would form a rotating disk around the CO rather than fall directly
into it. To intuitively understand this process, 
we take gas accretion onto a planet embedded in a proto-planetary disk as an 
analogy \citep[e.g.][]{Kimura21b}. As shown in simulations 
\citep{Ayliffe09, Tanigawa12}, 
a nearly Keplerian rotating circum-planetary disk forms around the planet; 
the majority of mass is concentrated in the outer region of the disk, mainly 
around the circularization radius $r_{\rm{cir}}$ of the accreted gas (see
 below). Back to the AGN 
disk case, the initially circular gas disk would then go through viscous evolution 
and effective accretion to the CO on viscous timescale $t_{\rm{vis}}$ 
\citep[e.g.][]{Kato08}\footnote{When the initial mass inflow rate to CO is
extremely large, as in the AGN disk environment we discuss, 
the infalling gas potentially follows an alternative structure called 
“zero-Bernoulli accretion” flows suggested by \cite{Coughlin14}; the inflow inflates 
nearly to the poles and forms a weakly bound, quasi-spherical structure, 
rather than a large-scale accretion disk. The criteria of 
which structure occurs under the specific
environment parameters, and the distinctions of the relevant 
outflow feedback will be studied in another work.}.

The initial inflow mass rate around the disk outer boundary 
can be described by the Bondi-Holye-Lyttleton (BHL) formulation 
\citep[see][for a review]{Edgar04}; 
under the corrections of the SMBH gravity effect in radial 
direction and the finite height of the AGN disk
in vertical direction, the gas inflow rate is estimated as 
follows \citep{Kocsis11, Pan21b}:
\begin{equation}
\dot{M}_{\rm {inflow}}=\dot{M}_{\rm{BHL}} \times \min \left\{1, \frac{H}{r_{\rm{BHL}}}\right\} \times \min \left\{1, \frac{r_{\rm {Hill }}}{r_{\mathrm{BHL}}}\right\} ,
\end{equation}
and the outer boundary radius of the circum-CO disk $r_{\rm{obd}}$ can be approximate to
$\sim r_{\rm{cir}}$, which can be estimated from the angular 
momentum conservation 
of the infalling gas, i.e.,
\begin{equation}
\sqrt{G m_{\rm{CO}} r_{\rm{cir}}}=v_{\rm{rel}}(r_{\rm{rel}} ) r_{\rm{rel}}  , \label{robd}
\end{equation}
where $m_{\rm{CO}}$ is the mass of CO, 
$r_{\rm{rel}}:=\min \{r_{\rm{BHL}}, r_{\rm{Hill}}\}$ is the radius of 
CO gravity sphere, $r_{\rm{Hill}}=(m_{\rm{CO}} /3M)^{1/3}
R_{\rm{CO}}$ is the Hill radius, and
$r_{\rm{BHL}}=G m_{\rm{CO}}/(v_{\rm{rel}}^{2}+\tilde{c}_{s}^{2} )$ 
is the BHL radius; the canonical equation of the BHL accretion rate is
\begin{equation}
\dot{M}_{\rm{BHL}}=\frac{4 \pi G^2 m_{\rm{CO}}^2 \rho_{\rm{CO}}}{(v_{\rm{rel}}^{2}+\tilde{c}_{s}^{2} )^{3/2}}  , \label{BHL}
\end{equation}
where $\rho_{\rm{CO}}$ is the density of AGN disk gas surrounding the CO, 
and $v_{\rm{rel}}$ is the relative velocity between the CO and the gas, which 
contains components of the gas relative bulk motion and
shear rotation. To estimate the shear 
velocity $v_{\rm{shear}}$, we linearize the velocity 
disparity at the location of CO $R_{\rm{CO}}$ and the boundary 
of the CO gravity sphere $(R_{\rm{CO}} \pm r_{\rm{rel}} )$, i.e., 
\begin{equation}
v_{\rm{rel}}(r_{\rm{rel}} )=V_{\rm{K}}(R_{\rm{CO}} )-V_{\rm{K}}(R_{\rm{CO}}+r_{\rm{rel}} )\approx \frac{1}{2} r_{\rm{rel}}\tilde{\Omega}_{\rm K}  ,
\end{equation}
where we consider the CO co-rotates with the Keplerian AGN disk 
at its midplane, 
$V_{\rm{K}}=\sqrt{G M/R}$, in 
which case the gas bulk motion can be neglected 
\citep{Kocsis11}. In fact, the CO can hold large bulk 
velocity $v_{\rm{bulk}}$ w.r.t. the ambient disk gas, 
as briefly discussed in Section \ref{subs46}. 
Here we primarily investigate the long-term evolution of COs 
in the AGN disk, which are 
probably on the Keplerian co-rotating orbits; 
the COs with large $v_{\rm{bulk}}$ may escape from the disk, 
or the gas dynamical and accretion frictions reduce the relative 
velocity of CO, leading to the subsequent co-rotation
\citep[e.g.][]{Ostriker99, Bartos17, Secunda19, Fabj20, Nasim22}. 

When living in the AGN disk, the CO can interact with the 
surrounding disk through gravity, inducing gravitational 
torque to additional exchange energy and 
angular momentum with the AGN disk (the disk–satellite
interactions would be analogous to a giant planet
in a proto-planetary disk; see \citealt{Armitage07},
\citealt{Baruteau14}, as the reviews). 
The ambient gas has the tendency to be repelled from the CO, and thus 
gas density is reduced in the disk annulus at  
$R_{\rm{CO}}$. The gas density and half-width of the reduced region 
are estimated as 
$\rho_{\rm d,gap}/\rho_{\rm d}=1/[1+0.04(m_{\rm{CO}}/M)^2
(H/R_{\rm{CO}} )^{-5}\alpha^{-1} ]$ 
and $R_{\rm d,gap}/R_{\rm{CO}}=0.21(m_{\rm{CO}}/M)^{1/2}(H/R_{\rm{CO}} )^{-3/4}\alpha^{-1/4}$
\citep{Kanagawa15, Kanagawa16, Tanigawa16}. 
We set the density used to calculate the 
BHL accretion rate, i.e. in Equation (\ref{BHL}), as
\begin{equation}
\rho_{\rm{CO}}=
\begin{cases}
\rho_{\rm{d}}  ,  & R_{\rm d,gap} < r_{\rm{rel}} \\ 
\rho_{\rm{d,gap}} . & R_{\rm d,gap} > r_{\rm{rel}}
\end{cases}
\end{equation}

For the circum-CO accretion disk with extremely high inflow mass rate 
and distant outer boundary, the outer disk region would be self-gravity 
unstable, accordingly the gas captured by the CO fragments, resulting in 
the reduction of mass inflow rate \citep[e.g.][]{Pan21b, Tagawa22}. 
The Toomre parameter 
of the circum-CO disk can be approximatively expressed as 
$Q_{\rm{CO}}=\Omega_{\rm K} {c}_{s}/\pi G \Sigma \sim 2 
\alpha_{\rm{CO}}h^3 v_{K}^3/G\dot{M}_{\rm {inflow}}$, 
where $\Omega_{\rm{K}}$ is the Keplerian 
angular velocity, ${c}_{s} \sim h v_{\rm{K}}$ is 
the sound speed, $\Sigma=\dot{M}_{\rm {inflow}}/2 \pi r v_{\rm{r}}$ is the 
disk surface density and $v_{\rm{r}}=\alpha_{\rm{CO}} h^2 v_{\rm{K}}$ is 
the inflow radial velocity, $\alpha_{\rm{CO}}$ is the 
viscosity parameter, $h=H_{\rm{CCOD}}/r$ is the disk height ratio, 
$H_{\rm{CCOD}}$ is the vertical scale height of the disk,
and $v_{\rm{K}}=\sqrt{Gm_{\rm{CO}}/r}$;
the stabilization of circum-CO disk requires $Q_{\rm{CO}}\geq 1$,
so the modified mass inflow rate is expressed by
\begin{equation}
\dot{M}_{\rm {obd}}=
\begin{cases}
\dot{M}_{\rm {inflow}} , & Q_{\rm{CO}}(r_{\rm {obd}} )\geq 1 \\ 
2 \alpha_{\rm{CO}}h^3 v_{K}^3/G . & Q_{\rm{CO}}(r_{\rm {obd}} )< 1
\end{cases} \label{obdinflow}
\end{equation}

In short, when a CO accretes the AGN disk gas, a circum-CO disk 
would form, with the initial radius $r_{\rm {obd}}$ 
and the mass inflow rate $\dot{M}_{\rm {obd}}$ given by Equations 
(\ref{robd}) and (\ref{obdinflow}).

\begin{figure*}
	\begin{center}
		\includegraphics[width=0.32\textwidth]{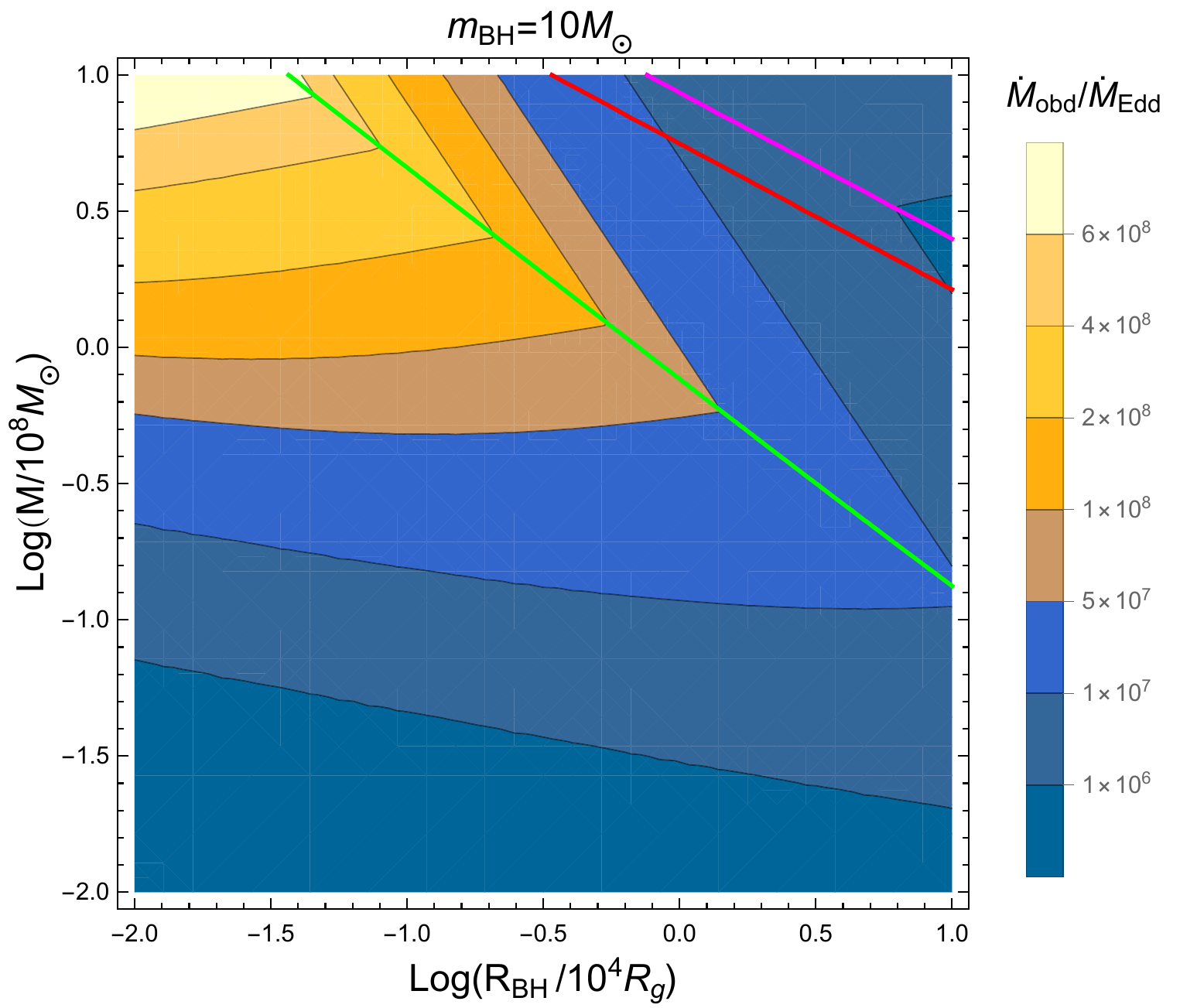}
		\includegraphics[width=0.32\textwidth]{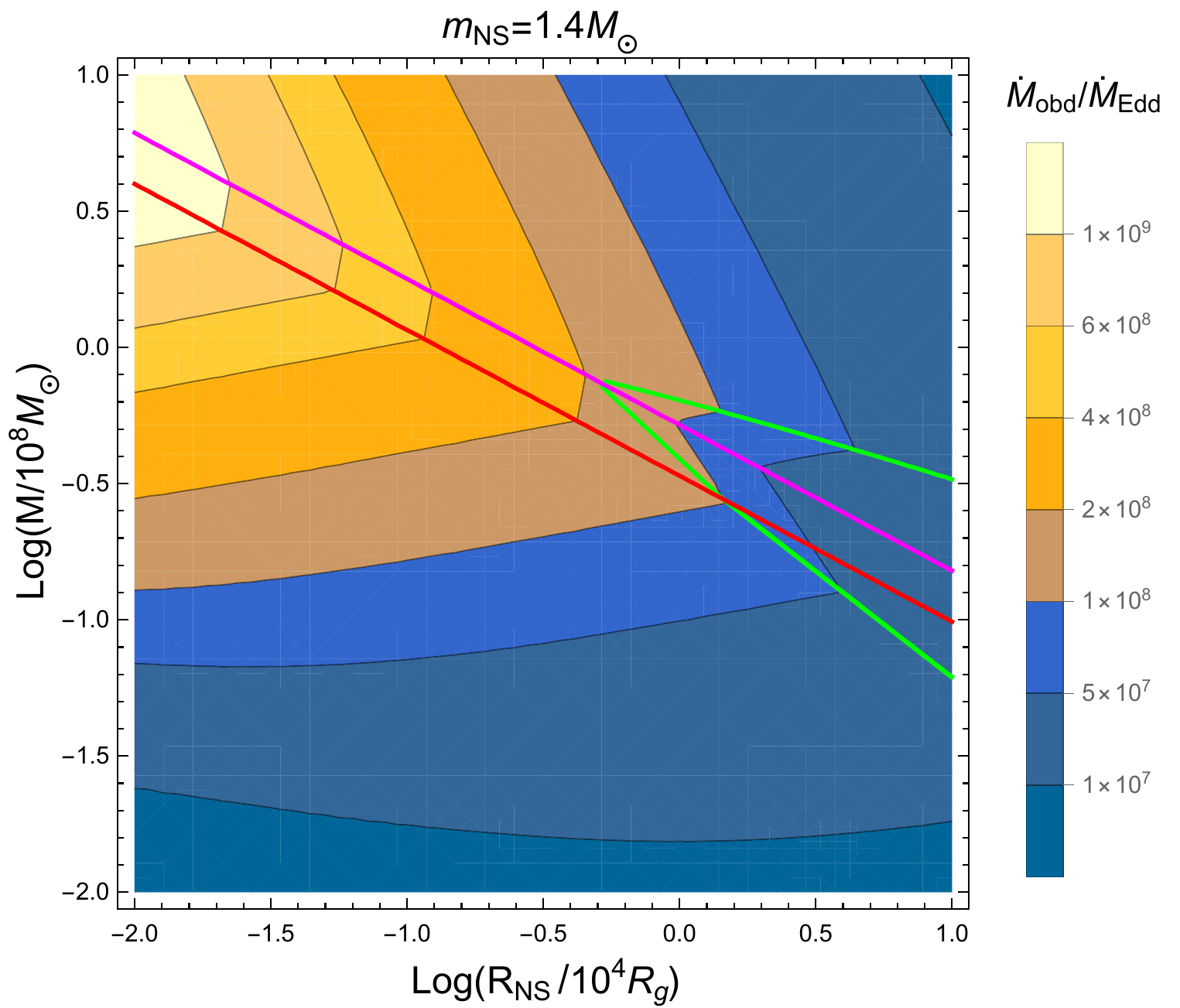}
	\end{center}
	\caption{The initial mass inflow rates at the outer boundary 
	of the circum-CO disks accreting the AGN disk gas shown in the 
	$R_{\rm{CO}}-M$ plane. The fiducial values of the parameters are 
	$\alpha=0.1$, $\dot{\mathcal{M}}=0.5$, $\alpha_{\rm{CO}}=0.1$, 
	$h=0.5$. Left-panel shows a case of $m_{\rm{BH}}=10 M_\odot$ BH 
	accretion, and right-panel shows a case of $m_{\rm{NS}}=1.4 M_\odot$ 
	NS accretion, where the magenta, red lines and above regions represent 
	the boundary of $r_{\rm{Hill}}\geqslant 
	r_{\rm{BHL}}$ and $H \geqslant r_{\rm{BHL}}$. 
	The green lines represent the boundary of $Q_{\rm{CO}}=1$, where the 
	above region in left panel and the contained region in right panel 
	show $Q_{\rm{CO}}<1$.}	
    \label{Fig:Mobd}
\end{figure*}

\begin{figure*}
	\begin{center}
		\includegraphics[width=0.32\textwidth]{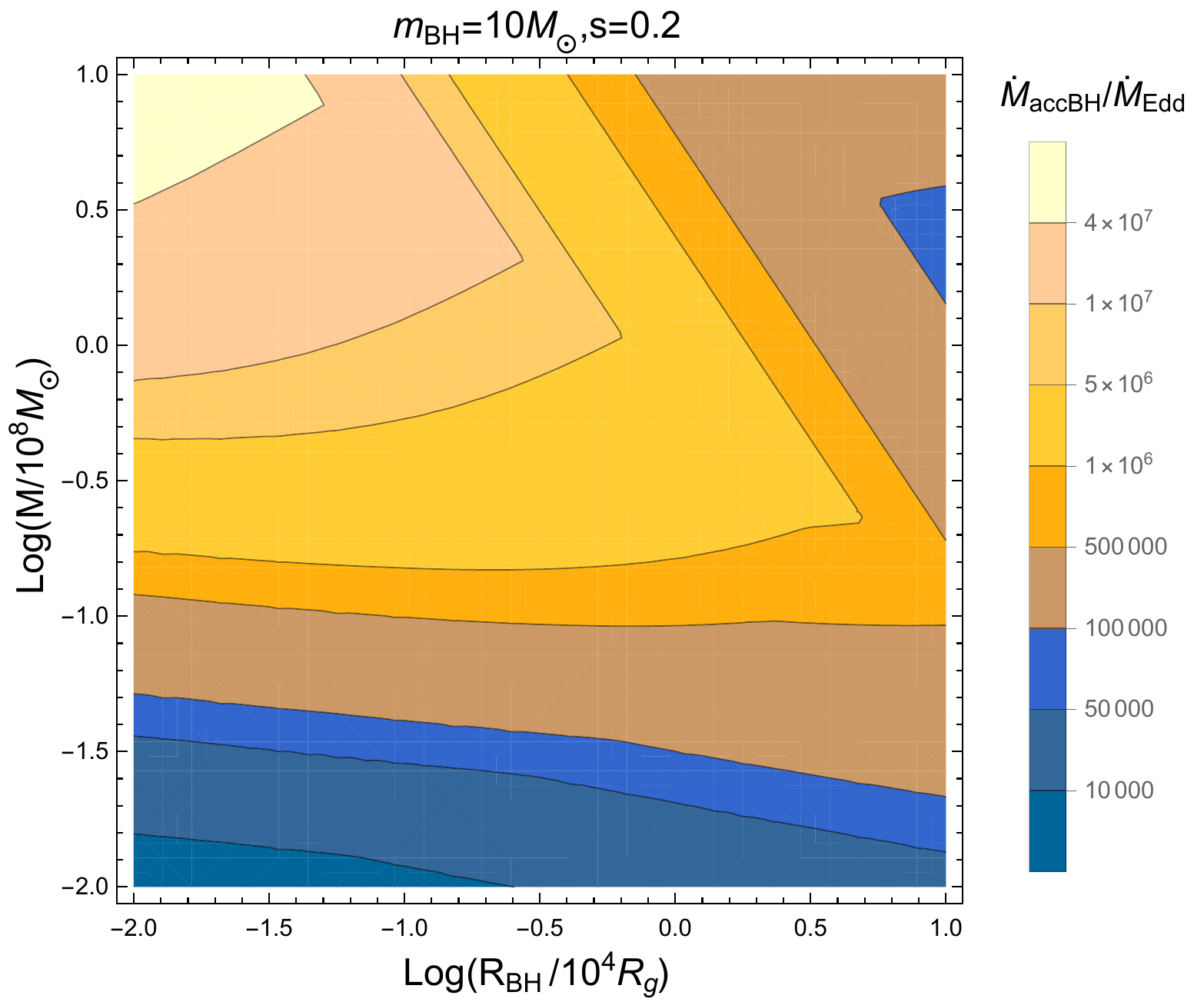}
		\includegraphics[width=0.32\textwidth]{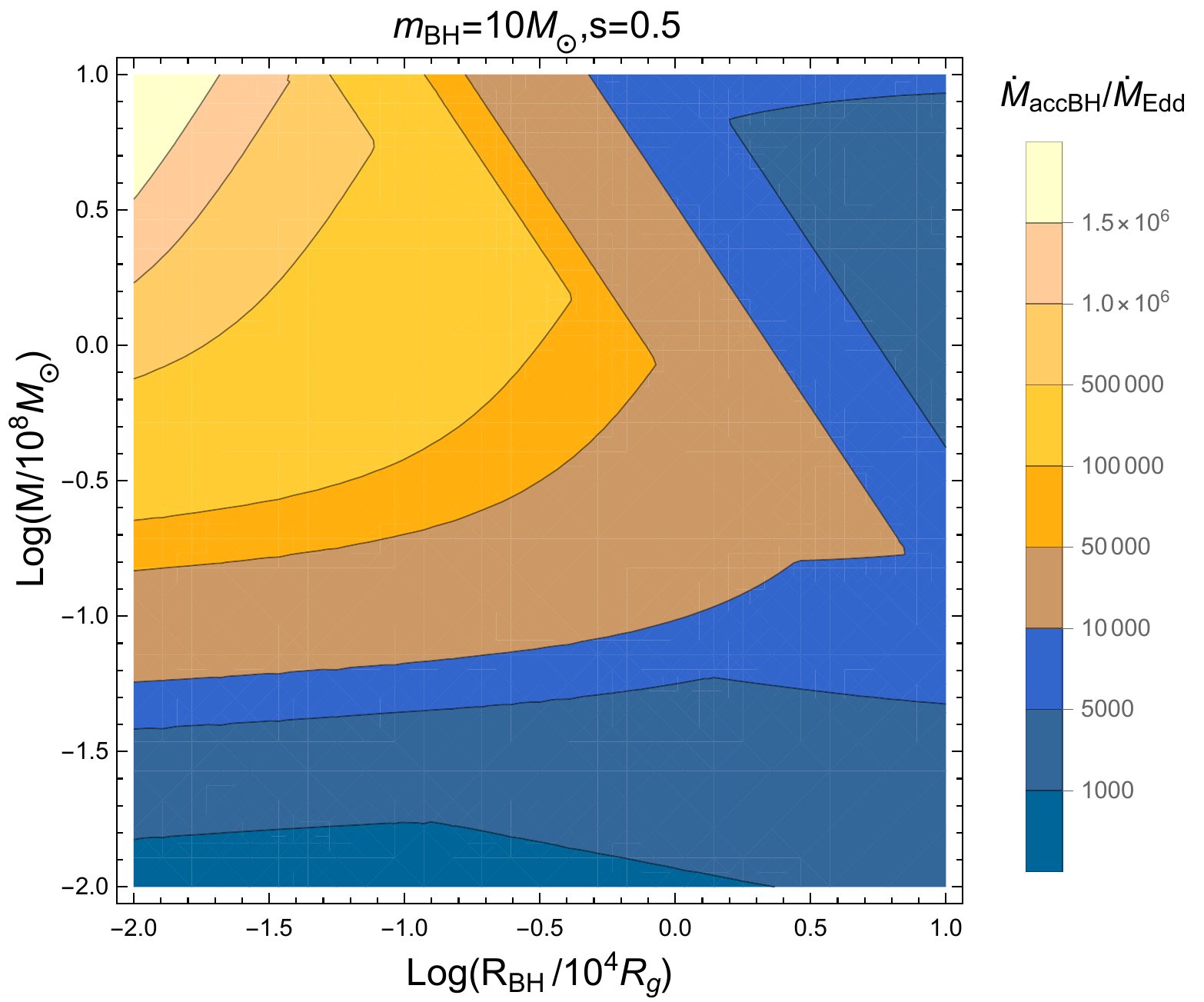}
		\includegraphics[width=0.32\textwidth]{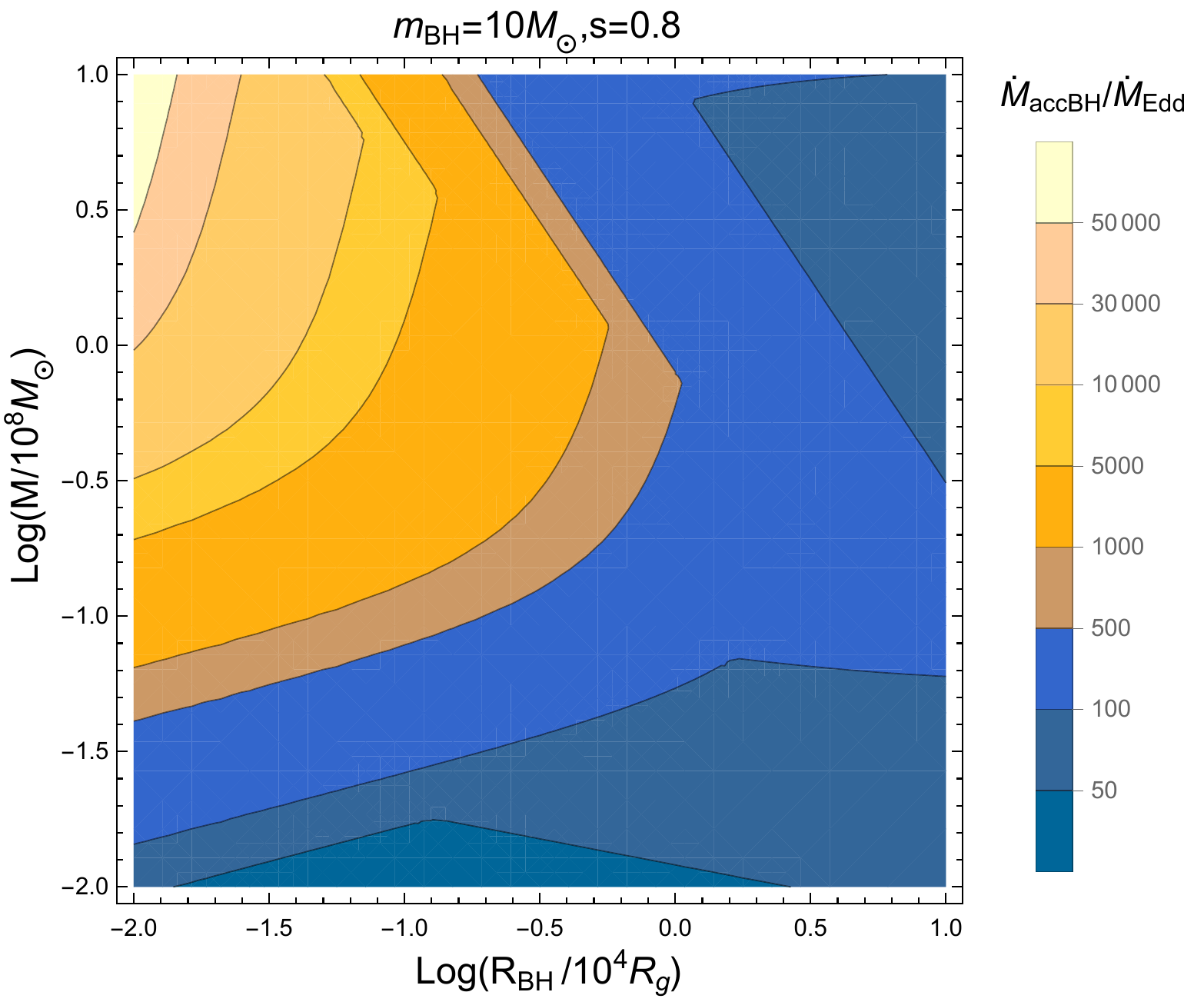}
	\end{center}
	\caption{The mass accretion rates of a $m_{\rm{BH}}=10 M_\odot$ BH ignoring 
		the accompanying outflow feedback effects. Three panels respectively show 
		the cases of circum-BH disk with different outflow strength, representing 
		as $s=0.2$, $s=0.5$, $s=0.8$. The BH generally undergoes 
		hyper-Eddington accretion processes in the AGN disk environment.}	
	\label{Fig:maccbh}
\end{figure*}

\subsection{Hyper-Eddington Accretion Disk and Outflow}
\label{2-2}
The initial mass inflow rates of COs are 
extremely hyper-Eddington, i.e. 
$\dot{M}_{\rm {obd}} \gg \dot{M}_{\rm{Edd}} $,
as shown in Figure \ref{Fig:Mobd};
a specific property of the accretion is photon 
trapping, i.e. the photons and radiation energy are 
trapped and advected inward with the dense gas inflow because of the 
photon diffusion timescale exceeding the disk accretion timescale. 
The radiation-effective flow converts to be radiation-ineffective 
nearly at the trapping radius \citep[e.g.][]{Kato08, Kitaki21}:
\begin{equation}
r_{\rm{tr}}=3\dot{m}hr_g, \label{rtr}
\end{equation}
where $\dot{m} \equiv \dot{M}_{\rm {obd}}/\dot{M}_{\rm{Edd}}$ and 
$r_g=Gm_{\rm{CO}}/c^2$. Inside $r_{\rm{tr}}$, the disk radiation pressure 
efficiently drives outflow, leading to 
reduction of the circum-CO disk mass inflow rate; the radius-dependent 
rate $\dot{M}_{\rm {in}}$ can be written as \citep{Blandford99}:
\begin{equation}
\dot{M}_{\rm {in}}(r)=
\begin{cases}
\dot{M}_{\rm {obd}}(r/r_{\rm{tr}} )^{s}, & r<r_{\rm{tr}} \\ 
\dot{M}_{\rm {obd}}, & r>r_{\rm{tr}}
\end{cases} \label{inflow}
\end{equation}
where the power-law index $s$ is a free parameter and 
its value is taken to be $0<s\leqslant1$;
numerical simulations show that the value lies within $\sim 0.4 - 1.0$ 
\citep{Yang14}, but $\sim 0$ out to $O(10^2)r_g$ for the case of 
modestly circularized gas inflow \citep{Kitaki21}. Under the uncertainty, 
we set three values $s=0.2$, $0.5$, $0.8$ to study the outflow feedback.  
For $r_{\rm{tr}}<r_{\rm {obd}}$, the circum-CO disk is advection-dominated 
within $r_{\rm{tr}}$, and is radiation-dominated between $r_{\rm{tr}}$ and $r_{\rm{obd}}$; 
but when $r_{\rm{tr}}>r_{\rm {obd}}$, the disk is totally advective 
accompanied with the outflow formation\footnote{But the mass inflow rate is not 
large enough to become neutrino-cooling dominated, see \cite{Kohri05} and 
Equation (20) in \cite{Kumar08}.}, and the mass inflow rate is written as
\begin{equation}
	\dot{M}_{\rm {in}}(r)=\dot{M}_{\rm {obd}}(r/r_{\rm{obd}} )^{s} . \label{inflow2}
\end{equation}

The radiation-pressure-driven outflow can take away a considerable 
fraction of viscous heating of the circum-CO disk, as discussed in 
Appendix \ref{Appendix-disk}. The local viscous heating rate per unit 
area is given by Equation (\ref{Qvis}), so the total disk generated heat
is
\begin{align}
L_{\rm{d}}=&\int^{r_{\rm{out}}}_{r_{\rm{in}}} Q_{\rm{vis}}(r)2 \pi r {\rm{d}} r \nonumber\\
&=\frac{3}{2}
\left(1-\frac{l^2 s}{s+1/2}\right)\dot{M}_{\rm {out}} c^2 \frac{r_g}{r_{\rm{in}}}
\left(\frac{r_{\rm{out}}}{r_{\rm{in}}}\right)^{-s} \nonumber\\
&\times\left[\frac{1-(r_{\rm{out}}/r_{\rm{in}})^{s-1}}{1-s}-
\frac{1-(r_{\rm{out}}/r_{\rm{in}})^{-3/2}}{3/2}\right] , \label{Ld}
\end{align}
where $\dot{M}_{\rm {out}}$ is the mass inflow rate at $r_{\rm{out}}$;
$r_{\rm{out}}=r_{\rm {obd}}$ for $r_{\rm{tr}}>r_{\rm {obd}}$ case, 
and $r_{\rm{out}}=r_{\rm{tr}}$ for $r_{\rm{tr}}<r_{\rm {obd}}$ case, 
for which we ignore the heat generated in the outer radiation-dominated 
region $r_{\rm{tr}}<r<r_{\rm {obd}}$
because only tiny gravitational energy is released via accretion at large 
radius.
We take $r_{\rm{in}}\sim 10r_g$ as the inner radius outside which 
the self-similar solutions hold and the outflow cooling is efficient, because 
when $r \lesssim 10r_g$, the relativistic effects of the central CO
can not be neglected \citep{Pan21b}, the outflow is dragged by the 
strong gravitational 
force to be weak \citep{Jiao15} and the cooling is dominated by 
advection rather than outflow \citep{Wu22}.
We roughly assume that a 
constant fraction $f_w$ of the heat is taken away by 
the disk wind and radiation, i.e., 
\begin{align}
L_{\rm{w}}=&f_w\dot{M}_{\rm {out}} c^2 \frac{r_g}{r_{\rm{in}}}
\left(\frac{r_{\rm{out}}}{r_{\rm{in}}} \right)^{-s} \nonumber\\
&\times\left[\frac{1-(r_{\rm{out}}/r_{\rm{in}} )^{s-1}}{1-s}-
\frac{1-(r_{\rm{out}}/r_{\rm{in}} )^{-3/2}}{3/2} \right] , \label{Lw}
\end{align}
where $f_w$ contains the constant coefficients reflecting the 
wind property in Equation (\ref{Ld}); we set $f_w=0.5$ as the 
fiducial value.

The accretion features of the NS are sightly different 
from those of the BH due to the existence of a stellar 
magnetic field and hard surface \citep{Takahashi18}. 
A strong magnetic field may truncate the circum-NS 
disk at a radius $r_{\rm{m}}$ where the magnetic stress of NS 
magnetosphere matches the disk stress
(e.g. \citealt{Zhang09}; and \citealt{Lai14}, \citealt{Romanova15}, as reviews). 
For a dipole field, 
equaling magnetic pressure $B(r_{\rm{m}} )^2/8\pi$ 
to disk pressure $\rho_{\rm{CNSD}}(r_{\rm{m}} )c_s(r_{\rm{m}})^2$, 
where the density $\rho_{\rm{CNSD}}$ of the circum-NS disk is 
given by Equation (\ref{B1}), 
$r_{\rm{m}}$ can be estimated as
\begin{equation}
	 r_{\rm{m}}=k\left(\frac{\mu^{4} r_{\rm{out}}^{2s}}
	 {Gm_{\rm{NS}} \dot{M}_{\rm {out}}^{2}} \right)^\frac{1}{7+2s}, \label{eqrm}
\end{equation}
where $k=0.4[(\alpha_{\rm{CO}}/0.1)^2 (h/0.5)^2/4]^{1/(7+2s)}$, which is 
slightly lower than the typical value $0.5-1$ \citep[e.g.][]{Ghosh79, Chashkina19},
the distinction may derive from the properties of hyper-Eddington 
accretion disk; for simplicity, we leave the specifically analysis of the
disk structure and truncation for a future work, and take $k=0.5$ briefly. 
When magnetic field is weak, the disk can extend to the NS surface
$r_{\rm{NS}}$; so in the NS accretion case, we take 
$r_{\rm{in}} = max\{r_{\rm{m}},r_{\rm{NS} },10r_g \}$ to determine the 
rate of mass eventually flowing onto the NS, where $10r_g$ is set 
due to the effects of NS's gravity as explained above (but we ignore the 
potential effects of the energy released near the NS surface on the accretion 
structure and properties, e.g. \citealt{Takahashi18}). In addition, 
the accretion flow would release additional energy with luminosity 
$L_{\rm{acc}} \simeq G \dot{M}_{\rm {in}}(r_{\rm{in}} )m_{\rm{NS}}/r_{\rm{NS}}$ 
when ultimately hitting the NS hard surface,  
where we crudely ignore the neutrino 
cooling processes around the NS surface.
So the total energy injected into the wind outflow contains the disk 
and the accretion component \citep{Chashkina19}, i.e.,
\begin{equation}
	L_{\rm{wNS}}= L_{\rm{w}} + L_{\rm{acc}}. \label{eqlns}
\end{equation}

The direction of the outflow significantly affects its feedback on the 
environment. Many numerically 
simulations and analytical solutions of the super-critical accretion 
flow around CO predicted that the emitted winds are focused in the 
high latitude region, i.e. the outflow is anisotropic 
\citep[e.g.][]{Jiao15, Sadowski152}, the feedback 
as well \citep[e.g.][]{Takeo20, Tagawa22}. These works mainly 
investigate disk parameter regions of the mass inflow rate 
$ \lesssim O(10^3)\dot{M}_{Edd}$ and the radial dimension 
$ \lesssim O(10^4)r_{g}$ (see Table 1 in \citealt{Kitaki21}, 
which lists the recent simulation studies), 
where the outflow is automatically
collimated by the geometrically thick disk and the optically 
thin polar funnel, but these parameters are 
much smaller than the case of 
CO accretion in the AGN disk. Also, when studying the feedback, 
only the kinetic energy of the outflow has been considered and the winds 
are set to spread over a limited solid angle, thereby, the 
initial setup of outflow is anisotropic. 

For the hyper-Eddington accretion process of CO, winds 
initially take along the radiation energy when just launching. 
Because the outflow and the AGN disk surroundings (the optical depth from 
the disk midplane $\tau_d \sim \kappa_{ff} \rho_d H = 1.7
\times10^4 \alpha_{-1}^{-4/5} M_{8} ^{1/5}\dot{\mathcal{M}}_{0.5}^{1/5} \gg 1$)
are both extremely dense, 
photons can not effectively diffuse away but are trapped in the outflow, 
coupled with and continually accelerating the optically thick gas 
\citep{Hashizume15,Sadowski152}.
Thus, the outflow contains not only kinetic but also thermal energy. 
We can extrapolate the properties of winds generated by the large-scale 
hyper-Eddington accretion disk from the simulations and the self-similar 
solutions shown in Appendix \ref{Appendix-disk}.
As discussed in \cite{Begelman12} 
and can be seen by Equations (\ref{inflow}) and (\ref{Lw}), 
most of the inflow mass is turned into the
outflow at large radius comparable to $ r_{\rm out}$, conversely, most 
accretion energy is released closer to $ r_{\rm in}$, carried by winds 
over a wide solid angle \citep{Sadowski161,Sadowski162}; winds 
successively loaded at different radius of the circum-CO disk would 
go through interactions and internal collisions \citep{Metzger12,Kremer19}, 
causing the gas thermalization during the outward propagation and mixture. 
So, we predict that the outflow kinetic energy is close to the 
radiation-domination thermal energy within the circum-CO disk scale.
Winds moving radially at bulk velocity $ v_{\rm w} $ would press and heat 
the circum-CO disk as well, where the thermal pressure  
$\sim \rho_{\rm w}c_{\rm s,w}^2 \sim \rho_{\rm w}v_{\rm w}^2 $.
When the sound speed of the heated disk gas exceeds 
the CO local escape velocity, the gas outside is 
ejected and hence the disk is truncated \citep[e.g.][]{Tagawa22}. The 
truncation radius $r_{\rm{tru}}$ can be estimated as
\begin{equation}
	\rho_{\rm{CCOD}}(r_{\rm{tru}} ) v_{\rm{K}}(r_{\rm{tru}} )^2 =
	\rho_{\rm w}c_{\rm s,w}^2 \sim 
	\rho_{\rm{w}}(r_{\rm{tru}} ) v_{\rm{w}}(r_{\rm{tru}} )^2,
\end{equation}
where $\rho_{\rm{CCOD}}$ is the density of circum-CO disk, whose 
expression is shown in Appendix \ref{appendix-disk-density}; 
the density of the outflow $\rho_{\rm{w}}$ is set as a simplified 
spherical distribution, i.e.,
\begin{equation}
    \rho_{\rm{w}} \simeq \frac{\dot{M}_{\rm{w}}}{4 \pi r^2 v_{\rm{w}}} 
    =\frac{\dot{M}_{\rm{in}}}{4 \pi r^2 v_{\rm{w}}}, \label{rhowind}
\end{equation}
in other words, we assume that the heavy gas would 
distribute roughly uniformly over 
a wide solid angle at large radius because of the gas thermal mixture, 
and we neglect the spatial existence of the circum-CO disk compressing the 
wind region when calculating the disk truncation, which may increase $\rho_{\rm{w}}$ 
but would not significantly affect the result because of the disk thickness 
$h \sim O(10^{-3})-O(10^{-1})$ not too large;
the velocity of the mixed disk wind can be obtained from
$\dot{M}_{\rm{w}} v_{\rm{w}}^2/2 \simeq L_{\rm{w}}/2$, i.e., 
\begin{equation}
	v_{\rm{w}}\simeq \left(\frac{ L_{\rm{w}}}{\dot{M}_{\rm{w}}} \right)^{1/2}
	=\left(\frac{ L_{\rm{w}}}{\dot{M}_{\rm{in}}} \right)^{1/2}, \label{vwi}
\end{equation}
in the NS case we replace the energy term with $L_{\rm{wNS}}$. 
After getting $r_{\rm{tru}}$, we can 
decide the disk outer boundary $r_{\rm{obd}}=
\min \{r_{\rm{cir}},r_{\rm{tru}}\}$, then 
calculate the modified $\dot{M}_{\rm{obd}}=\dot{M}_{\rm{in}}(r_{\rm{obd}})$ 
and the mass accretion rate of CO 
$\dot{M}_{\rm{CO}}=\dot{M}_{\rm{in}}(r_{\rm{in}})$ using Equations 
(\ref{inflow}) and (\ref{inflow2}). As examples shown in 
Figure \ref{Fig:maccbh}, we find that the mass accretion rate of 
CO is well hyper-Eddington when ignoring the effects of outflow feedback 
on the environment; and we confirm a comprehensible tendency that the 
mass rate increases with the decrease of index $s$, which controls 
the strength of disk wind.

Even if the outflow is 
initially anisotropic (collimated by the geometrically thick disk) 
when leaving the local circum-CO disk system, because 
\begin{equation}
\rho_{\rm w}v_{\rm w}^2 \sim \rho_{\rm w}c_{\rm s,w}^2 
\gg \rho_{\rm{CO}} \tilde{c}_{s}^2,
\end{equation}
where the double-greater-than symbol holds (we check numerically and find that 
the inequality assuredly holds for the regions 
$R_{\rm{CO}}\gtrsim O(10^2)R_g$ we interested) as 
$v_{\rm{w}}/v_{\rm{K}}(r_{\rm{obd}}) \sim f_w^{1/2}
(r_{\rm{obd}}/r_{\rm{in}})^{(1-s)/2} \gg 1$ with the 
supersonic accretion $v_{\rm{K}}(r_{\rm{obd}}) > \tilde{c}_{s}$ and 
$\rho_{\rm{w}} \sim \rho_{\rm{CO}}$ due to $\dot{M}_{\rm{w}} \sim \dot{M}_{\rm{in}}$,
the pressure of the environment gas is unable to prevent the outflow 
becoming asymptotically isotropic and spherically symmetric
(though there may still have anisotropy, with lower wind mass rate 
and larger wind velocity at higher latitude, e.g., \citealt{Kitaki21}) 
before sweeping a large amount of the 
AGN disk gas; the equatorial region's gas captured by the CO will 
also be pushed away. During the 
adiabatic expansion, most of heat in the outflow would be
converted to bulk kinetic energy and accelerate the winds with 
initial velocity of Equation (\ref{vwi}) \citep{Kremer19, Piro20}, 
so the asymptotic velocity, kinetic energy and momentum of the 
outflow are
\begin{equation}
\frac{1}{2}\dot{M}_{\rm{w}}v_{\rm{w}}^2 = L_{\rm{w}/\rm{wNS}} , \label{vwind}
\end{equation}
\begin{equation}
\dot{p}_{\rm{w}}=\dot{M}_{\rm{w}}v_{\rm{w}}  . \label{pwind}
\end{equation}

\subsection{Time Evolution of Circum-CO Disk}
For the moment, we ignore the detailed time-evolution of the circum-CO 
accretion disk, which may follow the works of \cite{Kumar08} and \cite{Shen14}, 
determining the specific evolution of mass rate accreted 
onto CO and the variation 
of outflow generated from the accretion system. For now, the structure 
of the hyper-Eddington accretion disk, much less the disk evolution, 
has not been well studied, as discussed in Appendix 
\ref{appendix-disk-density}; at the same time,
the qualitative properties of CO accretion can be 
approximatively described using the viscous timescale. So, for simplicity, we 
put aside the complicated time-evolution of the piecewise 
disk, instead we assume that the efficient accretion 
proceeds within $t_{\rm{vis}}$, after which the 
efficient accretion stops.

\section{Accretion and feedback of a CO in the AGN Disk} \label{sec:feedback}
In previous sections we have discussed the asymptotically 
isotropic properties of wind ejected from 
the circum-CO disk. Taking Equations (\ref{rhowind}) and 
(\ref{vwind}) as the initial injection, we study the interaction 
between outflow and AGN disk 
and its resultant effects on the CO accretion and its surroundings.

\begin{figure*}
	\begin{center}
		\includegraphics[width=0.32\textwidth]{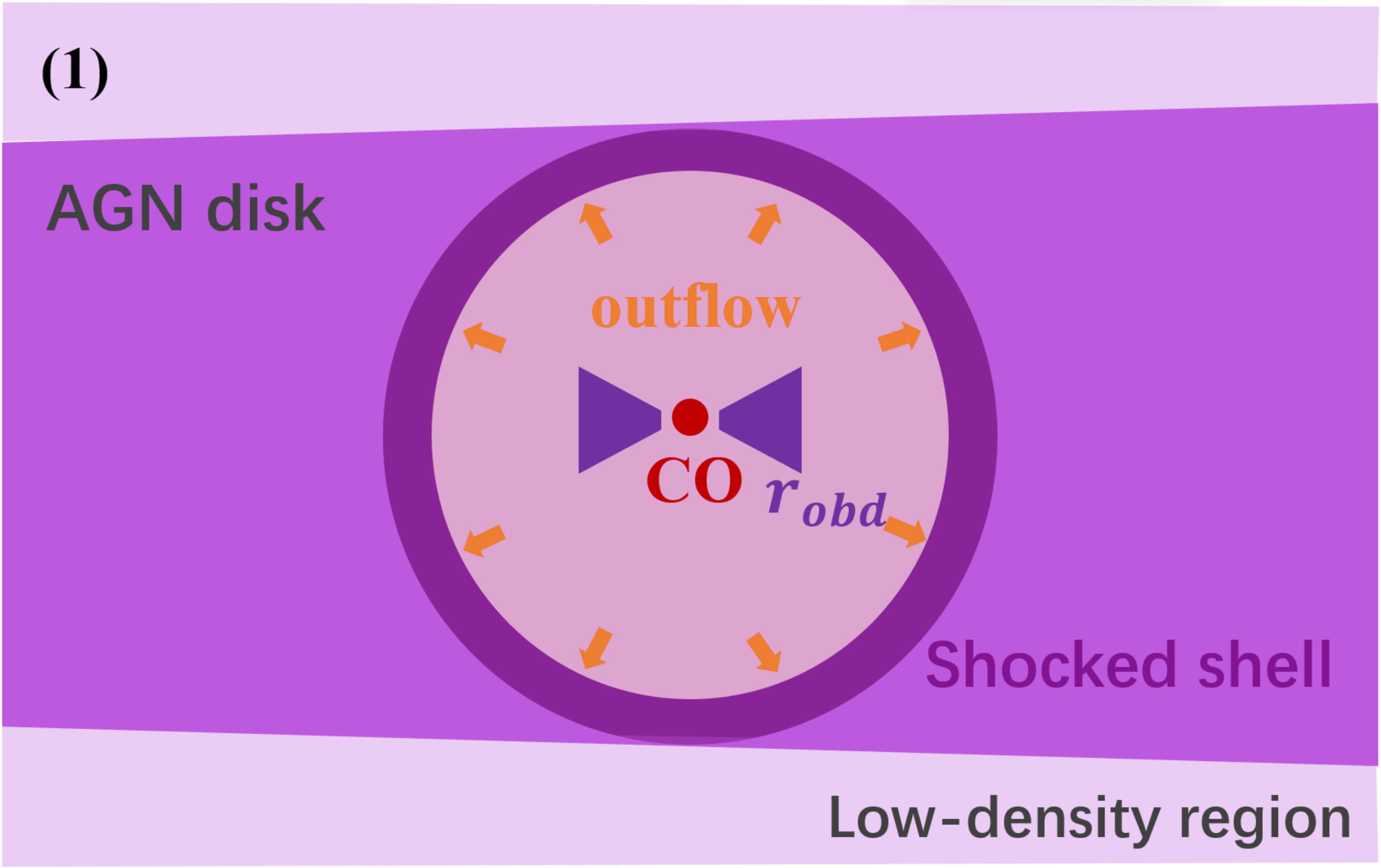}
		\includegraphics[width=0.318\textwidth]{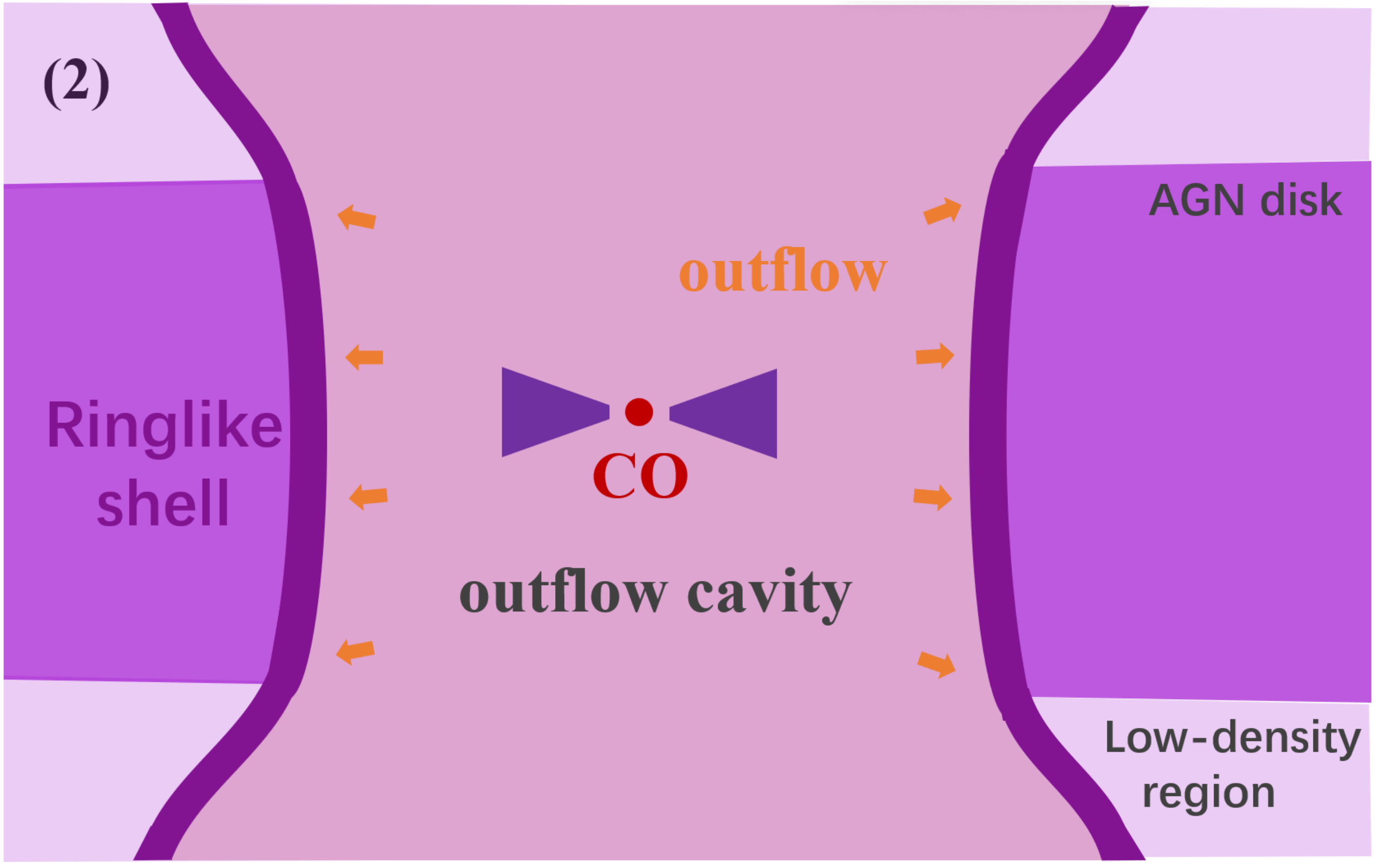}
		\includegraphics[width=0.318\textwidth]{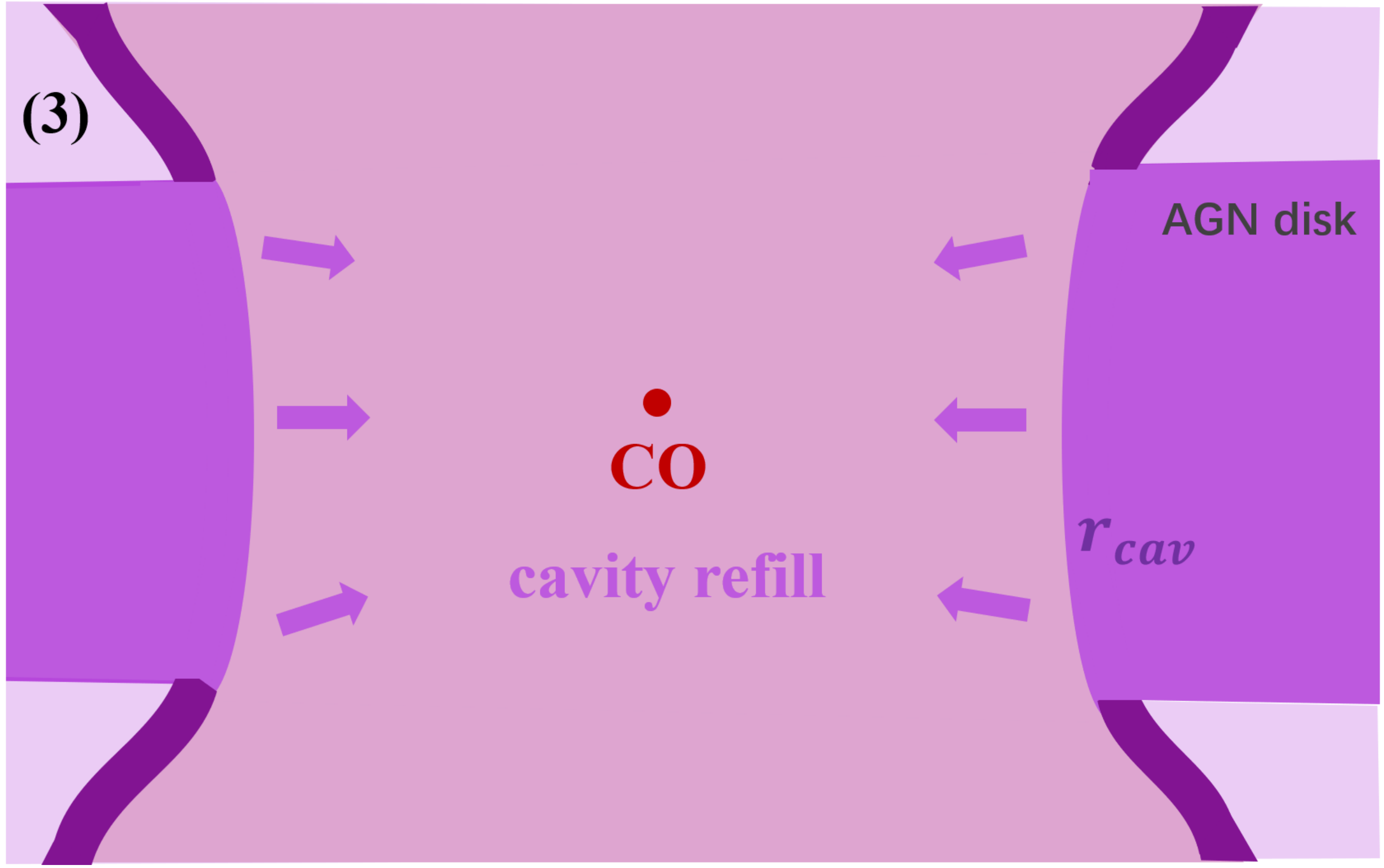}
	\end{center}
	\caption{Schematic diagram of the outflow ejected 
		from the circum-CO disk evolving in the AGN disk. (1) The 
		asymptotically isotropic outflow interacts
	    with the AGN disk gas to form a shocked shell; the shell expands 
	    to break the AGN disk and punches a cavity. (2) Outflow 
        persistently pushes the ringlike shell along AGN disk on CO 
        accretion timescale, then the shell undergoes a momentum 
        conservation snowplow phase. (3) After the shell becomes 
        transonic, the surrounding gas instead resumes refilling the 
        low-density cavity. The whole evolution processes 
        proceed circularly.}	
	\label{Fig:sch}
\end{figure*}

\subsection{Evolution of an Outflow in the AGN Disk} \label{3-1}
In analogy with the interaction between the stellar wind and 
interstellar medium \citep{Weaver77},
the wind from the circum-CO disk interacts with the surrounding AGN 
disk gas via a forward shock, forming a shell of shocked disk gas 
separated by the shocked outflow material; 
inside the shell, a low-density outflow cavity forms. We 
ignore the radial and vertical structure of the AGN disk, and
roughly set the disk with uniform density $\rho_{\rm{d}}$ 
and a concrete height $H$ at each $R_{\rm{CO}}$. A schematic diagram 
of outflow evolution in the AGN disk is shown in Figure \ref{Fig:sch}.

During the efficient accretion of CO sustaining timescale 
of $t_{\rm{acc}}=t_{\rm{vis}}(r_{\rm{obd}} )=\alpha_{\rm{CO}}^{-1} h^{-2} \Omega_{\rm{K}}^{-1}$, 
we approximately assume the properties of winds remain unchanged. 
The radius and velocity of the expanding shell evolve over 
time as \citep{Weaver77}:
\begin{equation}
r_{\rm{shell}}=0.88\left(\frac{L_{\rm{w}} t^{3}}{\rho_{\rm{d}}} \right)^{1/5} , \label{eqrshell}
\end{equation}
and
\begin{equation}
v_{\rm{shell}}=0.53\left(\frac{L_{\rm{w}}}{\rho_{\rm{d}}t^{2}} \right)^{1/5} . \label{eqvshell}
\end{equation}

When $r_{\rm{shell}}(t_{\rm{bre}} )=H$, 
the shell will vertically break out from the AGN disk and 
propagate in the above lower density region (e.g. broad-line region); 
meanwhile, along the AGN disk, the outflow continually interacts and 
pushes the shocked gas shell laterally, which now shows a ringlike structure, 
until the efficient accretion stops. Because the vertical expansion of the shell
is much more rapid
($\rho_{\rm{BLR}}\sim 10^{-17}\g\cm^{-3} \ll \rho_{\rm{d}}$), 
we assume that the shocked outflow undergoes an overall rapid 
depressurization after the breakout \citep[e.g.][]{Schiano85}
\footnote{\cite{Moranchel21} studied the supernova 
explosions in the AGN disk, and found that the analytic evolution 
(vertically and laterally) 
of the SNe-driven bubble under the assumption of bubble
receiving rapid depressurization after the breakout matches 
well with the numerical simulations.}, and then the ring moves
driven by the momentum of the long-lasting outflow 
($t_{\rm{acc}} \gg t_{\rm{bre}}$ holds within wide parameter regions, as 
shown in Figure \ref{Fig:cav}), i.e., 
\begin{equation}
\frac{d}{dt} \left(2 \pi H r_{\rm{shellb}}^2 
\rho_{\rm{d}} \dot{r}_{\rm{shellb}} \right) \simeq
4 \pi H r_{\rm{shellb}} \rho_{\rm{w}} v_{\rm{w}}^2, \label{eqmo}
\end{equation}
where we roughly assume that the ring is cylindrical 
and only moves laterally. Because
$\rho_{\rm{w}}\propto r_{\rm{shellb}}^{-2}$, the ringlike shell
evolves after the breakout as:
\begin{equation}
	r_{\rm{shellb}} \simeq r_{\rm{shell}}(t_{\rm{bre}})
	\left( \frac{t}{t_{\rm{bre}}} \right)^{1/2}, 
\end{equation}
and
\begin{equation}
	v_{\rm{shellb}} \simeq v_{\rm{shell}}(t_{\rm{bre}})
	\left( \frac{t}{t_{\rm{bre}}} \right)^{-1/2},
\end{equation}
where we simply ignore the specific depressurization process and link 
the outflow-energy and outflow-momentum driven phase directly. 
Lastly, when the CO accretion stops, the evolution of 
the ringlike shell along the AGN disk translates from the outflow-momentum 
driven phase to its own momentum conservation snowplow phase, 
i.e.,
\begin{equation}
r_{\rm{sh}}^{2} \dot{r}_{\rm{sh}} \simeq r_{\rm{shellb}}^{2}(t_{\rm{acc}} ) 
v_{\rm{shellb}}(t_{\rm{acc}} ) , \label{eqm}
\end{equation}
where $r_{\rm{sh}}$ and $\dot{r}_{\rm{sh}}$ represent the radius and 
velocity of shell during the snowplow phase; on both sides 
of the equation we have eliminated $\pi \Sigma_{\rm{d}}=2\pi\rho_{\rm{d}}H$, 
which approximatively represents the mass of the swept disk gas 
$\pi r_{\rm{sh}}^{2} \Sigma_{\rm{d}}$. With the continuous sweep, 
the shell will decelerate to a velocity similar to 
the environment sound speed, i.e., 
$\dot{r}_{\rm{sh}} \sim \tilde{c}_{\rm{s}}$, 
which ends up being transonic and is 
unable to propagate further as a shock. So the horizontal half-width of the 
outflow cavity in the AGN disk can be estimated by 
\begin{equation}
r_{\rm{cav}}= \left[\frac{r_{\rm{shellb}}^{2}(t_{\rm{acc}}) v_{\rm{shellb}}(t_{\rm{acc}} )}{\tilde{c}_{\rm{s}}} \right]^{1/2} .\label{eqrcav}
\end{equation} 
The timescale of the cavity formation can be calculated from Equation 
(\ref{eqm}), i.e., 
\begin{equation}
t_{\rm{cav}}= \frac{r_{\rm{cav}}^{3}-r_{\rm{shellb}}^{3} (t_{\rm{acc}} ) }
{3r_{\rm{shellb}}^{2}( t_{\rm{acc}} ) 
v_{\rm{shellb}}( t_{\rm{acc}} )} + t_{\rm{acc}} ,\label{eqtcav}
\end{equation} 
which is the sum of the efficient accretion timescale and the 
evolution timescale of the subsequent snowplow phase.

But if $t_{\rm{bre}}>t_{\rm{acc}}$, the CO would stop strong 
accretion before the successful shell breakout. In this case 
we roughly take the whole energy of the CO-accretion-generated 
winds, i.e., $E_{\rm{w}}=L_{\rm{w}}t_{\rm{acc}}$, 
as injection; the shell thus 
expands driven by $E_{\rm{w}}$, the radius and velocity 
are given as \citep[e.g.][]{Ostriker88}:
\begin{equation}
r_{\rm{shellE}}=\left(\frac{E_{\rm{w}} t^{2}}{\rho_{\rm{d}}} \right)^{1/5} , \label{eqrshellE}
\end{equation}
and
\begin{equation}
v_{\rm{shellE}}=0.4\left(\frac{E_{\rm{w}}}{\rho_{\rm{d}}t^{3}} \right)^{1/5} , \label{eqvshellE}
\end{equation}
where we ignore the mass of wind for simplicity. We can also achieve 
the breakout time from  $r_{\rm{shellE}}(t_{\rm{breE}} )=H$. After 
the breakout accompanied with the rapid cavity depressurization, the shell 
propagating along the AGN disk enters the snowplow phase, i.e., 
\begin{equation}
r_{\rm{shE}}^{2} \dot{r}_{\rm{shE}} \simeq r_{\rm{shellE}}^{2}(t_{\rm{breE}} ) 
v_{\rm{shellE}}(t_{\rm{breE}} ) , \label{eqmE}
\end{equation} 
the half-width $r_{\rm{cavE}}$ and the formation timescale 
$t_{\rm{cavE}}$ of the cavity can be solved similar to Equation 
(\ref{eqrcav}) and (\ref{eqtcav}).

The accretion rate of CO in the underdense cavity is too low to
continuously produce powerful outflow maintaining the cavity structure 
(see below); conversely the relatively dense AGN disk gas will refill 
the cavity roughly on a timescale given by \citep[e.g.][]{Wang21a}:
\begin{equation}
t_{\rm{ref}}=
\begin{cases}
r_{\rm{cav}}/\tilde{c}_{\rm{s}}=
[r_{\rm{shellb}}^{2}(t_{\rm{acc}} ) v_{\rm{shellb}}(t_{\rm{acc}} )/
\tilde{c}_{\rm{s}}^3]^{1/2}, \\ 
\qquad \qquad \qquad \qquad \qquad \qquad \qquad t_{\rm{bre}}<t_{\rm{acc}} \\ 
r_{\rm{cavE}}/\tilde{c}_{\rm{s}}=
[r_{\rm{shellE}}^{2}(t_{\rm{breE}} ) v_{\rm{shellE}}(t_{\rm{breE}} )/
\tilde{c}_{\rm{s}}^3]^{1/2}. \\ 
\qquad \qquad \qquad \qquad \qquad \qquad \qquad t_{\rm{bre}}>t_{\rm{acc}}
\end{cases} \label{eqref}
\end{equation}
Then the efficient CO accretion will be activated in the refilled 
cavity again; subsequently the cavity formation and refilling would 
take place alternately.
\subsection{Reduced Mass Accretion of BH} \label{3-2}
\begin{figure*}
	\begin{center}
		\includegraphics[width=0.31\textwidth]{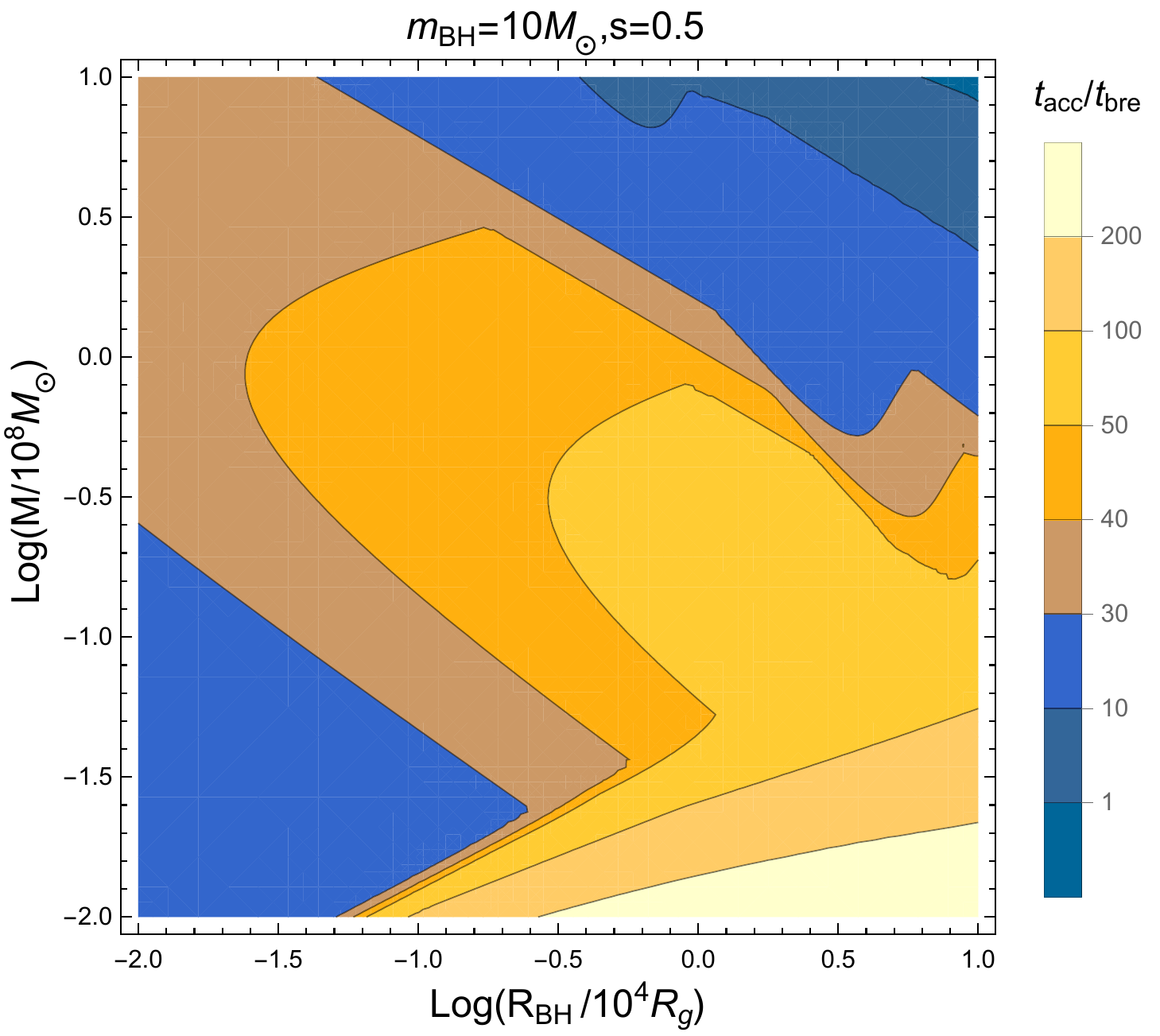}
		\includegraphics[width=0.32\textwidth]{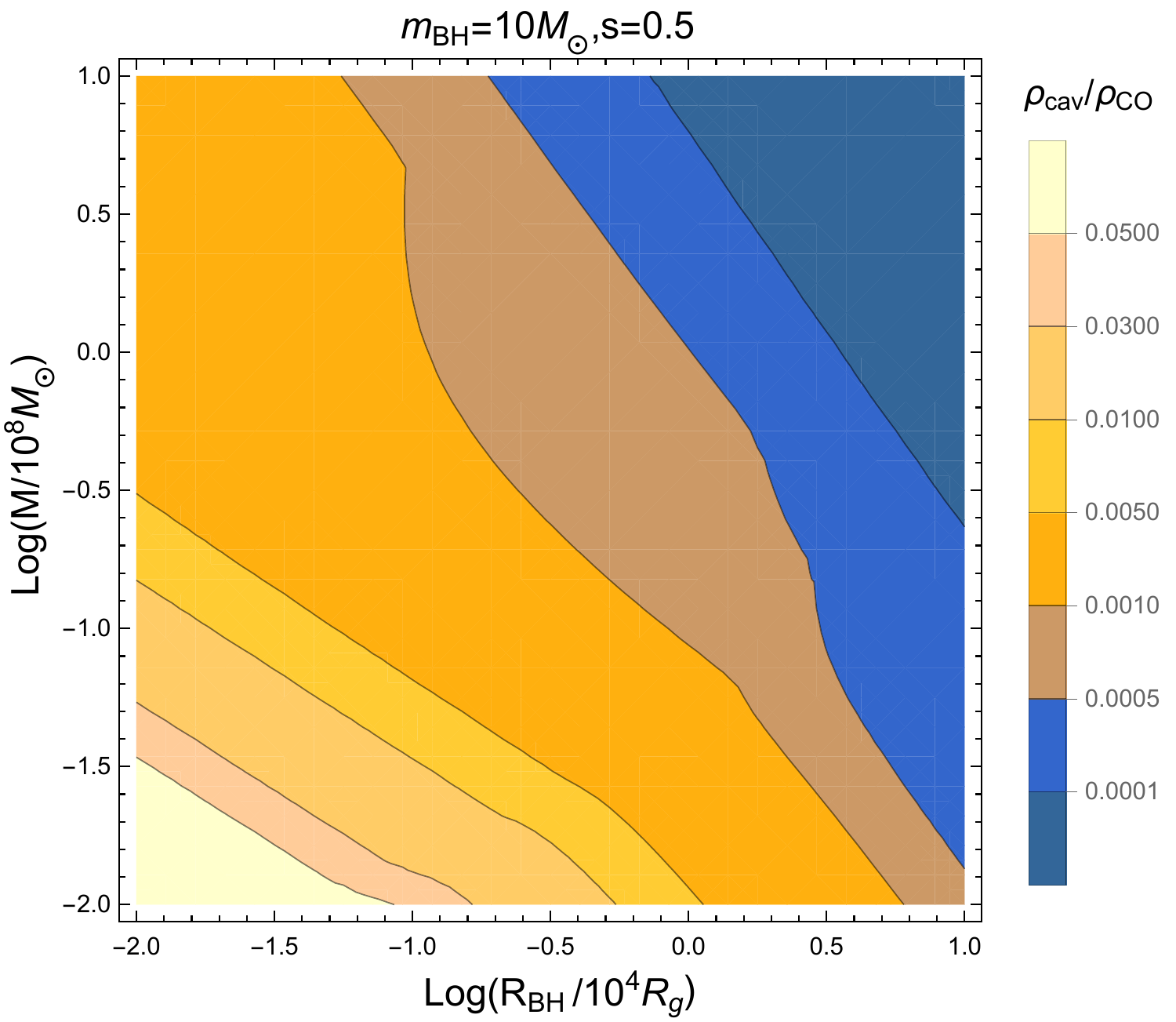}
		\includegraphics[width=0.32\textwidth]{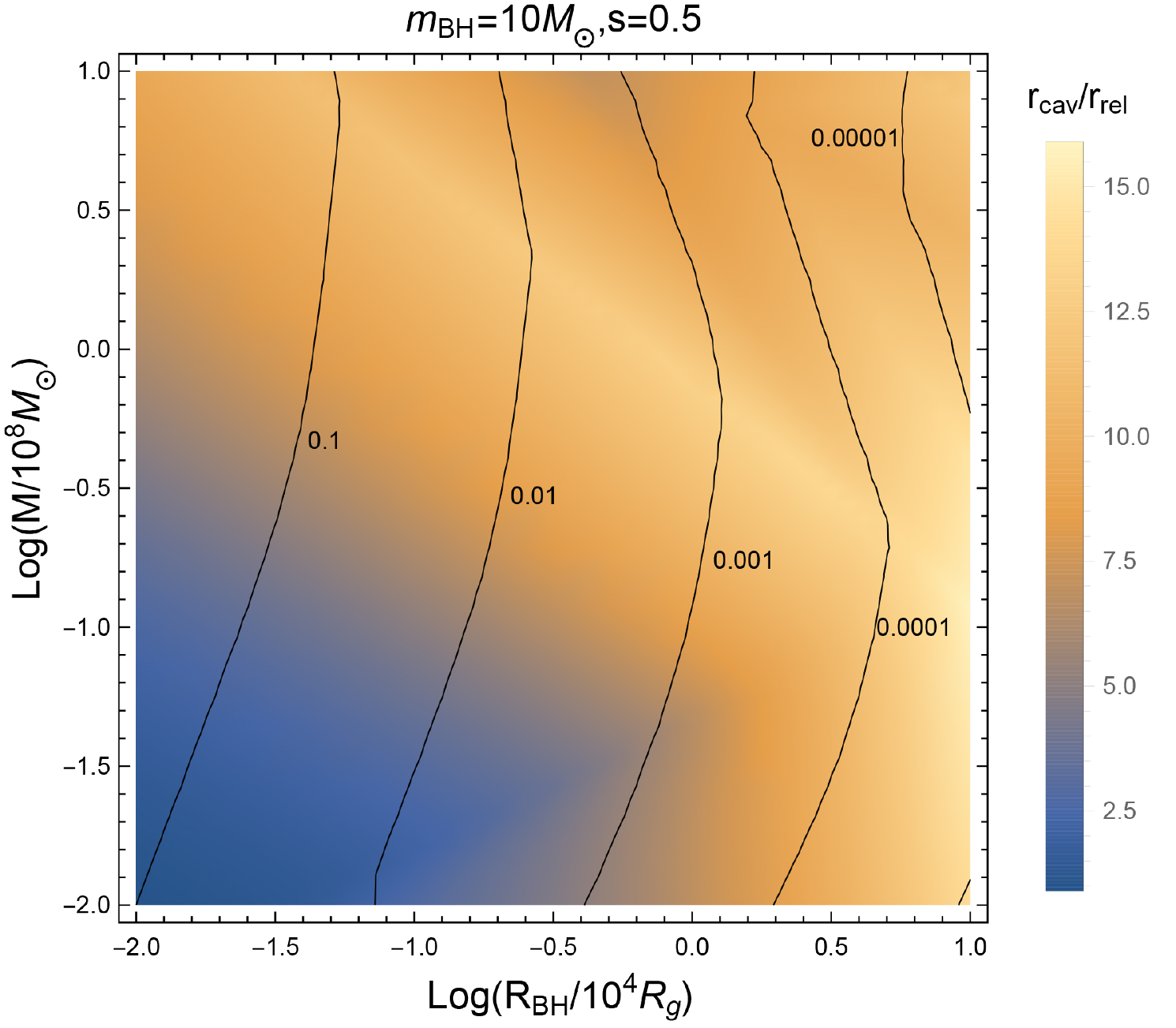}
	\end{center}
	\caption{Properties of the outflow feedback cavity for the 
		$m_{\rm{BH}}=10M_{\odot}$, $s=0.5$ case.
		Left panel shows the comparison between the BH accretion timescale
		and the shell breakout timescale $t_{\rm{acc}}/t_{\rm{bre}}$,
		with $t_{\rm{acc}} \gg t_{\rm{bre}}$ over wide parameter
		regions. Middle panel shows 
		the comparison of density between the cavity and the initial 
		environment $\rho_{\rm{cav}}/\rho_{\rm{CO}}$, indicating that
		the gas density surrounding the BH is significantly reduced. 
		Right panel shows the size of the cavity 
		$r_{\rm{cav}}/r_{\rm{rel}}$, 
		which is much larger than the BHL and Hill radius of the 
		BH; and shows the trapping radius $r_{\rm{tr,cav}}/r_{\rm{obd}}$ 
		of the circum-BH disk accreting the cavity gas by contour-lines, 
		$r_{\rm{tr,cav}}\ll r_{\rm{obd}}$
		implies a significant reduction of BH mass accretion rate 
		in the low-density cavity.}
	\label{Fig:cav}
\end{figure*}

We take $m_{\rm{BH}}=10M_{\odot}$ and $s=0.5$ as a typical case to study 
the influence of outflow feedback on the BH accretion in the AGN disk 
environment, of which the specific processes have been discussed in 
Section \ref{3-1}. We can calculate the 
$r_{\rm{cav}}$(or $r_{\rm{cavE}}$) under variable AGN disk parameters, 
and then estimate the density of outflow cavity as 
\begin{equation}
	\rho_{\rm{cav}}=\frac{M_{\rm{wind}}}{V_{\rm{cav}}}=
	\frac{\dot{M}_{\rm{w}} t_{\rm{acc}}}{4/3 \pi r_{\rm{cav}}^3} ,
\end{equation}
where we treat the cavity as spherical, which overestimates the 
density of cavity because the shell expands much 
faster above the AGN disk causing a much larger volume. 
Then we use Equation (\ref{obdinflow}) 
to get the mass accretion rate of BH in the cavity, 
replacing $\rho_{\rm{CO}}$ with $\rho_{\rm{cav}}$. 
Relevant properties of the cavity and BH accretion are shown in 
Figure \ref{Fig:cav}. We find that the size of outflow-induced 
cavity is much larger than the BHL and Hill radius, i.e., 
the size of BH gravity sphere, which suppresses 
the BH further capturing the AGN disk gas.
The density of cavity is significantly reduced compared to the 
initial unperturbed environment.
We calculate the trapping radius $r_{\rm{tr,cav}}$ of BH 
accretion in the cavity using Equation (\ref{rtr}), 
and find that it is well smaller than the circum-BH disk outer boundary, 
reflecting that the BH mass accretion rate is significantly reduced, 
and hence the induced outflows are not 
powerful enough to notably affect or maintain the 
cavity structure. For the weak accretion in cavity, 
we simply neglect the specific evolution of the BH mass 
accretion rate and set a factitious value 
of $\dot{M}_{\rm{BH,cav}}=10\dot{M}_{\rm{Edd}}$, which is 
the approximate minimum mass inflow rate at the inner boundary of the 
advection-dominated disk \citep[e.g.][]{Pan21b}.

\begin{figure*}
	\begin{center}
		\includegraphics[width=0.32\textwidth]{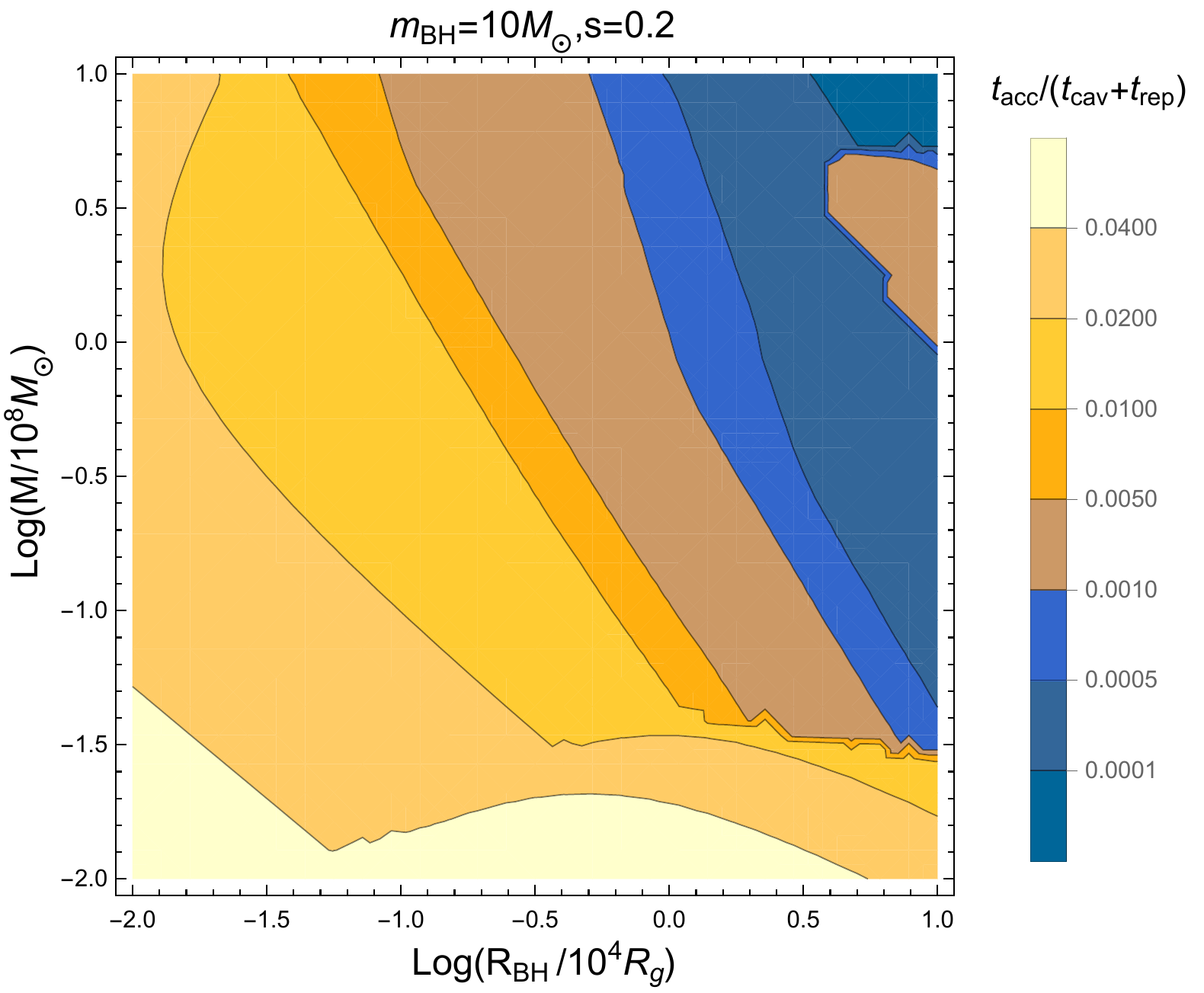}
		\includegraphics[width=0.32\textwidth]{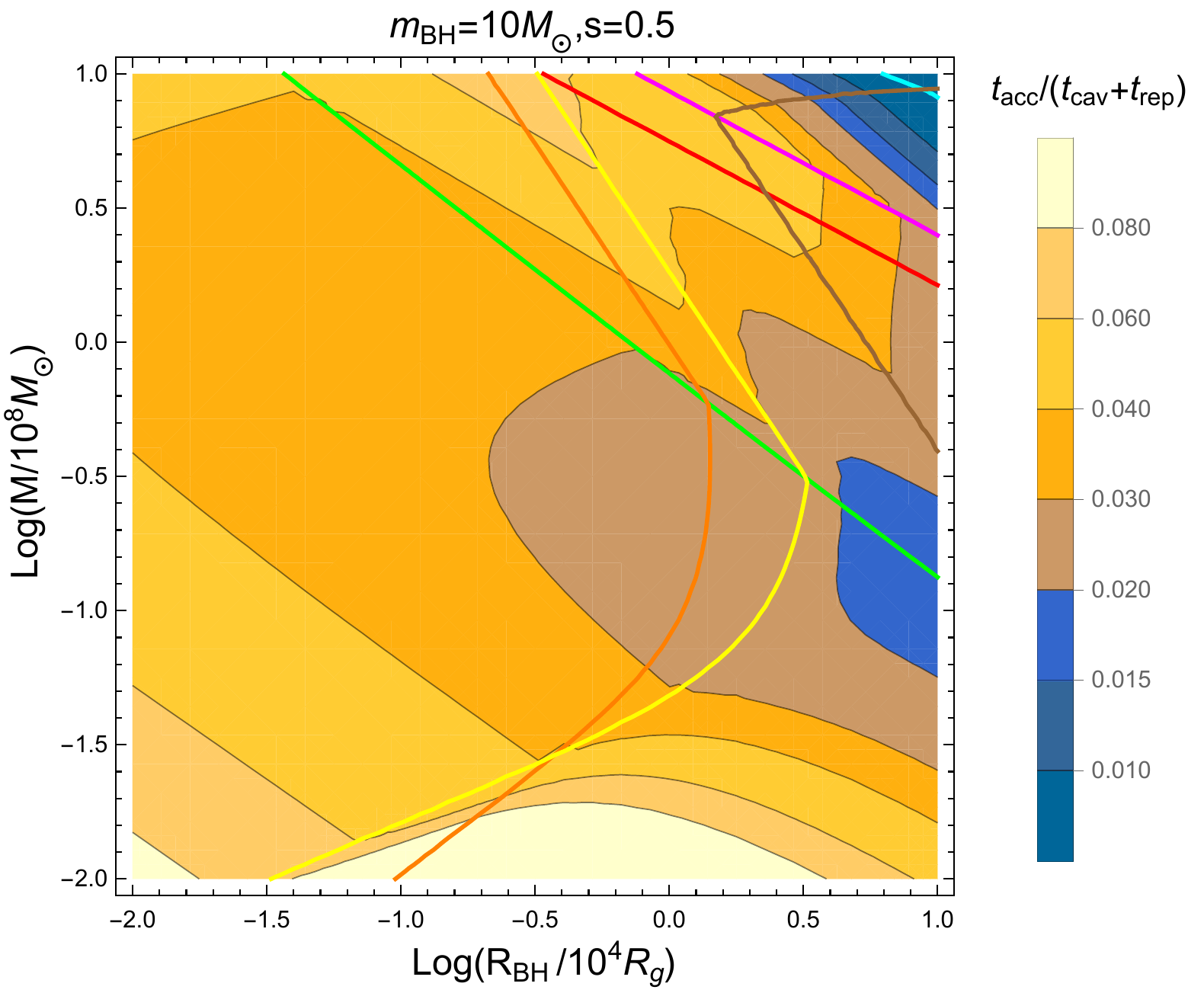}
		\includegraphics[width=0.32\textwidth]{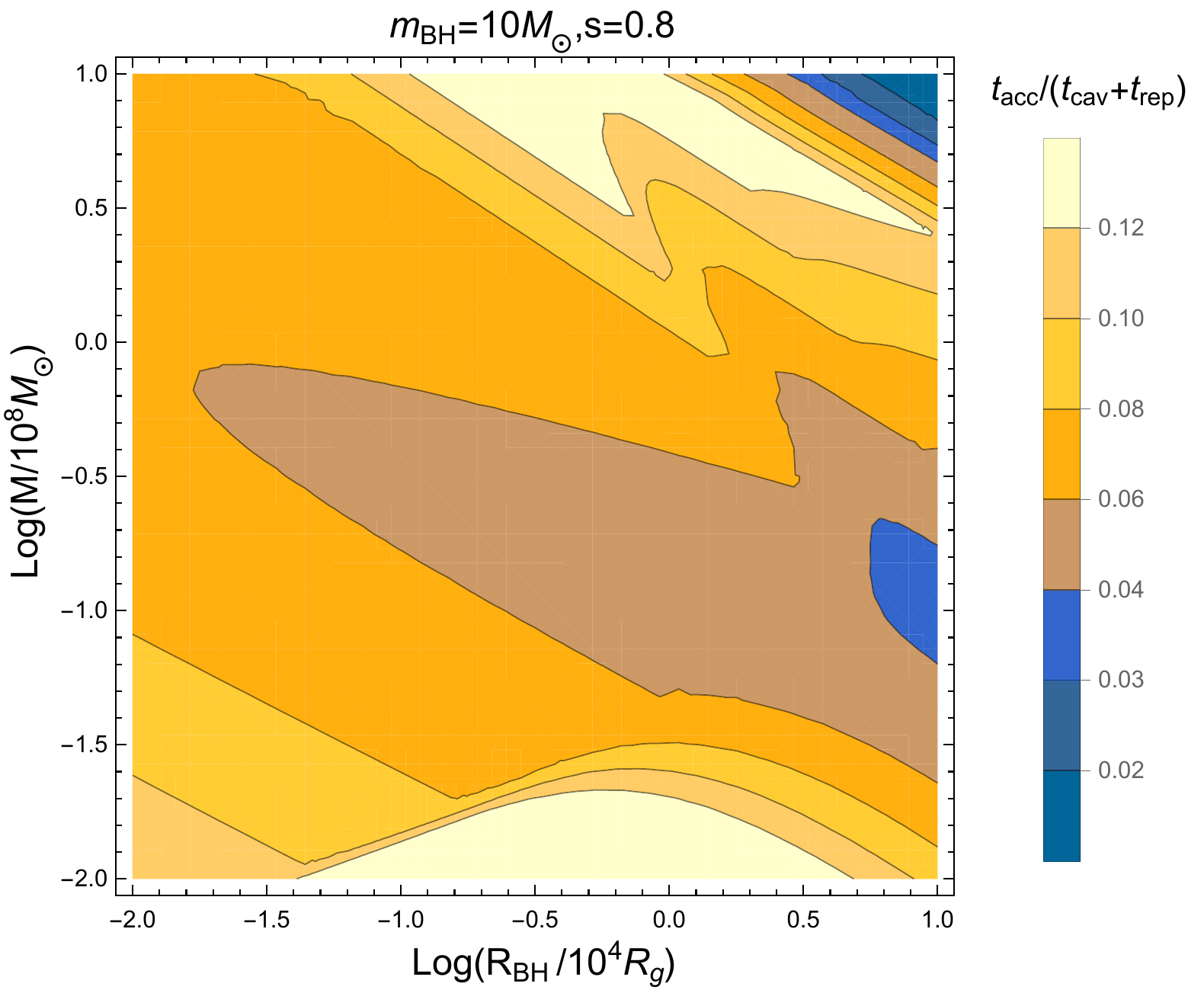}
		\includegraphics[width=0.32\textwidth]{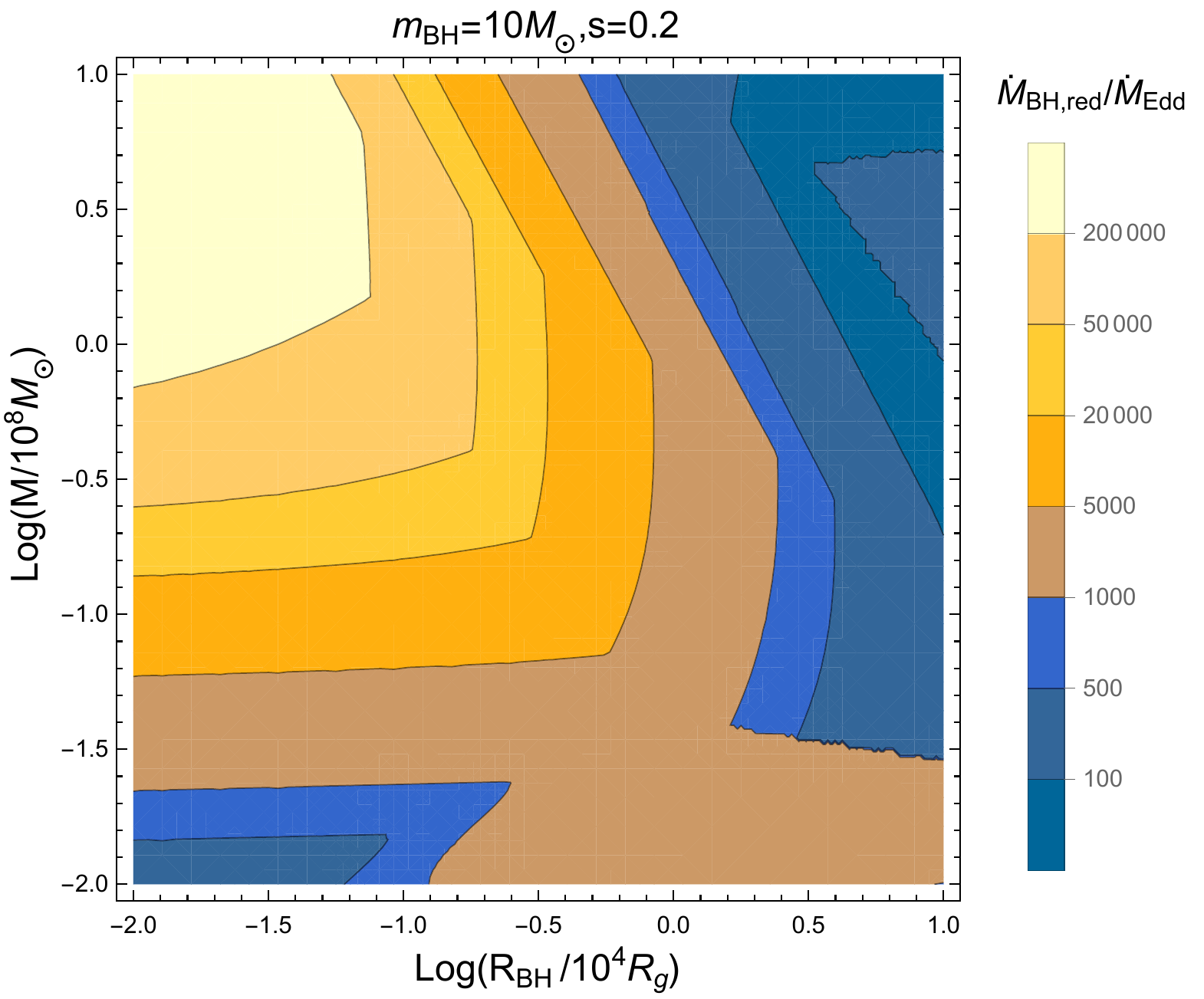}
		\includegraphics[width=0.32\textwidth]{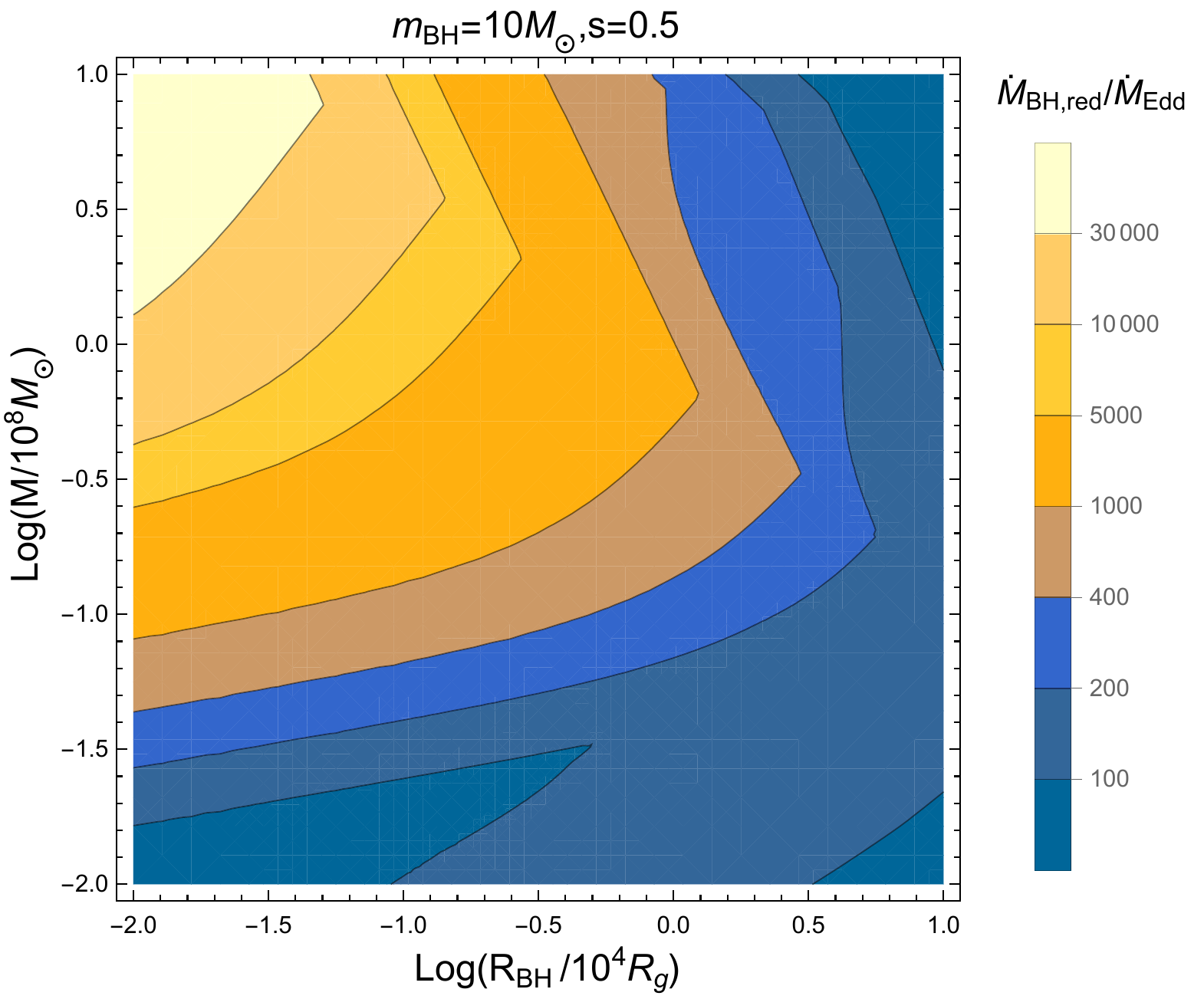}
		\includegraphics[width=0.32\textwidth]{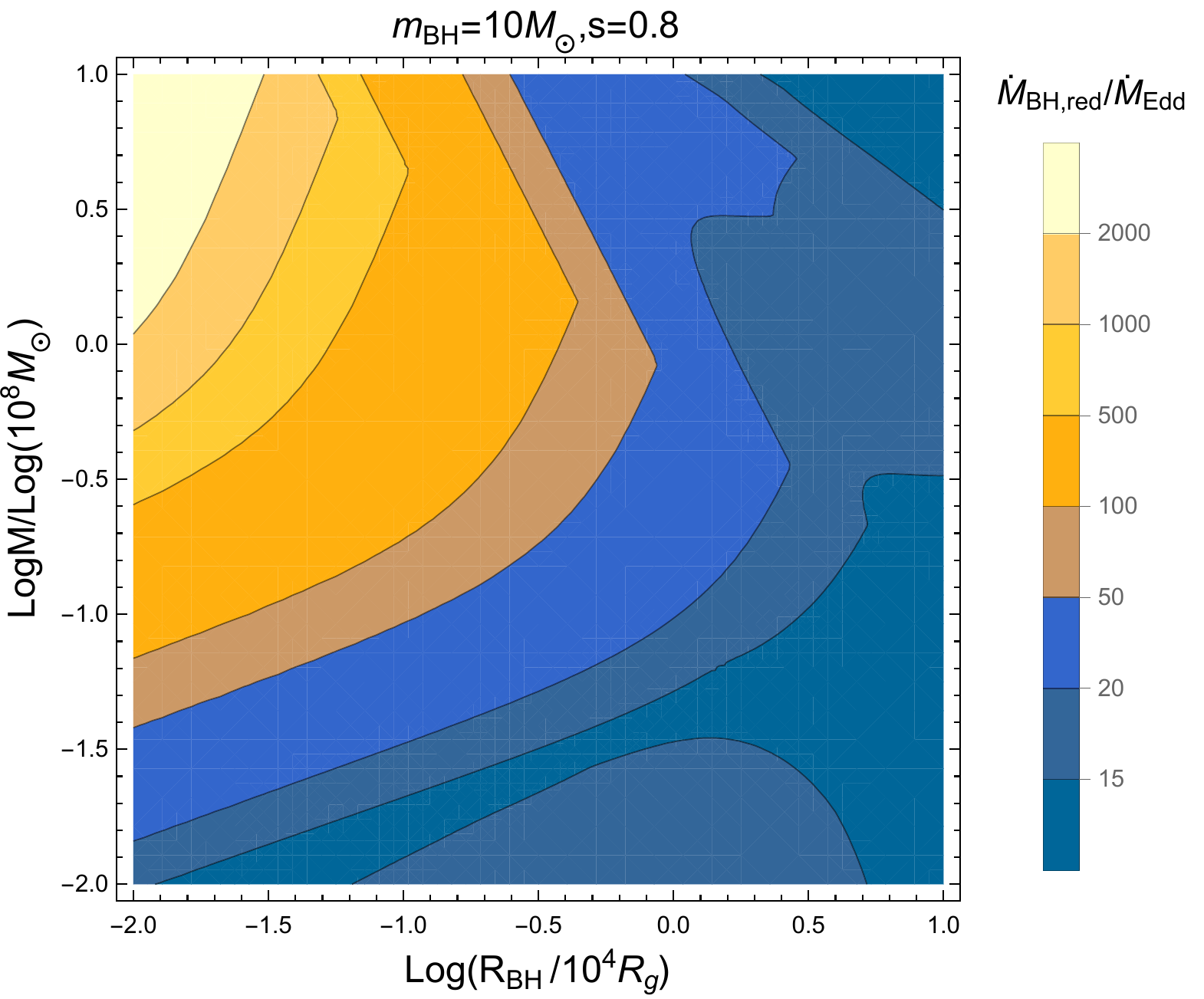}
	\end{center}
	\caption{Comparison between the BH accretion timescale and the 
	cavity evolution timescale, and the corresponding reduced BH 
    mass accretion rate. Top rows show  
    $t_{\rm{acc}}/(t_{\rm{cav}}+t_{\rm{ref}} )$ and bottom rows 
    show  $\dot{M}_{\rm{BH,red}}/ \dot{M}_{\rm{Edd}}$.
    Three columns show the 
    $s=0.2$, $s=0.5$, $s=0.8$ cases with
    $m_{\rm{BH}}=10M_{\odot}$, respectively. In the top-middle 
    panel, the magenta, red and green lines represent the same 
    boundaries as in Figure \ref{Fig:Mobd};
    the yellow, orange, brown lines and left regions 
    represent the boundaries of 
    $r_{\rm{obd}} \leqslant r_{\rm{tr}}$, 
    $r_{\rm{obd}} \leqslant r_{\rm{rad-gas}}$
    and $r_{\rm{obd}} \leqslant r_{\rm{es-ff}}$, respectively; 
    the cyan line and above region represent 
    the boundary of $t_{\rm{acc}} \leqslant t_{bre}$.}
	\label{Fig:reduced}
\end{figure*}

The timescale $t_{\rm{acc}}$ of efficient accretion is 
well shorter than the duration time $t_{\rm{cav}}+t_{\rm{ref}}$ 
of the cavity evolution, as shown in Figure \ref{Fig:reduced}.
A smaller index $s$ results in a overall more powerful 
outflow feedback, accordingly larger size and longer 
duration time of the cavity.
The variation trend of the contour map relies on various effects 
($s=0.5$ case as an example in Figure \ref{Fig:reduced}); 
first, a general trend of $t_{\rm{acc}}/(t_{\rm{cav}}+t_{\rm{ref}} )$ 
being larger for smaller BH location $R_{\rm{BH}}$ 
comes from the AGN disk gas environment being denser when closer 
to SMBH, which hinders the expansion of cavity;
second, the varied outer boundary radius of the circum-BH disk results in 
different features of the major disk region, 
thereby different dependence of the disk accretion timescale variation on 
the AGN disk parameters; third, the gap-depth is larger for lighter 
and closer SMBH cases, or the self-gravity instability in circum-BH disk 
inhibits the mass inflow rate, leading to lower mass accreted onto BH and 
weaker outflow feedback, accordingly, the size and evolution timescale of 
the cavity are smaller; in addition, the relative size of BHL radius, Hill 
radius and AGN disk height, and the unsuccessful breakout also affect 
the accretion and feedback properties of the BH.

Because the accretion of BH is inefficient in the outflow 
cavity, though hyper-Eddington during $t_{\rm{acc}}$, 
the averaged mass accretion rate onto the BH is well reduced in 
a whole circulation $t_{\rm{cav}}+t_{\rm{ref}}$. 
We can estimate the reduced mass accretion rate as: 
\begin{equation}
	\dot{M}_{\rm{BH,red}}=
	\frac{\dot{M}_{\rm{in}}(r_{\rm{in}} ) t_{\rm{acc}}
		+\dot{M}_{\rm{BH,cav}}(t_{\rm{cav}}-t_{\rm{acc}}+t_{\rm{ref}} )}
	{t_{\rm{cav}}+t_{\rm{ref}}} , \label{mbhred}
\end{equation}
some specific cases are shown in Figure 
\ref{Fig:reduced}. Indeed, the BH accretion rate receives a markedly 
overall reduction, which can be directly indicated by comparing 
the cases ignoring circum-BH disk outflow feedback as
shown in Figure \ref{Fig:maccbh}. The general variation trend 
of the reduced mass rate depending on the AGN disk parameters
has no significant change from the omitting-feedback cases. 
The reduction effect is more prominent for smaller index $s$, which produces
more powerful disk winds; but meanwhile, the initial mass accretion 
rate is higher, the combined effects bring about a relatively larger 
mass rate compared with the larger $s$ cases.

In a word, the circum-BH disk winds significantly impact the accretion 
processes of BH located in the AGN disk, the mass accretion 
rate of BH is markedly reduced, and the BH's surroundings 
is observably changed in the form of a long-standing, 
underdense cavity, or to say, feedback of the outflow 
generated by the hyper-Eddington disk can not be neglected. 
Moreover, the mass rate accreted onto BH relies heavily 
on parameters $M$ and $R_{\rm{BH}}$, and thus varies during 
the BH evolution, rather than remains as a fixed value.

\subsection{Reduced Mass Accretion of NS}
Compared to the BH accretion case, the NS accretion is relatively 
more efficient because the magnetosphere potentially truncates the 
circum-NS disk at a larger radius $r_{\rm{m}}$, Equation (\ref{eqrm}), to bring 
more mass accreted onto NS; meanwhile, feedback from the NS 
accretion system is more powerful because 
the existence of hard surface releases additional energy, 
Equation (\ref{eqlns}), to inject into the outflow.

\begin{figure*}
	\begin{center}
		\includegraphics[width=0.32\textwidth]{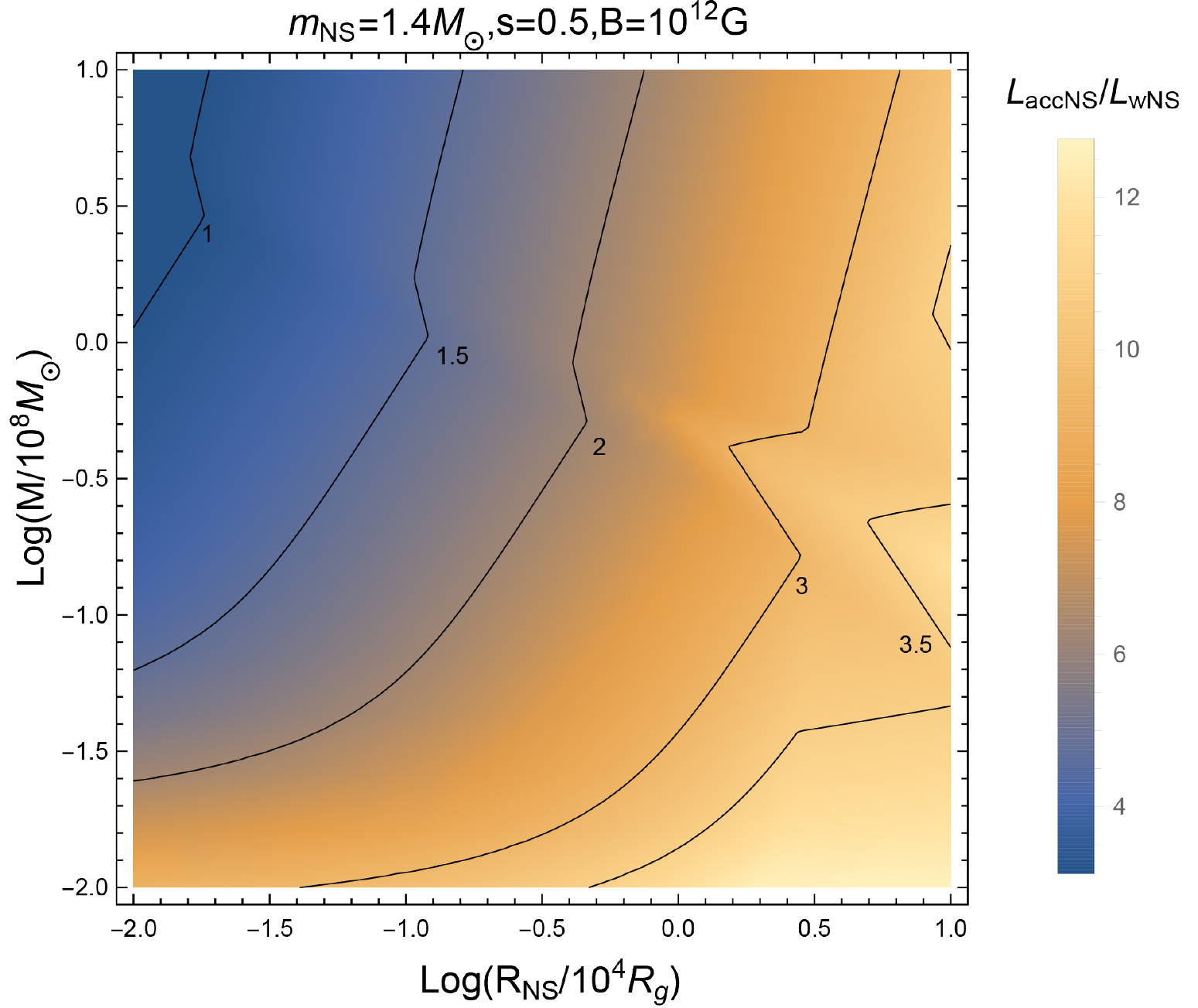}
		\includegraphics[width=0.32\textwidth]{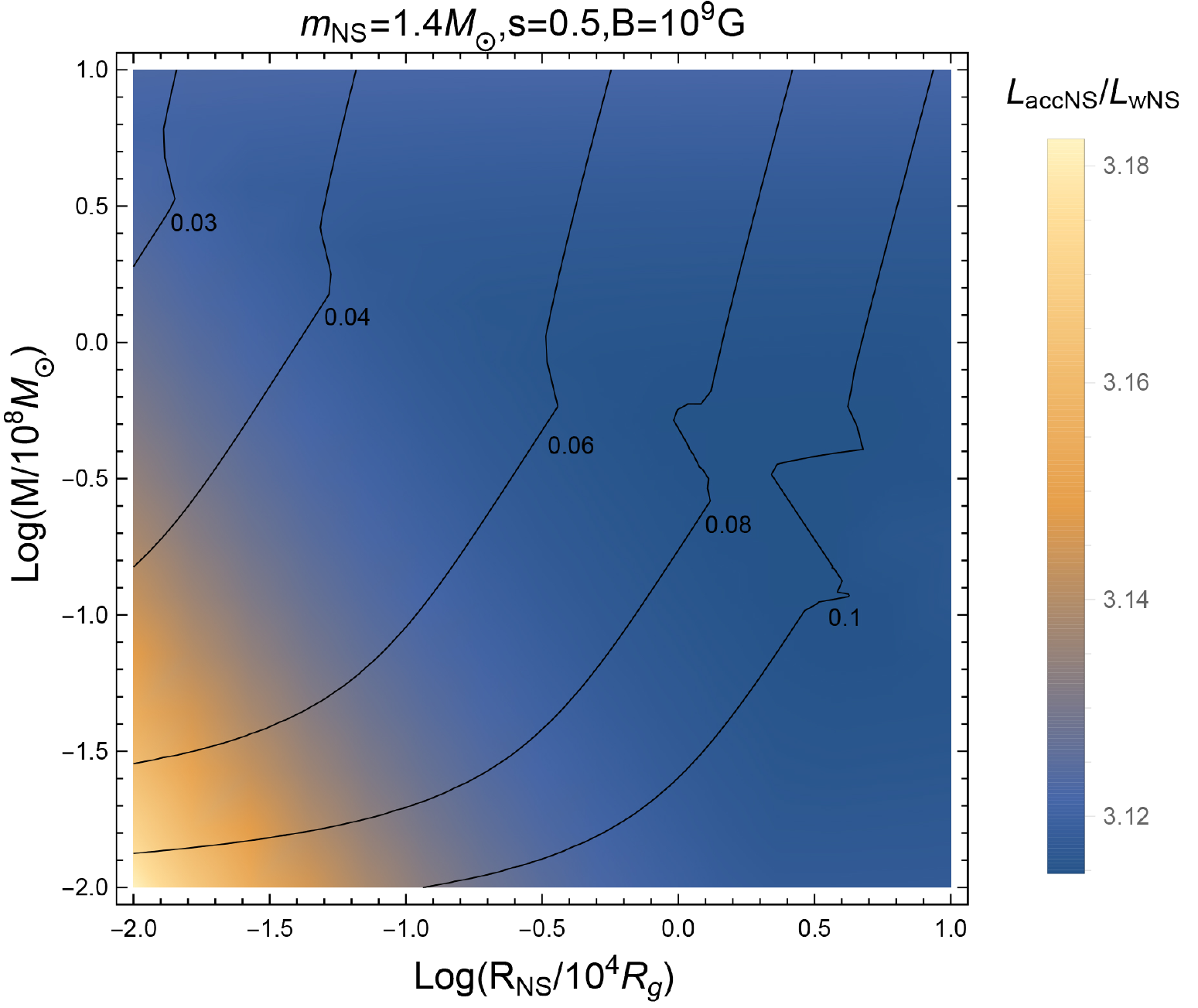}
		\includegraphics[width=0.32\textwidth]{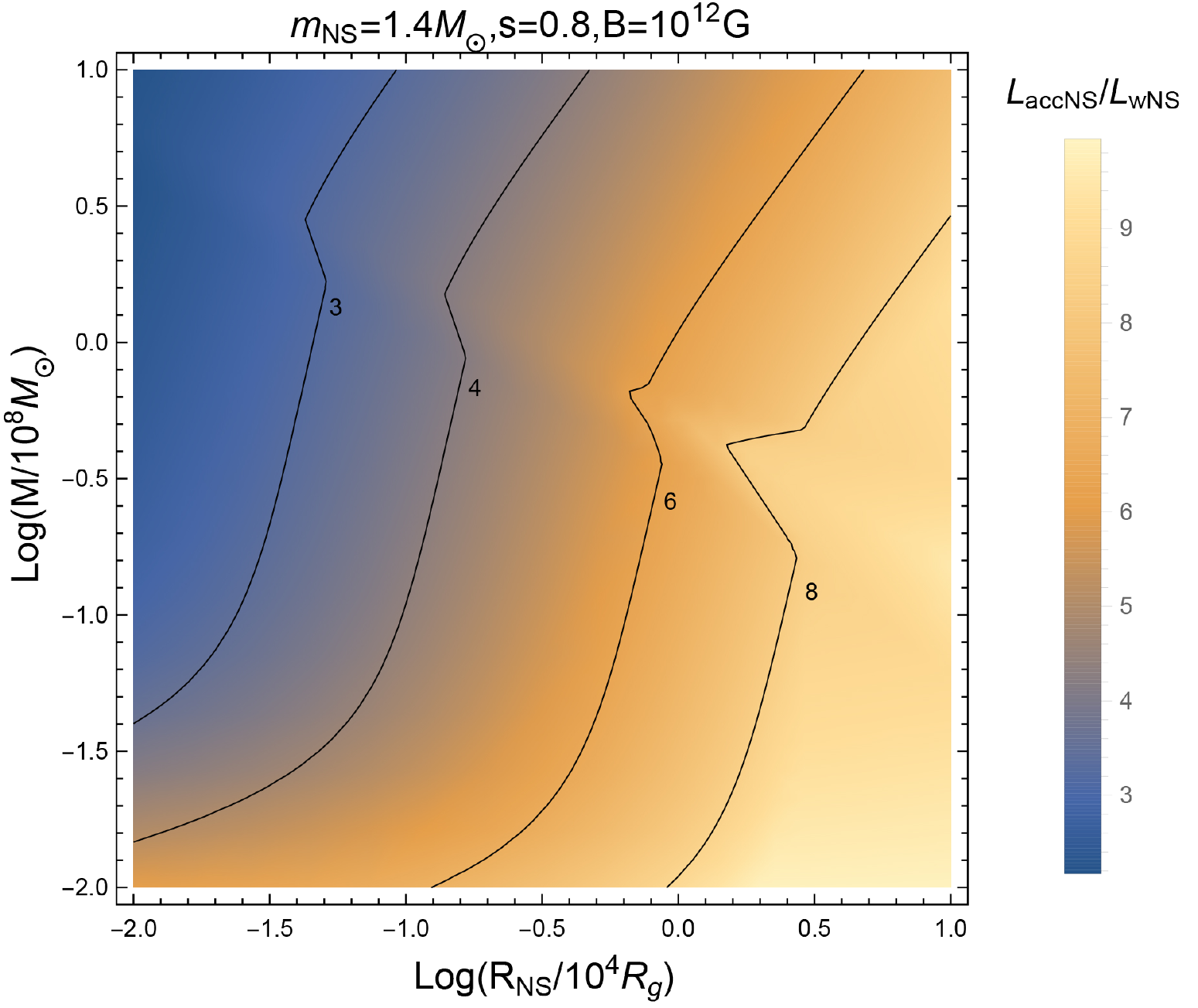}
	\end{center}
	\caption{Properties of the accretions for  
		$m_{\rm{NS}}=1.4M_{\odot}$, $r_{\rm{NS}}=10 \km$ NSs
		with different magnetic fields and indices 
		$s$. Each panel synchronously shows relative magnitude between
		the inner truncation radius $r_{\rm{m}}$ and the otherwise  
		inner disk boundary $r_{\rm{in}}=10r_g$, i.e., $r_{\rm{m}}/10r_g$, 
		by contour-lines, and comparison between the energy released 
		via the accreted mass hitting the NS surface $L_{\rm{acc}}$
		and the energy released from the circum-NS disk $L_{\rm{w}}$.
		$L_{\rm{acc}}$ is always larger than $L_{\rm{w}}$, indicating 
		that the NS-accretion liberated energy dominates the energy 
		injection into the outflow.}
	\label{Fig:NSdisk}
\end{figure*}

\begin{figure*}
	\begin{center}
		\includegraphics[width=0.32\textwidth]{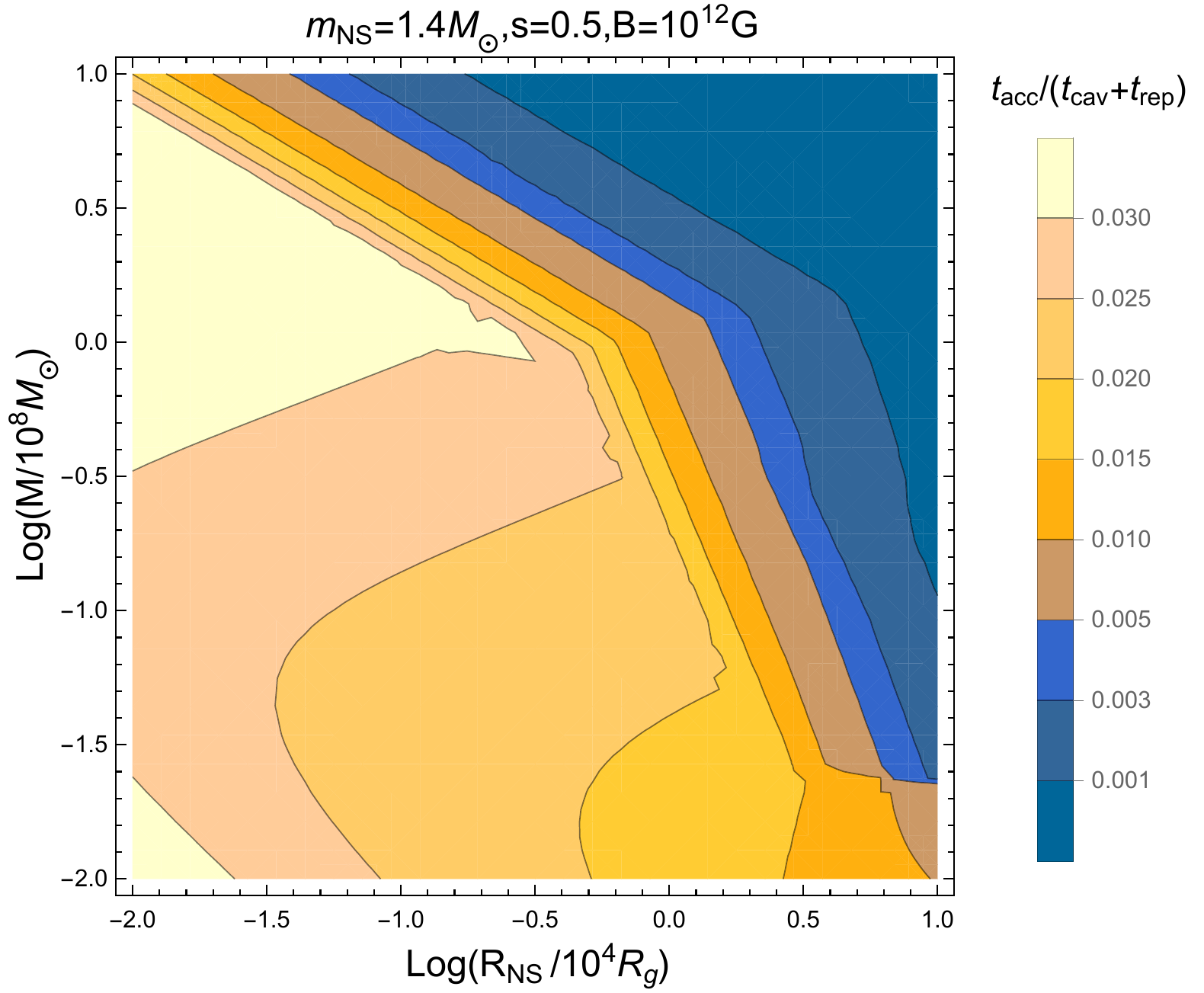}
		\includegraphics[width=0.32\textwidth]{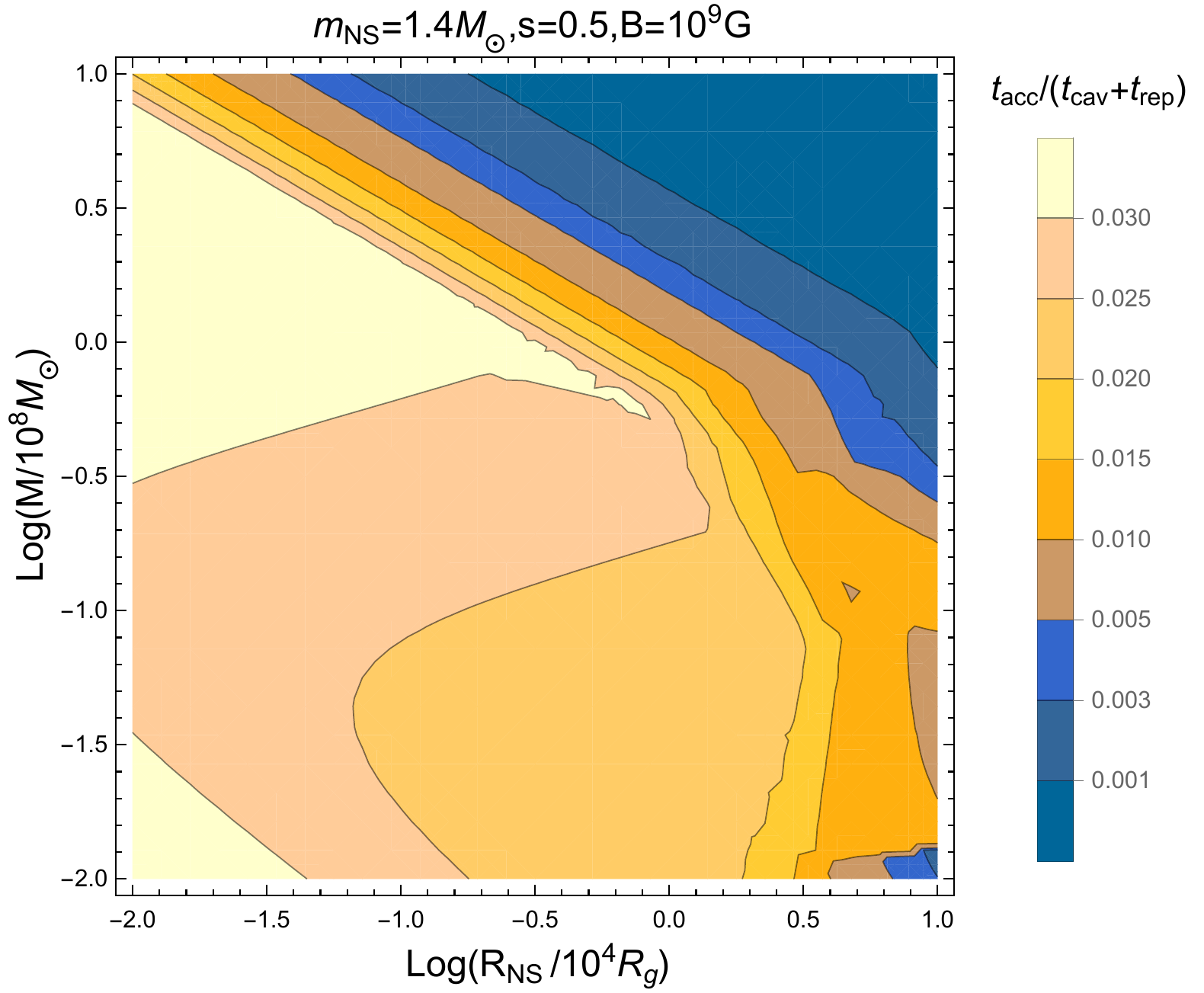}
		\includegraphics[width=0.32\textwidth]{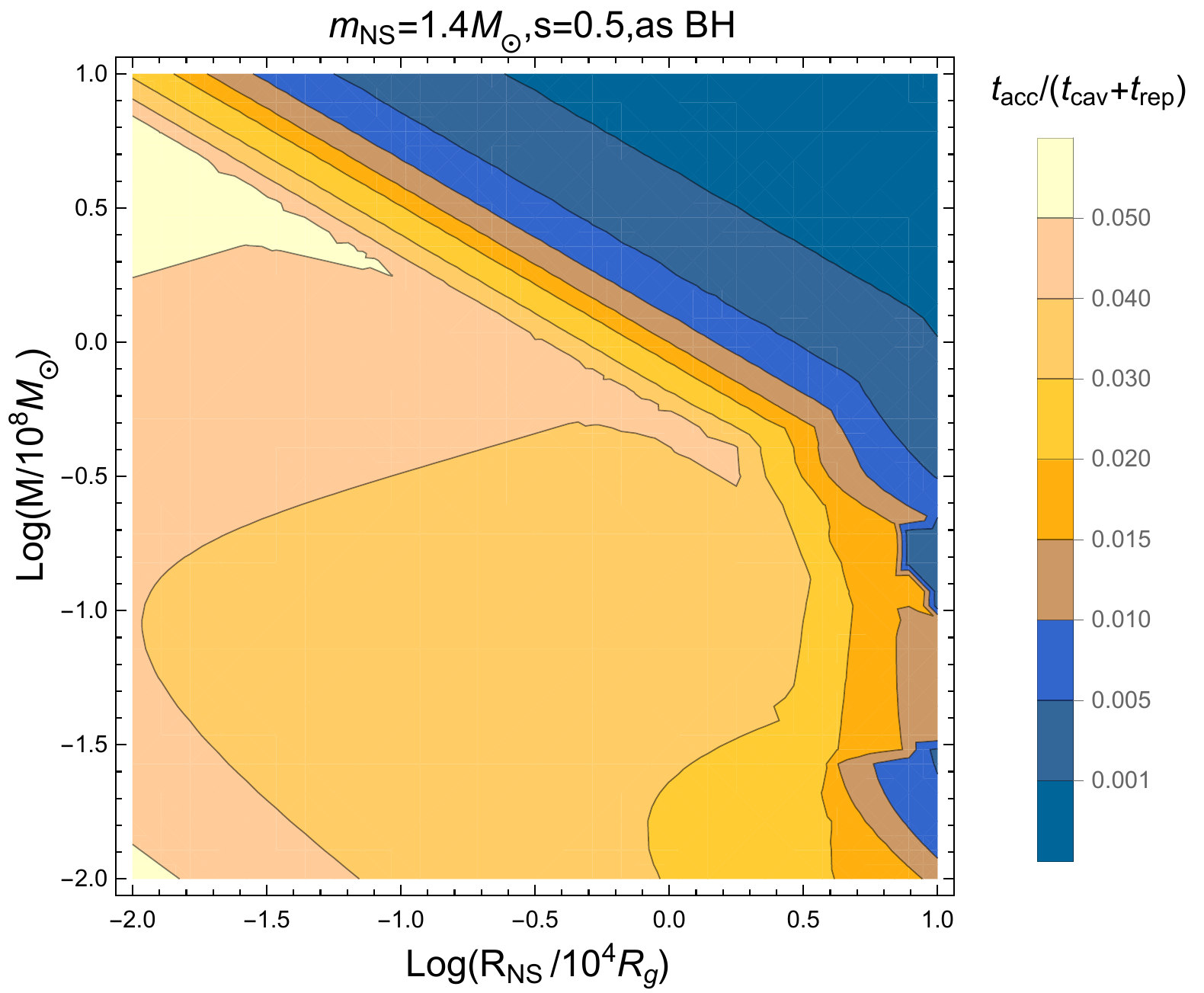}
		\includegraphics[width=0.32\textwidth]{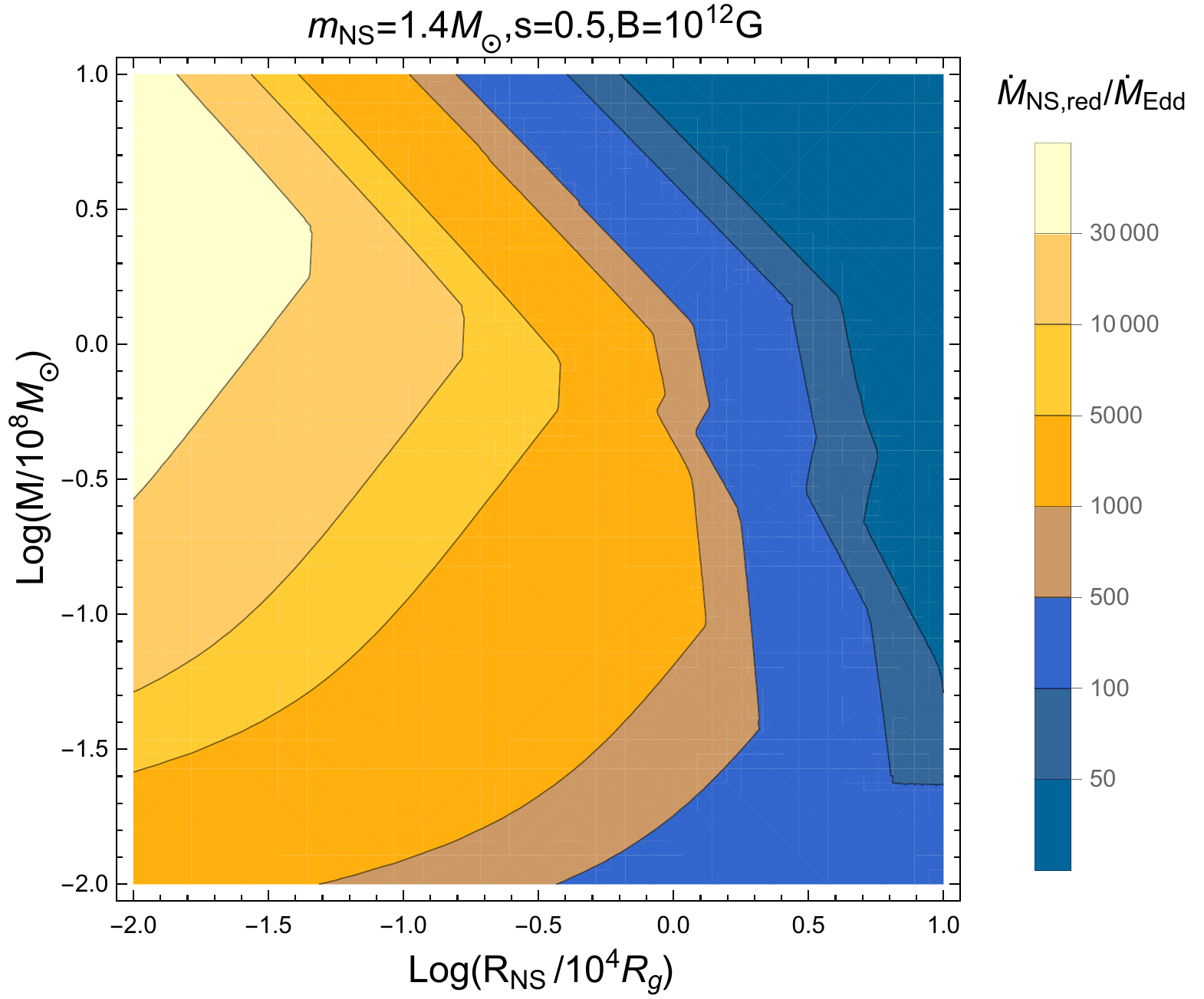}
		\includegraphics[width=0.32\textwidth]{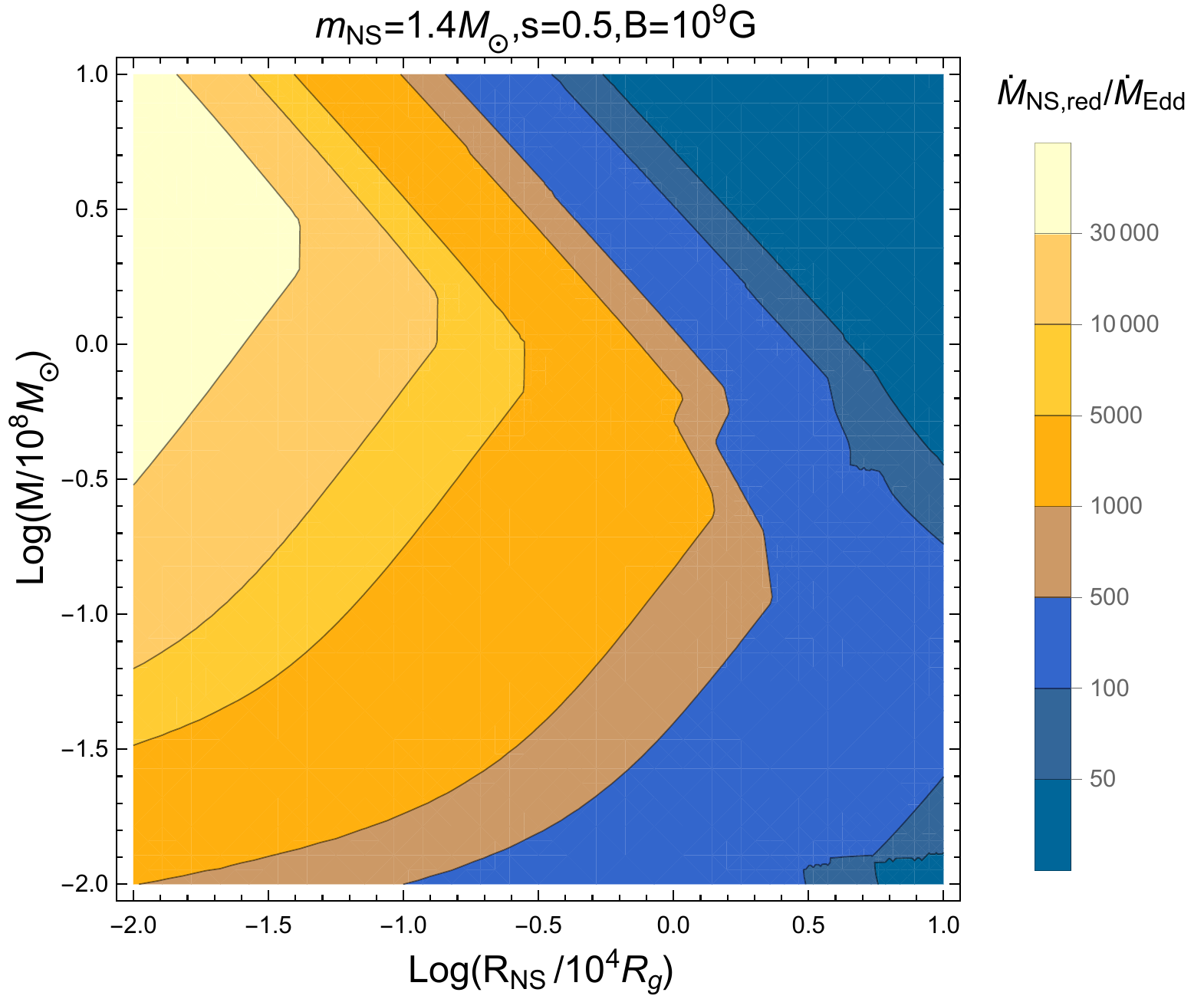}
		\includegraphics[width=0.32\textwidth]{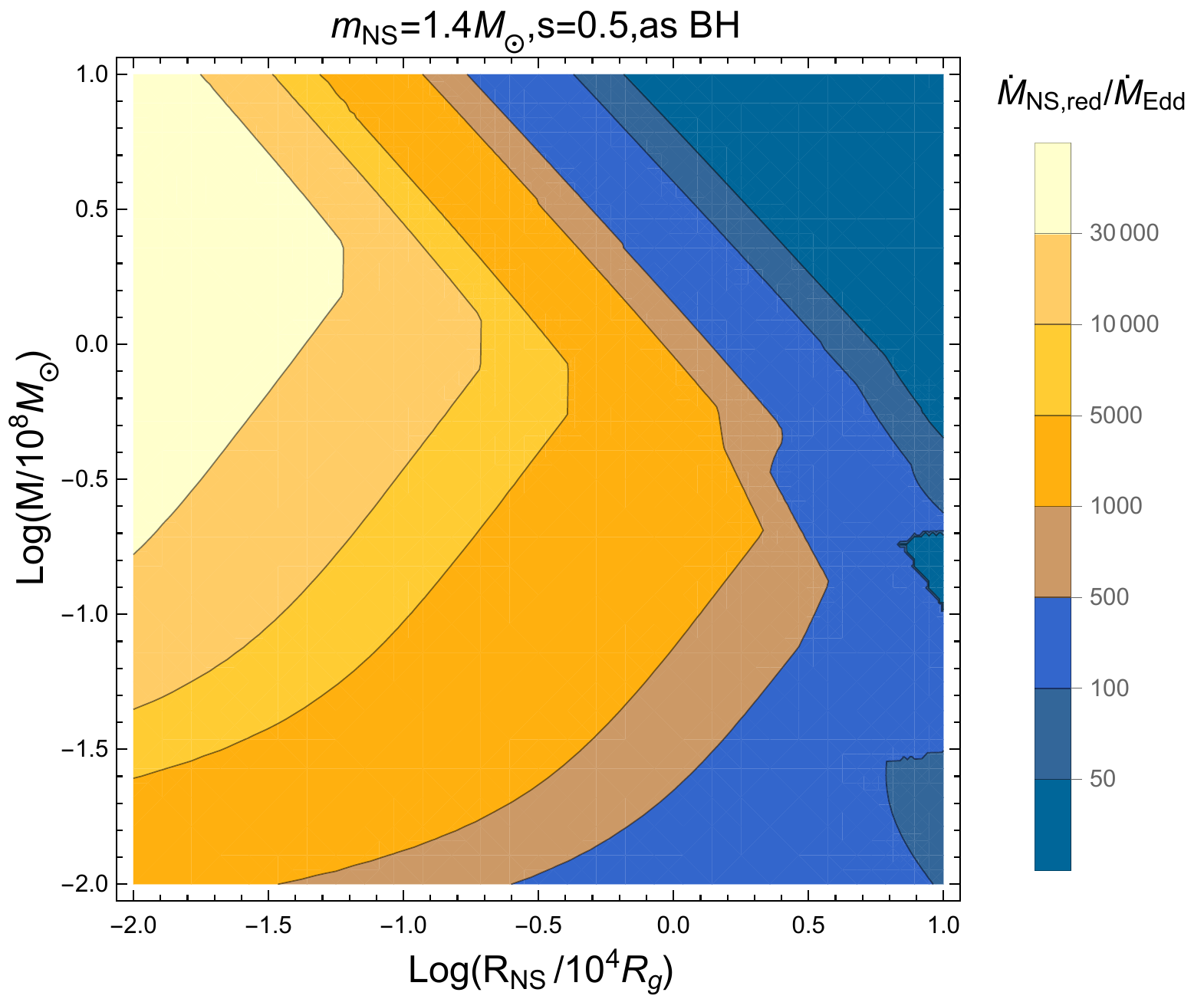}
	\end{center}
	\caption{Comparison between the NS accretion timescale and 
		the cavity evolution timescale, and the corresponding reduced NS 
		mass accretion rate. 
		Top rows show  
		$t_{\rm{acc}}/(t_{\rm{cav}}+t_{\rm{ref}} )$ and bottom rows 
		show $\dot{M}_{\rm{NS,red}}/ \dot{M}_{\rm{Edd}}$.
		Three columns show the cases of
		$m_{\rm{NS}}=1.4M_{\odot}$ and
		$s=0.5$, with $B=10^{12}G$, $B=10^{9}G$, 
		and treated as BH, respectively. The case treated as BH 
		means that the NS is set to behave as a BH without the hard 
		surface and magnetic field, and the only energy source is the 
		circum-NS disk.}
	\label{Fig:NSreduced}
\end{figure*}

The specific properties of NS accretion are shown 
in Figure \ref{Fig:NSdisk}. We find that only the NS with a strong 
magnetic field (e.g. $\sim 10^{12}G$) can give rise to  
magnetospheric disk truncation, 
on account of the large mass inflow rate of circum-NS disk leading 
to large disk pressure, which can only be resisted by large magnetic 
pressure; by contrast, when NS magnetic field is weak,
the disk extends to the NS surface instead
(comparing first and second panels in Figure \ref{Fig:NSdisk}). 
And we find that the NS-accretion energy $L_{\rm{acc}}$ 
is generally larger than the disk-generating energy $L_{\rm{w}}$ 
(seeing all three panels in Figure \ref{Fig:NSdisk}), indicating 
that the NS surface liberation dominates the outflow energy 
injection, which is a remarkable feature in comparison to the 
BH accretion. Also, the nature of the circum-NS 
disk affects the accretion and the induced outflow feedback.
The effects of NS mainly work in the cases of larger 
$R_{\rm{NS}}$ and smaller $M$, where the NS mass 
inflow rate is relatively small, and thereby
the disk is weak to confront the magnetosphere.
Larger index $s$ results in more disk mass losing 
into wind and hence weaker disk pressure at 
the inner region, leading to 
a more outside magnetospheric truncation; but even so, 
fewer mass eventually flows onto NS surface, releasing less 
accretion energy compared with the smaller $s$ case
(comparing first and third panels in Figure \ref{Fig:NSdisk}). 

Similar to BH, the outflow cavity also forms in the
NS accretion system. As shown in Figure \ref{Fig:NSreduced}, 
NS spends most of time in the underdense cavity going through
inefficient accretion, and hence the averaged mass accretion rate is 
markedly reduced, of which the general trend is akin to the BH case. 
As discussed above, more energy is injected into the outflow, 
which results in more powerful feedback on the AGN disk 
environment and longer duration of the cavity evolution. 
Accordingly, the mass rate reduction is indeed more prominent with 
the existence of a NS hard surface and a strong magnetic field (comparing all 
three columns in Figure \ref{Fig:NSreduced}). The enhanced 
reduction is more effective at larger $R_{\rm{NS}}$, 
where the AGN disk density is lower to block the cavity
expansion, i.e., an additional energy injection would lead to a punchier 
expansion; on the contrary, when the NS is closer to SMBH, 
the environment is denser to block the cavity expansion, hence the cavity 
size would not change significantly though injecting additional energy.
In addition, the more powerful $L_{\rm{acc}}$ for the cases of 
small $M$ and large $R_{\rm{NS}}$ causes the outflow feedback and 
the mass accretion rate of NS with strong magnetic fields a remarkable change.
(comparing first and second columns in Figure \ref{Fig:NSreduced}).

\section{Discussion} \label{sec:Discussion}

\begin{figure*}
	\begin{center}
		\includegraphics[width=0.32\textwidth]{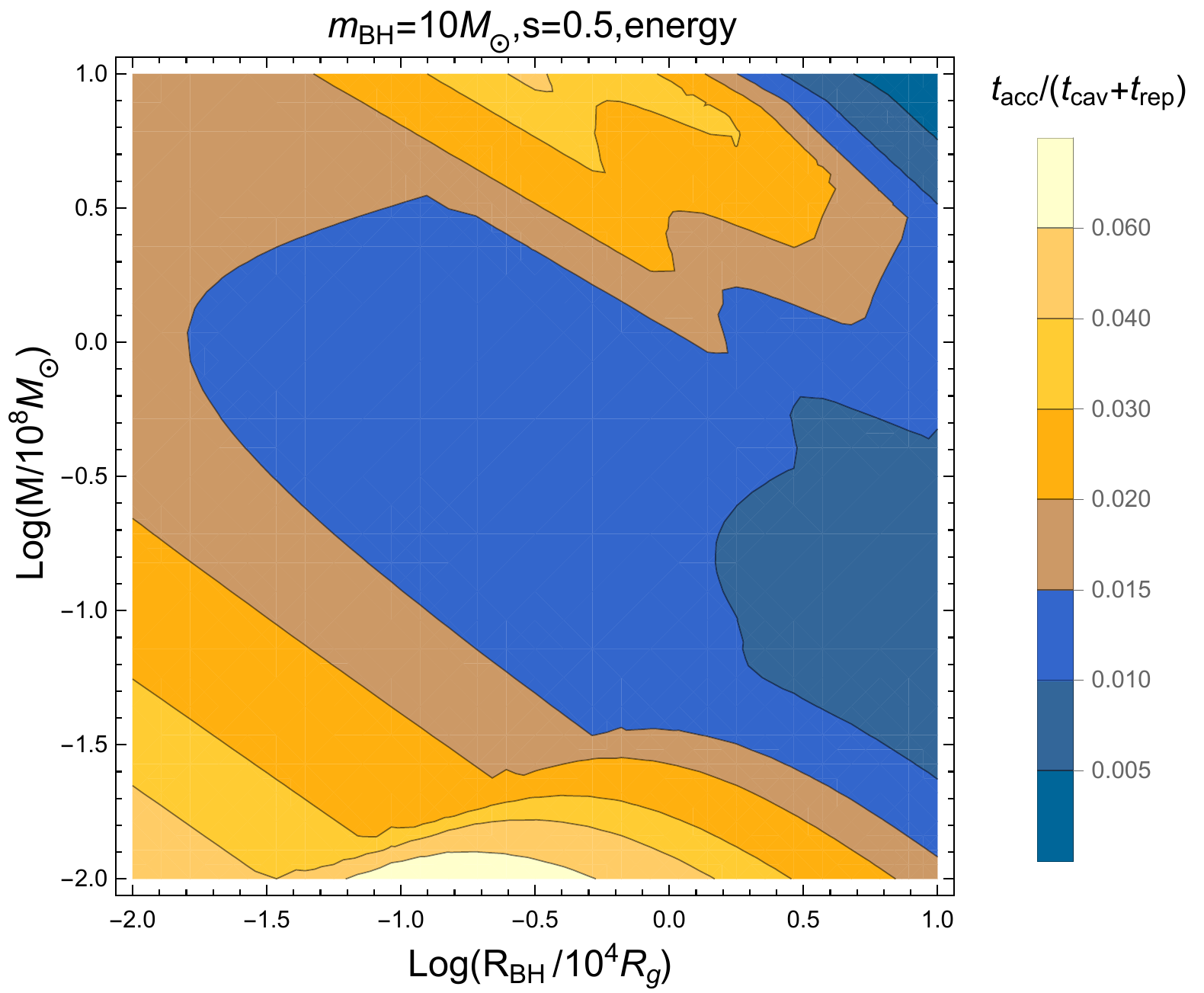}
		\includegraphics[width=0.32\textwidth]{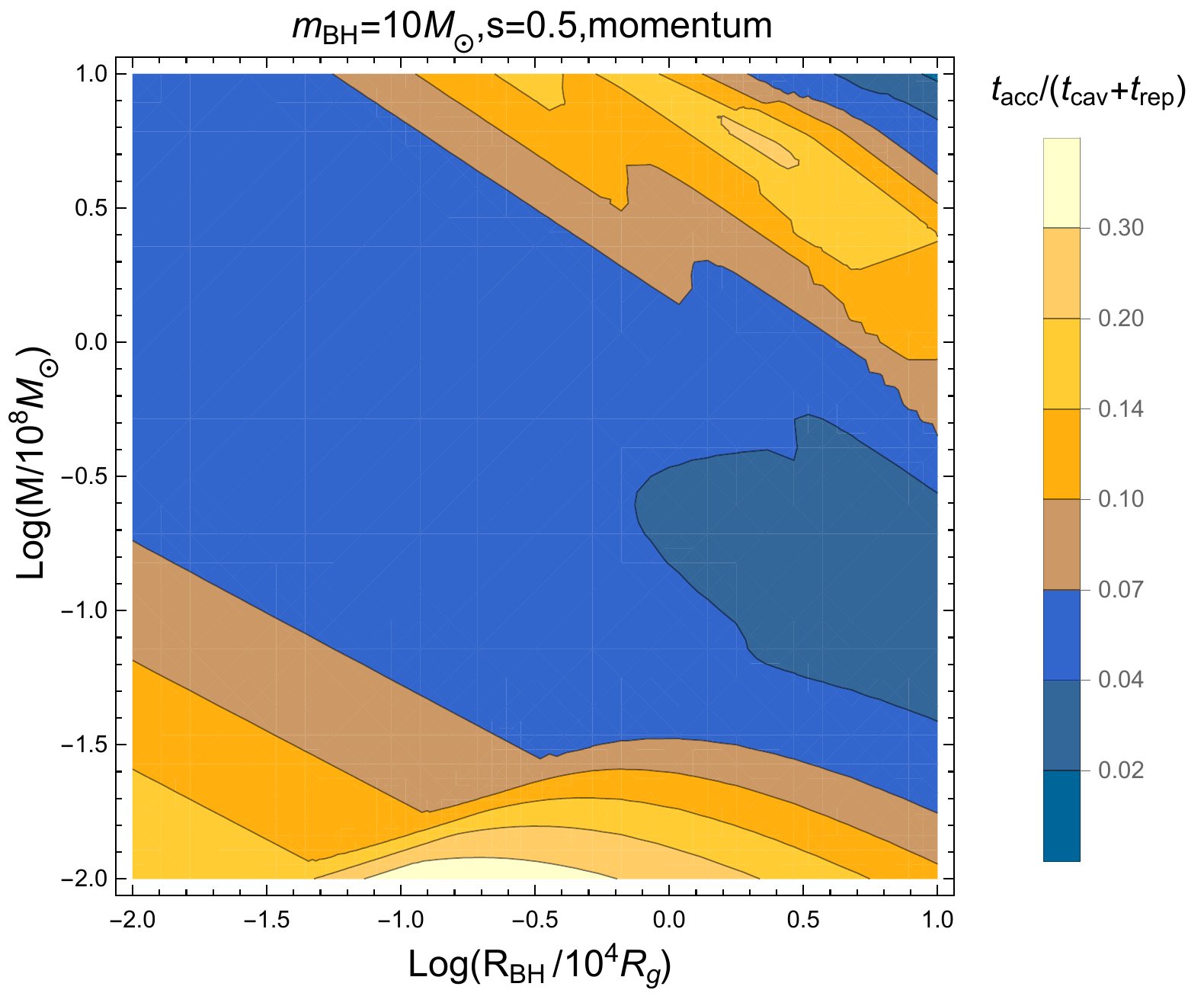}
		\includegraphics[width=0.32\textwidth]{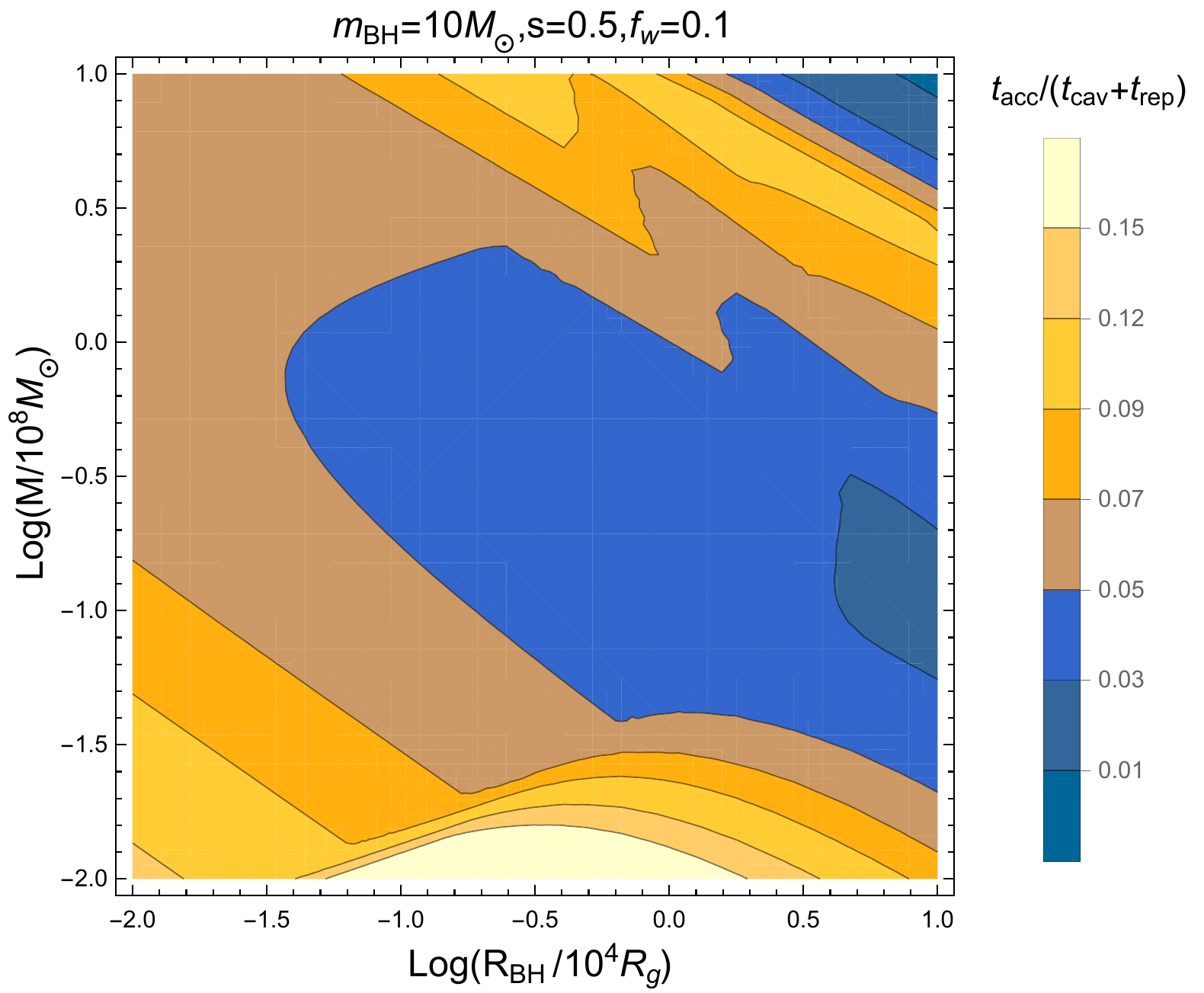}
	\end{center}
	\caption{Comparison between the BH accretion timescale and the 
		cavity evolution timescale under the different treatments of
		shell expansion. Left and middle panels represent the 
		extremely entire adiabatic and momentum 
		conservation evolution of the shell, respectively.  
		Right panel represents the shell evolution as described in 
		Section \ref{3-1}, but set $f_{w}=0.1$ to denote a weaker 
		outflow. We set $s=0.5$ as the example, verifying but 
		not showing for brevity, the properties of other $s$ value 
		cases are similar.}
	\label{Fig:Com}
\end{figure*}

\begin{figure*}
	\begin{center}
		\includegraphics[width=0.32\textwidth]{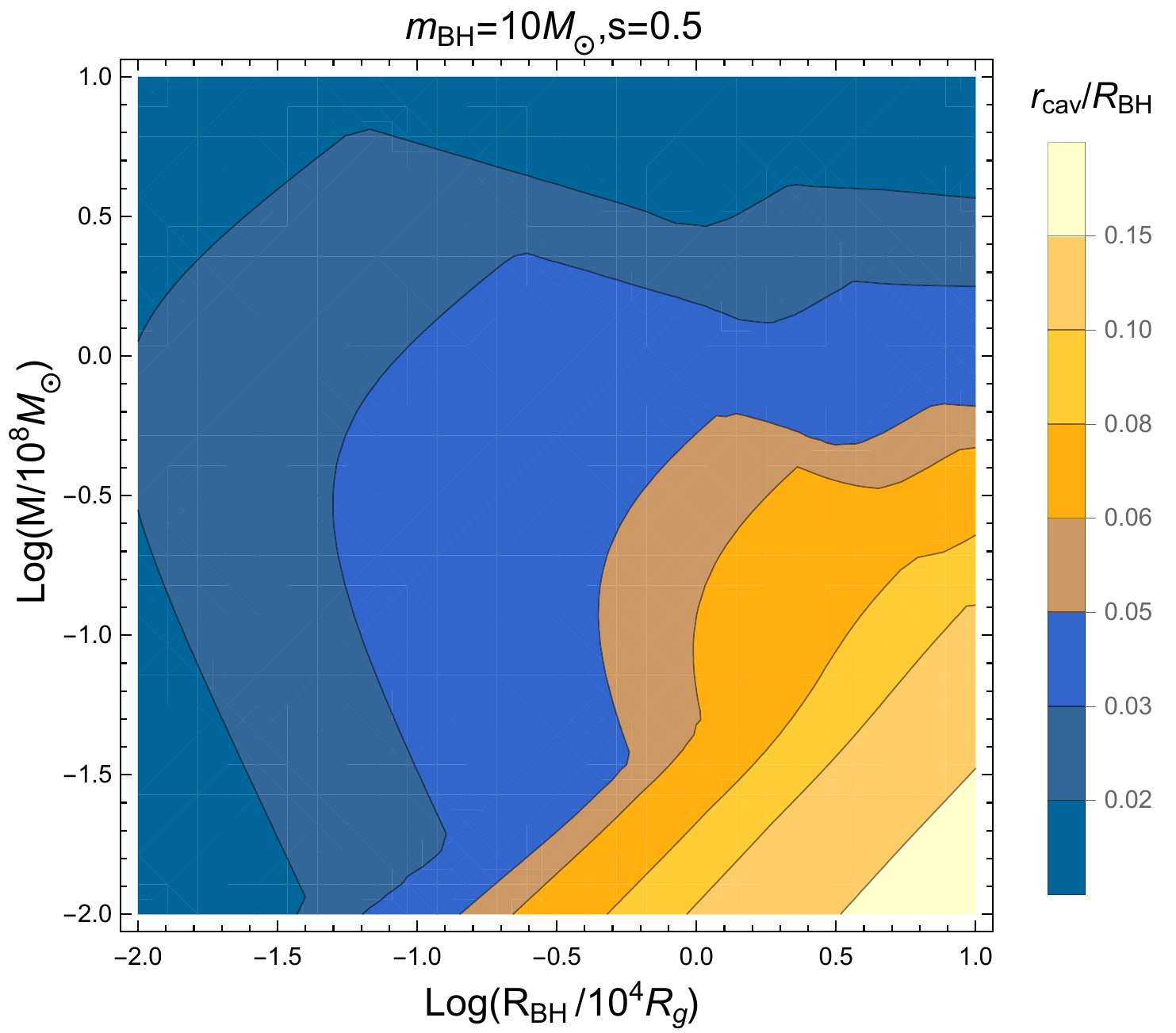}
		\includegraphics[width=0.312\textwidth]{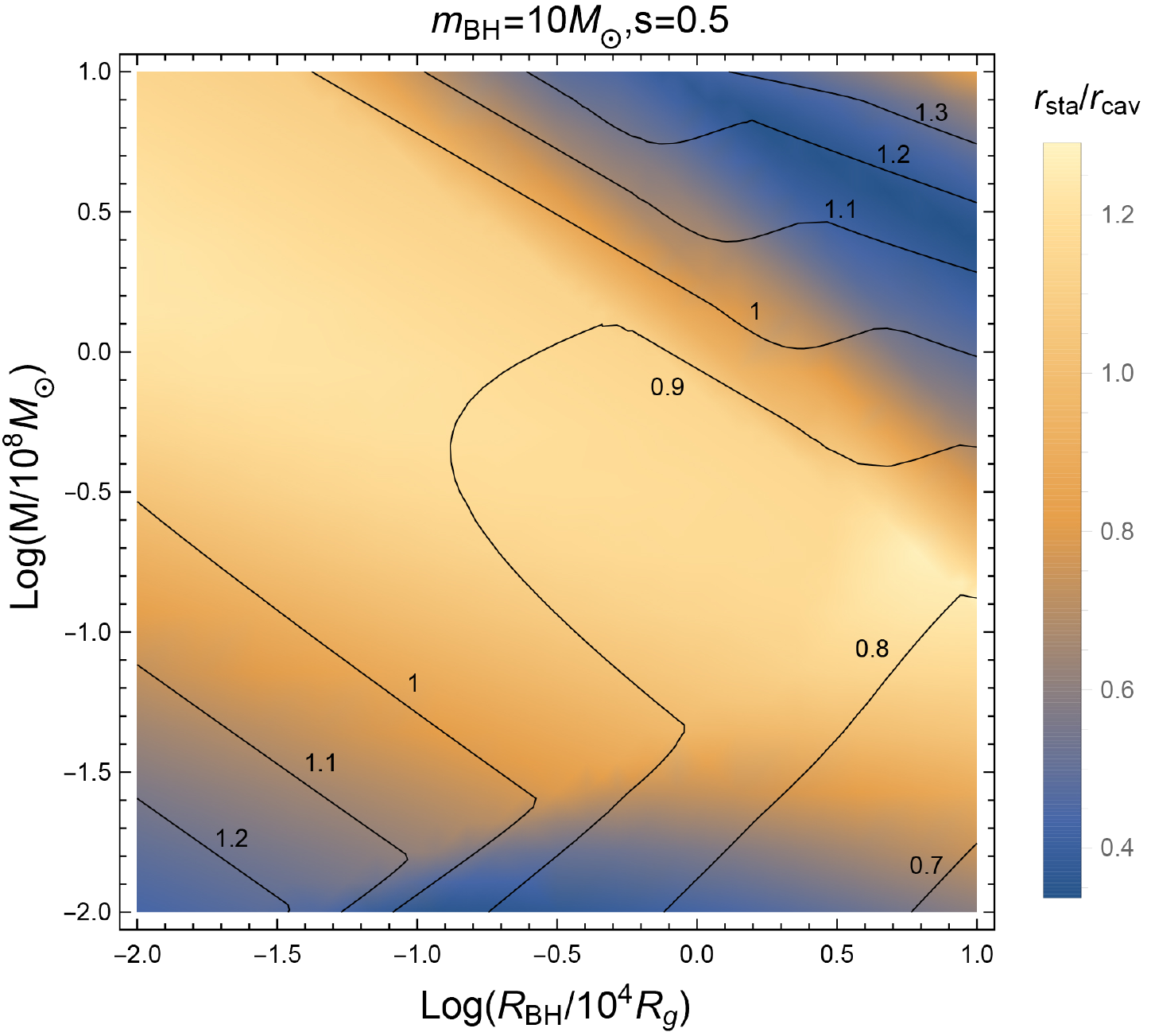}
	\end{center}
	\caption{Properties of the cavity half-widths to check the model 
		validity. Left panel shows comparison between the half-width of the outflow 
		cavity and the location radius of a BH $r_{\rm{cav}}/R_{\rm{BH}}$. Right 
		panel shows the wind stand-off radius $r_{\rm{sta}}/r_{\rm{cav}}$; and 
		shows by contour-lines the cavity maximum half-width dominated by the 
		shear motion of the AGN disk gas $r_{\rm{shear}}/r_{\rm{cav}}$. }
	\label{Fig:rcav}
\end{figure*}

\begin{figure*}
	\begin{center}
		\includegraphics[width=0.298\textwidth]{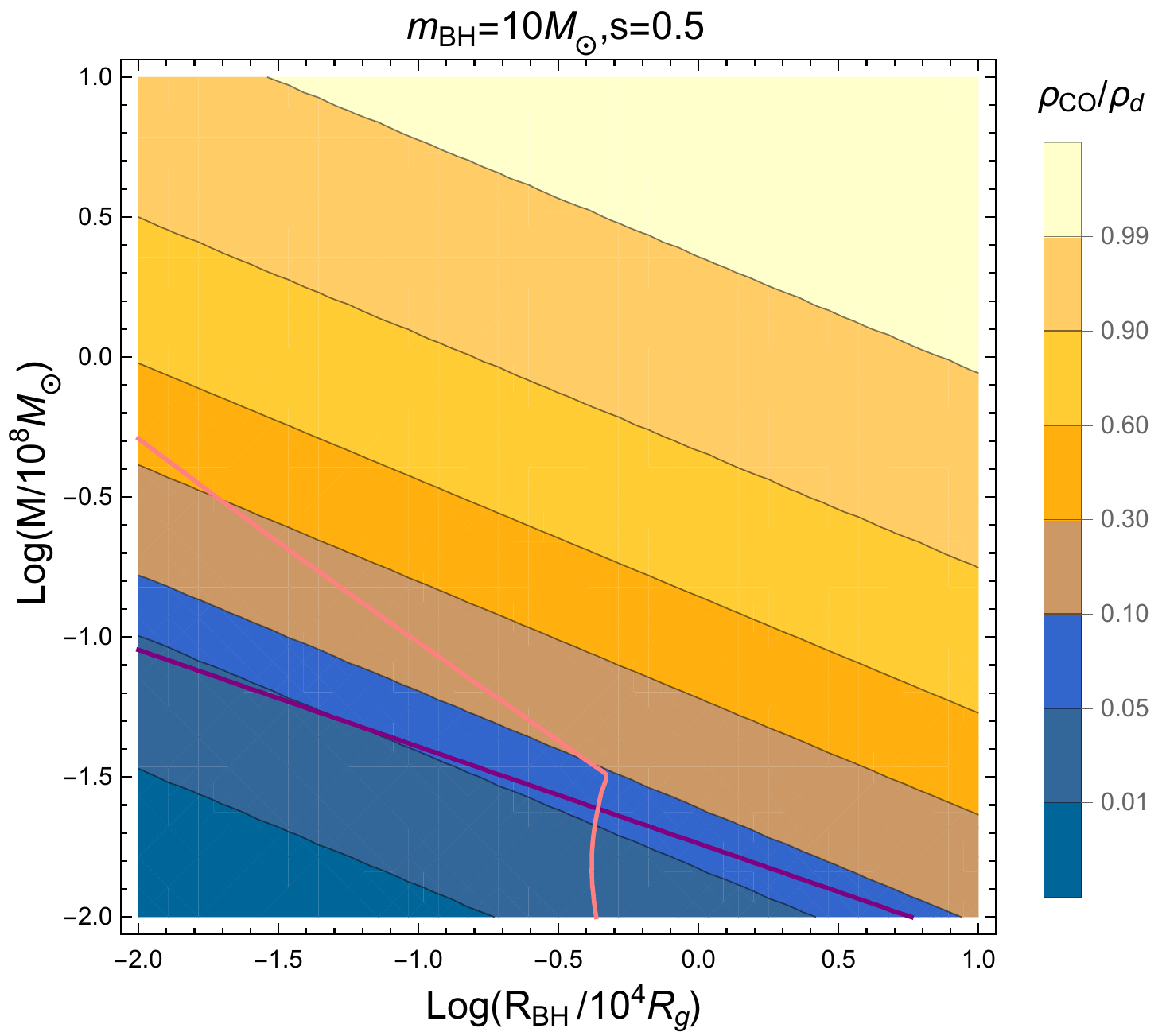}
		\includegraphics[width=0.32\textwidth]{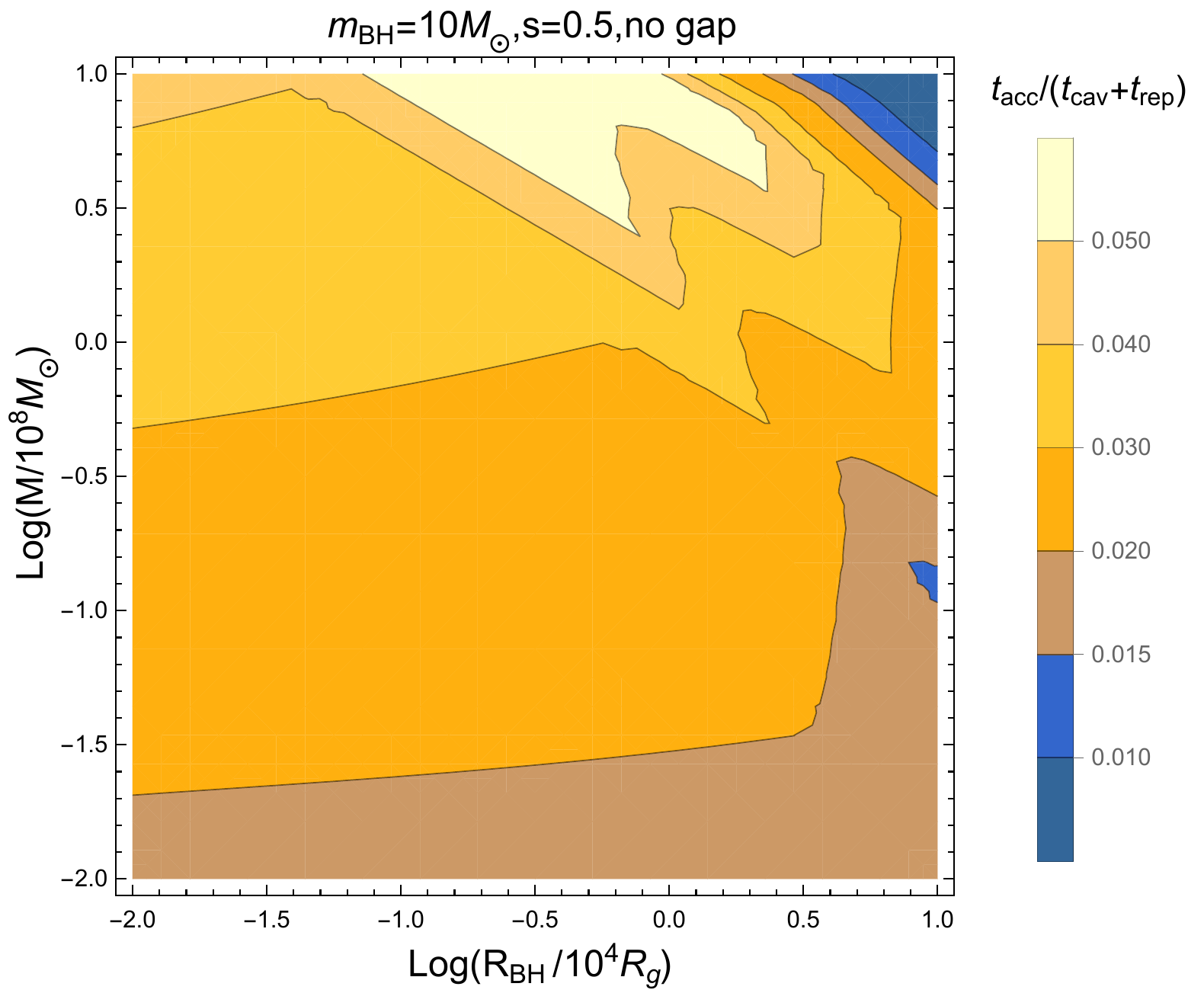}
	\end{center}
	\caption{Properties of the CO-gravity induced gap, and the outflow 
		feedback neglecting the gap effects of the 
		$m_{\rm{BH}}=10M_\odot$ case. Left panel shows 
		comparison between the gap density and the unperturbed AGN disk 
		gas density $\rho_{\rm{CO}}/\rho_{\rm{d}}$; the pink 
		line and left region represent 
		$R_{\rm{d,gap}}\geqslant r_{\rm{cav}}$; 
		the purple line and left region represent 
		$t_{\rm{gap,g}}<t_{\rm{ref}}$, as a contrast, 
		$t_{\rm{gap,v}}>t_{\rm{ref}}$ always holds within the 
		panel's parameter space. Right panel shows
		comparison between the BH accretion timescale and the cavity evolution 
		timescale for the gap unopened case, i.e. $\rho_{\rm{CO}}=\rho_{\rm d}$.}
	\label{Fig:gap}
\end{figure*}

\subsection{Different Treatments of Cavity Expansion} \label{subs41}
In Section \ref{3-1}, we assume that the shocked shell and outflow undergo 
rapid depressurization after the breakout from the AGN disk, and then 
the ringlike shell evolves driven by the momentum of wind and itself successively. 
However, the AGN disk usually holds plane-parallel stratified 
structure \citep[e.g.][]{Grishin21}, 
rather than a steep cutoff at the height $H$; even so, the shocked gas 
would break out and reduce pressure at $\sim H$ \citep[e.g.][]{Schiano85, Mac Low88, Olano09}, 
hence we think our assumption can roughly describe the processes of wind-surroundings 
interaction. The uncertainty lies at the duration of the hot gas pushing the 
shell, where we simplistically link the outflow-energy and -momentum driven phase 
directly at $t_{\rm{bre}}$. In fact, during the depressurization, 
the shell is still pushed laterally 
by the thermal pressure.  On the other hand, the shocked 
gas perhaps experiences effective radiation cooling soon before or after 
the shell breakout, or even
when the shell still expanding deeply in the AGN disk. Since the actual 
structure of the AGN disk is complex, and the 
shell evolution needs to contain the concrete 
cooling mechanisms of the shocked gas, which requires precise hydrodynamical 
simulations and is beyond the scope of this work, 
we instead consider two simplified and
extreme cases, evolution following the adiabatic or momentum conservation phase 
in the spherically symmetric medium all the while, to study the effects 
of the duration of depressurization and the gas cooling. 
Corresponding descriptions of the evolution are shown 
in Appendix \ref{appendix-extreme-case}.

We choose $t_{\rm{acc}}/(t_{\rm{cav}}+t_{\rm{ref}})$ to typically 
reflect the features of CO accretion and outflow evolution, as shown 
in Figure \ref{Fig:Com}. The cavity still comes into being 
in the two extreme cases, so we 
confirm that the formation of cavity and the reduction of averaged 
CO accretion rate are common and inevitable, regardless of the cooling
efficiency. Comparing different treatments, the pure momentum conservation 
case leads to the largest $t_{\rm{acc}}/(t_{\rm{cav}}+t_{\rm{ref}} )$, 
i.e., the weakest outflow feedback; correspondingly, cavity 
expansion is strongest in the entirely adiabatic case. Namely, the longer 
duration of the thermal energy pushing the shocked shell, the more powerful 
expansion of the cavity. 

When estimating the strength of the disk winds, i.e. Equation (\ref{Lw}), 
we set $r_{in} \sim 10r_g$ on account of the theoretical studies; but various 
numerical simulations show different inner radii where the outflow is
observably produced, ranging $10-100r_g$ \citep[e.g.][]{Jiang14, Yang14, 
Sadowski151, Kitaki21}. Setting the larger inner radius, the disk winds would be 
weaker, with $L_w'/L_w \sim (r_{in}'/r_{in})^{1-s}$. To study the selection 
of a larger $r_{in}$, we decrease $f_w$ as a substitution, as shown in Figure 
\ref{Fig:Com}, finding that the feedback is assuredly weaker, but still the 
cavity forms and the averaged CO accretion rate is significantly reduced.

We also verify the validity of our description on the outflow 
cavity evolution. We have assumed that the relevant environment 
parameters, mainly $\rho_{\rm{d}}$, are fixed during the 
outflow-AGN disk interaction; but if the 
size of cavity is too large, the AGN disk properties would vary 
with radius $R$ to affect the cavity 
evolution, our description are thus unfaithful. Safely, the 
cavity is always much smaller than the AGN disk radius, i.e. 
$r_{\rm{cav}} \ll R_{\rm{CO}}$, as shown in left panel of 
Figure \ref{Fig:rcav}, so the fixed-environment assumption 
holds and our description is a feasible estimation. Moreover, since 
the half-width of the outflow cavity exceeds the CO gravity radius, 
as shown in Figure \ref{Fig:cav}, the shear motion of the AGN disk 
gas would exert ram pressure to affect the shell evolution 
\citep[e.g.][]{Rozyczka95}, which limits the horizontal expansion 
width of the ring when its velocity decreases to the shear velocity, 
i.e., $\dot{r}_{\rm{sh}} \simeq (3/4) 
\tilde{\Omega}_{\rm K} r_{\rm{sh}}$ \citep{Moranchel21}.
Using Equation (\ref{eqm}), the maximum half-width
of the cavity is given by 
\begin{equation}
r_{\rm{shear}} \simeq 2.2\left[\frac{r_{\rm{shellb}}^{2}(t_{\rm{acc}}) 
	v_{\rm{shellb}}(t_{\rm{acc}} )}{\tilde{\Omega}_{\rm K}} \right]^{1/3} ,\label{eqrshear}
\end{equation}
where the width is multiplied by 2 from the results of 
numerical simulation  \citep{Moranchel21}. As shown in 
Figure \ref{Fig:rcav}, we find that $r_{\rm{shear}} \sim 
r_{\rm{cav}}$, hence we think taking the shear into account 
can markedly deform the cavity structure, but would not 
significantly change the results in Section \ref{sec:feedback}. 

It is worth mentioning that the late evolution phase of ring driven by 
momentum is simplified, i.e., we ignore the ambient pressure 
$\sim \rho_{\rm d} \tilde{c}_{s}^2$ and assume that the efficient accretion 
stops after $t_{\rm{acc}}$ in Equation (\ref{eqmo}) and (\ref{eqm}); 
in practice, the accretion of CO would last for a 
few $t_{\rm{acc}}$ and launch winds 
to continuously push the shell. Meanwhile,
there exists a stand-off radius defined by the pressure balance
$\rho_{\rm w}v_{\rm w}^2 = \rho_{\rm d} \tilde{c}_{s}^2$ \citep[e.g.][]{Schiano85}, 
i.e.,
\begin{equation}
r_{\rm{sta}}= \left( \frac{\dot{p}_{\rm{w}}}
{4 \pi \rho_{\rm d} \tilde{c}_{s}^2}\right)^{1/2},
\end{equation}
when the half-width of the cavity becomes larger than 
$r_{\rm{sta}}$, the ambient pressure 
would be the dominant external force to decelerate the shell. We compare the 
stand-off radius with the cavity half-width, as shown in Figure \ref{Fig:rcav}, 
and find that $r_{\rm{sta}} \sim r_{\rm{cav}}$, which implies that the wind 
is strong enough to push the shell, i.e., Equation (\ref{eqmo}), and the ring 
evolution is mainly driven by its own momentum at around 
$r_{\rm{cav}}(r_{\rm{sta}})$, i.e., Equation (\ref{eqm}); 
in other words, our simplified model can roughly describe the shell evolution.
As an aside, the dropping disk wind would hinder the ambient inflow and 
maintain a cavity with decreasing $r_{\rm{sta}}$ if the CO hyper-Eddington
accretion still persists during the cavity refilling phase, which can 
lengthen $t_{\rm{ref}}$ to further reduce the averaged accretion 
rate of CO. All in all, the long-term evolution of the circum-CO disk 
and the cavity are vital for the understanding of the CO accretion in 
the AGN disk, further researches are therefore needed. 

\subsection{Whether Gap Opens around CO in the AGN Disk} \label{subs42}

\begin{figure*}
	\begin{center}
		\includegraphics[width=0.32\textwidth]{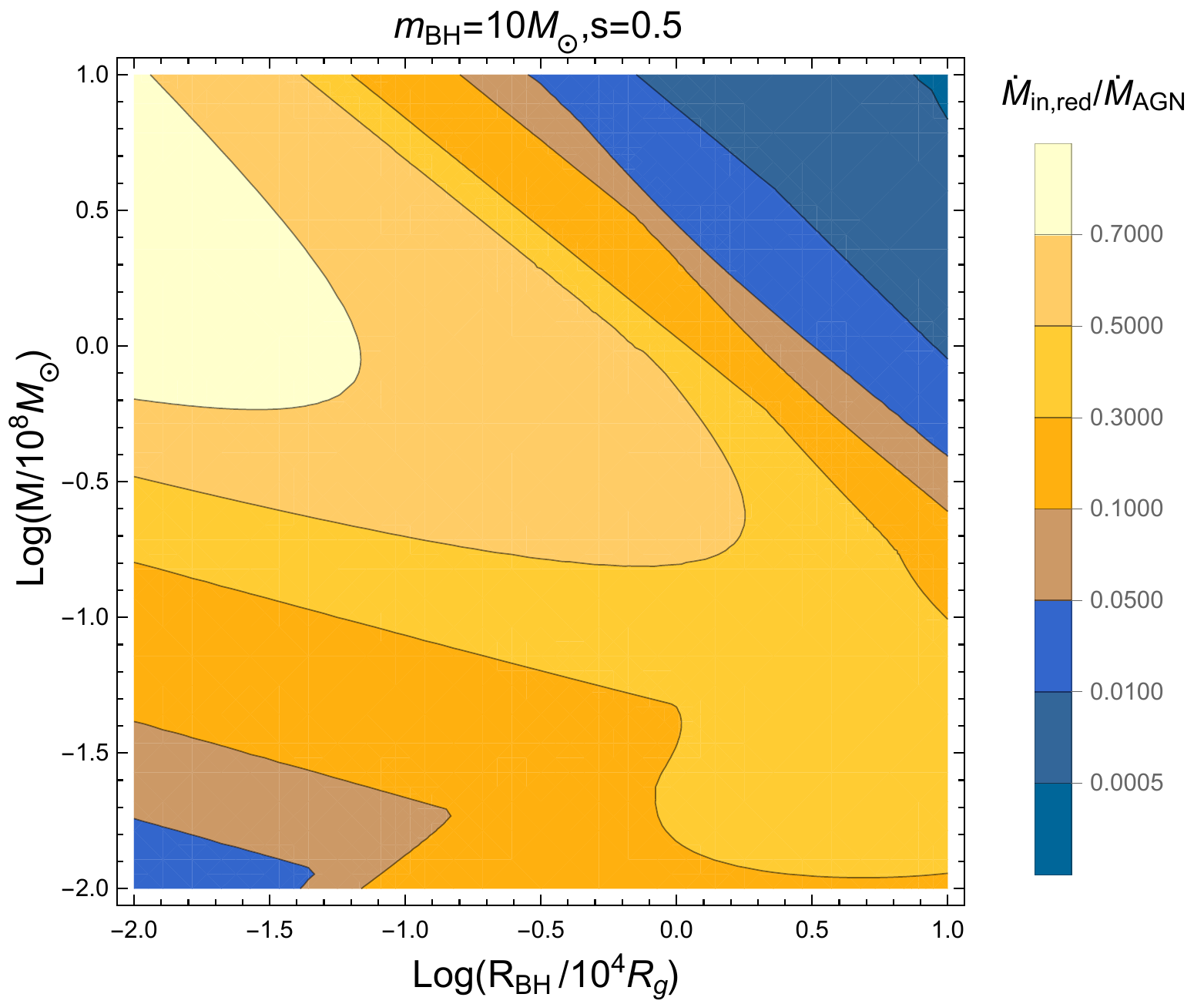}
		\includegraphics[width=0.30\textwidth]{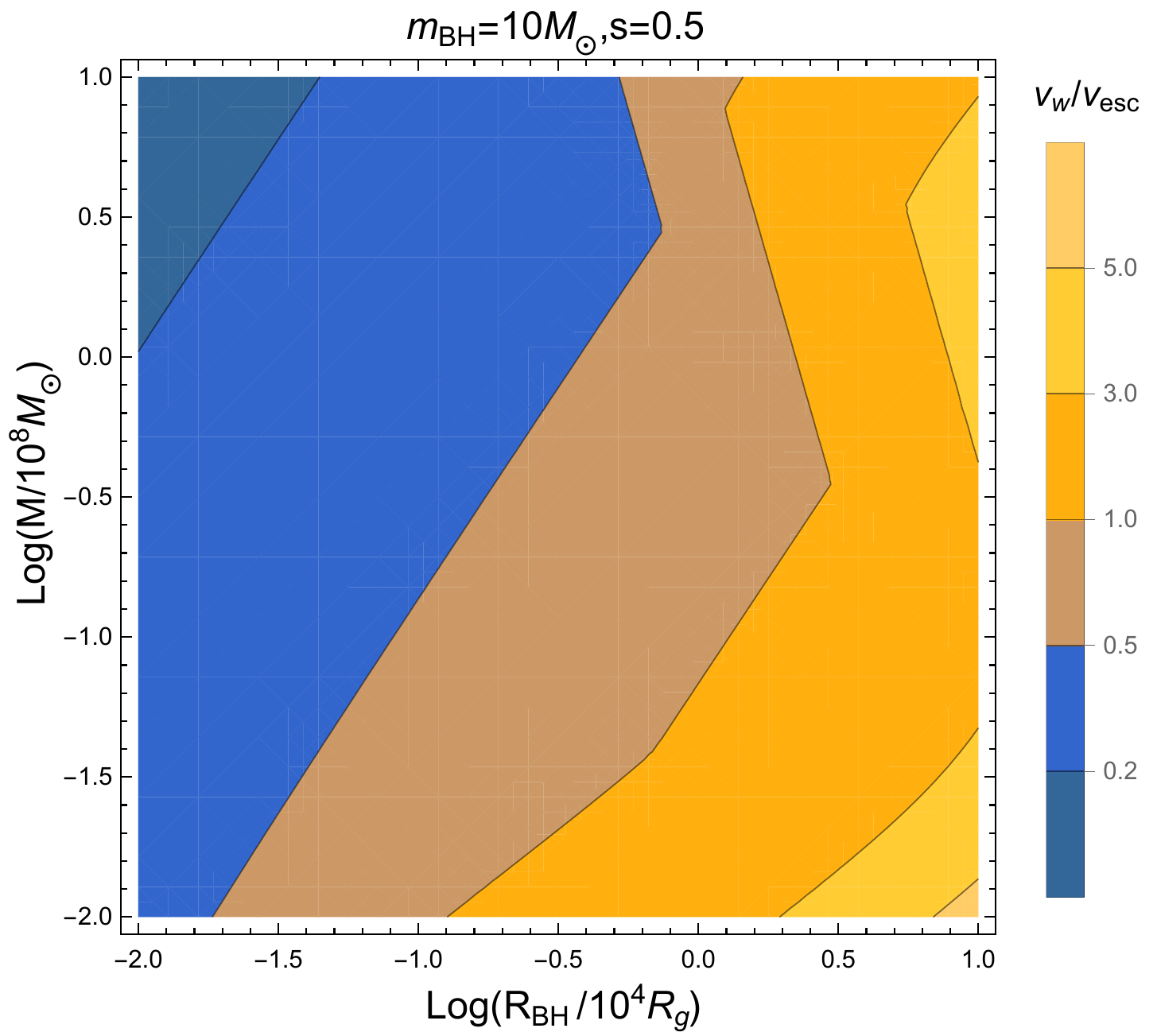}
	\end{center}
	\caption{Properties of the CO accretion used to 
	interpret the adequate supply of AGN disk gas. Left 
	panel shows comparison between the reduced 
	BH captured gas rate and the AGN disk inflow rate
    $\dot{M}_{\rm{in,red}}$/$\dot{M}_{\rm{AGN}}$ with 
    $\dot{M}_{\rm{AGN}}=0.5\dot{M}_{\rm{Edd}}$. Right panel shows 
    comparison between the launching outflow velocity and  
    the escape velocity of the SMBH $v_{\rm{w}}/v_{\rm{esc}}$.}
	\label{Fig:feedback}
\end{figure*}

We have considered the effect of underdense gap when calculating the BHL 
inflow rate; as shown in left panel of Figure \ref{Fig:gap}, the gas density 
is deeply reduced by the CO gravity for smaller SMBH mass and closer 
CO location cases. When studying the outflow feedback we briefly
ignore the gap structure, instead we assume the shell expanding 
within the unperturbed AGN disk environment, 
which may underestimate the size of cavity and the reduction of
averaged mass accretion rate because the shell can expand more easily in the
underdense gap. Nevertheless, since the size of cavity 
$r_{\rm{cav}}$ is general larger than the gap half-width
$R_{\rm{d,gap}}$ (the pink line in 
Figure \ref{Fig:gap}), the assumption is roughly suitable except for the 
very light SMBH and very close CO location cases.

Note that the structure of underdense gap is 
commonly studied in a stable state,
but the evolution of outflow cavity is dynamical and 
$r_{\rm{cav}}>R_{\rm{d,gap}}$ generally holds, so the properties of  
the gap environment around the CO may 
be affected by the outflow. 
The timescale of gap opening 
can be evaluated by the viscous timescale, 
i.e. $t_{\rm{gap,v}}\simeq 0.1 R_{\rm{d,gap}}^2/(\alpha \tilde{c}_s H)$ 
(where we estimate the deep gap opening timescale being one-tenth 
of the timescale for the whole gap to attain the steady state, as shown 
by Figure 13 in \citealt{Kanagawa17}), and the 
deep gap half-width can be set to $\Delta r\simeq 0.08(m_{\rm{CO}}/{M} )^{1/2}
(H/R_{\rm{CO}})^{-3/4}\alpha^{-1/4}R_{\rm{CO}}$ 
\citep{Kanagawa18}. 
\cite{Tagawa22} alternatively suggests the gap opening timescale can be calculated 
via the angular momentum of gas within the annular gap removed by the 
gravitational torque of CO, i.e. $t_{\rm{gap,g}}= \Delta J/T_{\rm{CO}}$, 
where 
$\Delta J\sim 2 \pi R_{\rm{CO}}^2 \Delta r^2 \tilde{\Omega}_{\rm{K}}\Sigma_{\rm{d}}$ 
and $T_{\rm{CO}}\sim (m_{\rm{CO}}/{M} )^{2}(H/R_{\rm{CO}} )^{-3} R_{\rm{CO}}^4
\tilde{\Omega}_{\rm{K}}^2\Sigma_{\rm{d}}$. Setting the cavity refilling 
timescale as $t_{\rm{ref}}=\Delta r/ \tilde{c}_s$,
we find that $t_{\rm{ref}}< t_{\rm{gap}}$ holds
within wide parameter space, i.e., 
$t_{\rm{gap,v}}/t_{\rm{ref}}=10 (m_{\rm{CO},1}/M_{8})^{1/2}(H/R_4)^
{-7/4}(\alpha_{-1})^{-5/4}$ or $t_{\rm{gap,g}}/t_{\rm{ref}}= 
110 (m_{\rm{CO},1}/M_{8})^{-3/2}(H/R_4)^
{13/4}(\alpha_{-1})^{-1/4}$,
so we boldly speculate that the deep gap around the CO 
has no enough time to be completely rebuilt to affect the  
efficient accretion of CO and 
$\dot{M}_{\rm {obd}}$ would be 
enhanced, because after the faster refilling of gas with 
density $\rho_{\rm{d}}$ from the AGN disk to the cavity within 
$t_{\rm{ref}}$, the circum-CO disk has already being built by 
the denser gas before the density being completely reduced 
to the lower $\rho_{\rm{d,gap}}$ within $t_{\rm{gap}}$;
conversely, the deep gap would be markedly rebuilt for the 
very light SMBH and very close CO location cases with 
$t_{\rm{gap,g}}<t_{\rm{ref}}$, as shown by the purple line 
in Figure \ref{Fig:gap}. 

Not only the outflow would impede the build of deep gap around CO, 
the gap itself may not open for different AGN disk models.
So we show the cases of which no gap opens in right panel of Figure
\ref{Fig:gap}. The most prominent feature is that the
outflow feedback is more effective when $M$ is smaller, and hence more
dramatically reduces the averaged CO mass accretion 
rate, which is in contrast to the 
gap-opening case as shown in Figure \ref{Fig:reduced}, 
because the gap effect is 
more noteworthy in these parameter regions. Overall, the outflow feedback 
likely impacts the gap around CO;
we suggest that the CO accretion system in the AGN disk carried by 
lighter SMBH are useful to investigate whether the gap opens and the 
properties of the potential gap.

\subsection{Comparison between CO-related Mass Rate and AGN Disk Inflow Rate} \label{subs43}

Since the density of environment is extremely large, the mass rate 
captured by the CO set in the AGN disk is hyper-Eddington,
as shown in Figure \ref{Fig:Mobd}, which even would be  
higher than the accretion rate of the SMBH itself; in other words, 
just one embedded BH (NS) can exhaust the mass supply of AGN disk if
the BH (NS) accretion persists the AGN lifetime.
\cite{Tagawa22} highlighted a trouble as \textquotedblleft \ Depletion 
problem \textquotedblright, i.e., 
the AGN disk gas can be depleted by COs if there is no feedback.

We argue that the AGN disk would not be starved by 
the accretion of embedded COs after the outflow 
feedback processes as described in Section 
\ref{3-1} taken into account.
First, due to the existence of outflow, 
the mass accretion rate onto CO, as shown in Figure \ref{Fig:maccbh}, 
which is the actual loss of the AGN disk mass, is far below the initial 
mass inflow rate of circum-CO disk and the AGN accretion rate. 
Meanwhile, the existence of outflow cavity reduces the averaged
circum-CO disk mass inflow rate akin to the reduction of averaged CO 
mass accretion rate, i.e., 
\begin{equation}
	\dot{M}_{\rm{in,red}}\simeq
	\frac{\dot{M}_{\rm{in}}(r_{\rm{obd}} ) t_{\rm{acc}}}
	{t_{\rm{cav}}+t_{\rm{ref}}} . \label{minred}
\end{equation}
As shown in left panel of Figure \ref{Fig:feedback}, $\dot{M}_{\rm{in,red}}$ 
is generally lower than AGN accretion rate with 
nearly all inflow ejected into outflow 
$\dot{M}_{\rm{w}}\sim \dot{M}_{\rm{in,red}}$, thereby even though all 
the circum-CO disk winds escape from the SMBH, the AGN disk is still capable to 
feed the SMBH. Besides, as shown in right panel of Figure \ref{Fig:feedback}, 
the wind velocity $v_{\rm{w}}$ just leaving 
the circum-CO disk, i.e., Equation (\ref{vwind}), 
is already generally lower than the escape 
velocity of the SMBH, 
$v_{\rm{esc}}=\sqrt{2 G M/ R_{\rm{CO}}}$. The parts of 
outflow moving along the AGN disk are eventually 
re-injected into the disk; the parts moving vertically 
would be dramatically decelerated by the dense environment, 
though breaking out, velocity of which 
is conceivable to be lower than $v_{\rm{w}}$, 
thus would eventually fallback into the AGN disk. To sum up, 
CO accretion with induced outflow feedback would just 
result in few mass loss, thus the AGN disk can live 
for a long time.

\begin{figure*}
	\begin{center}
		\includegraphics[width=0.319\textwidth]{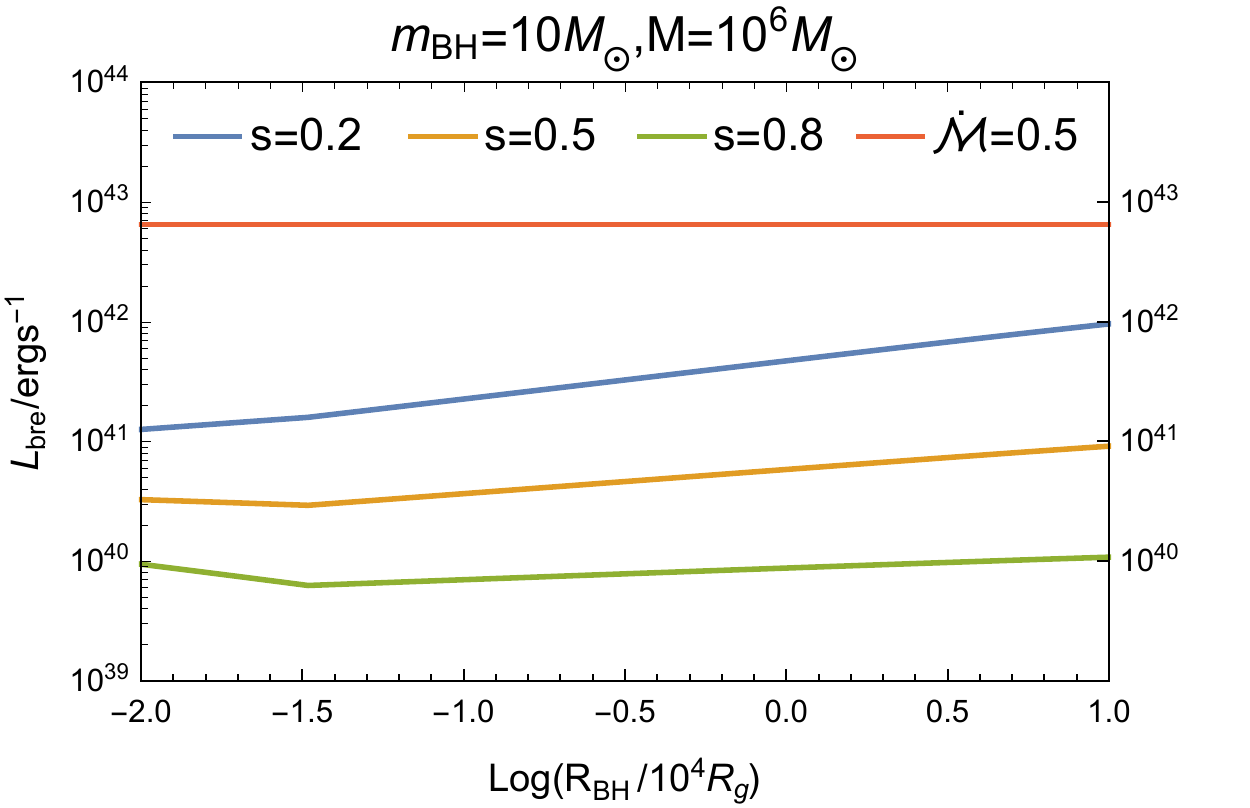}
		\includegraphics[width=0.320\textwidth]{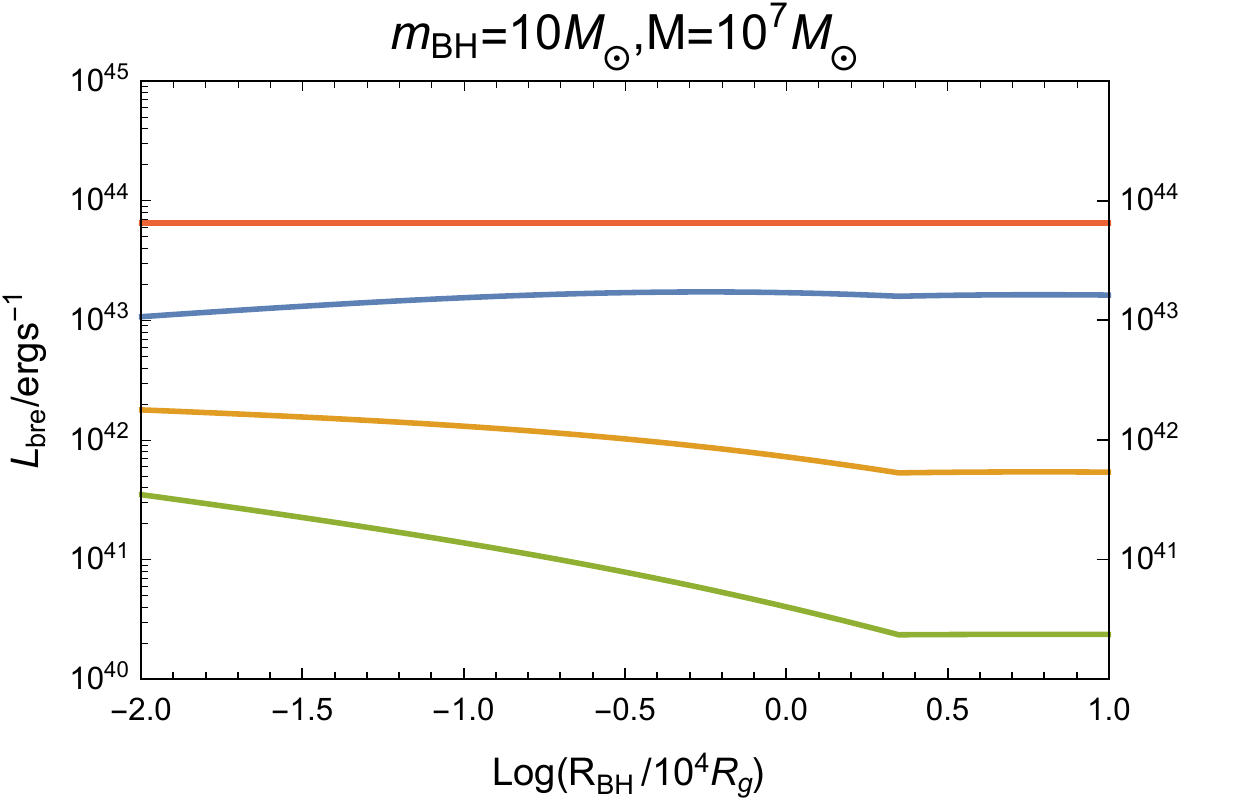}
		\includegraphics[width=0.320\textwidth]{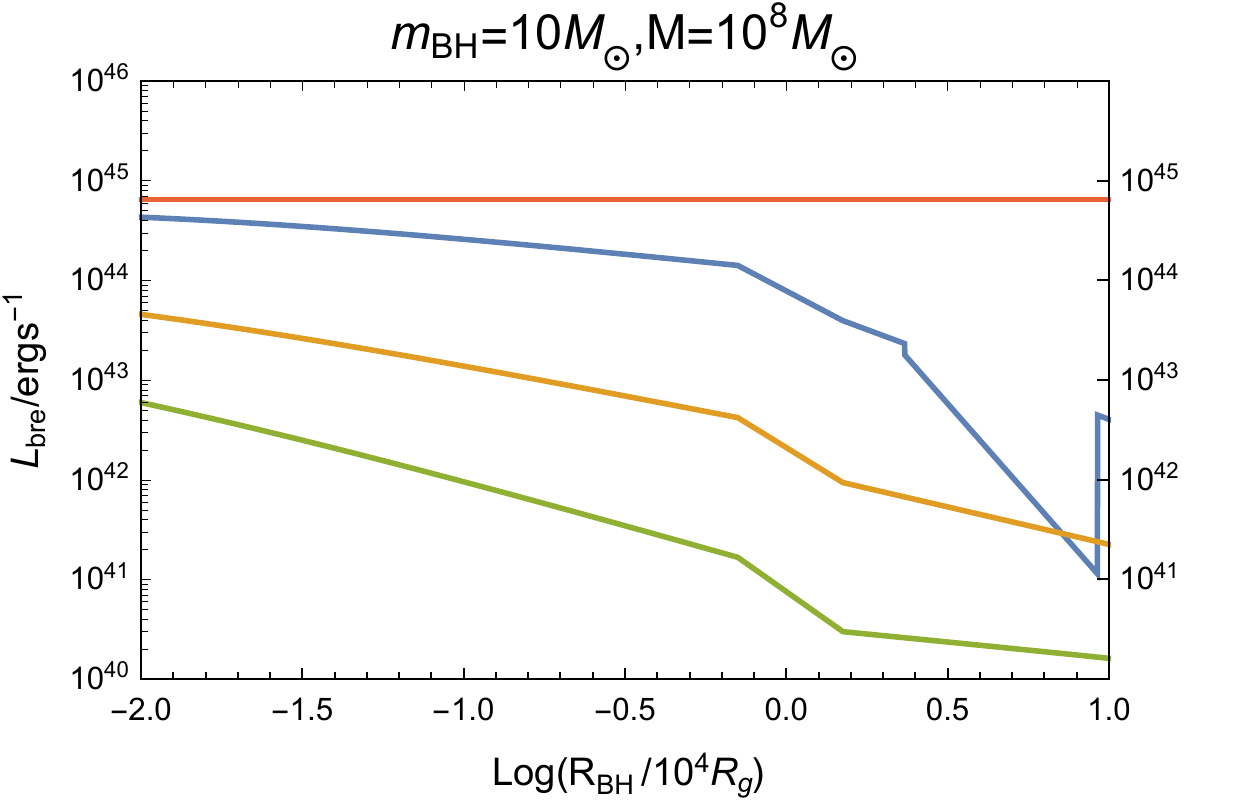}
		\includegraphics[width=0.325\textwidth]{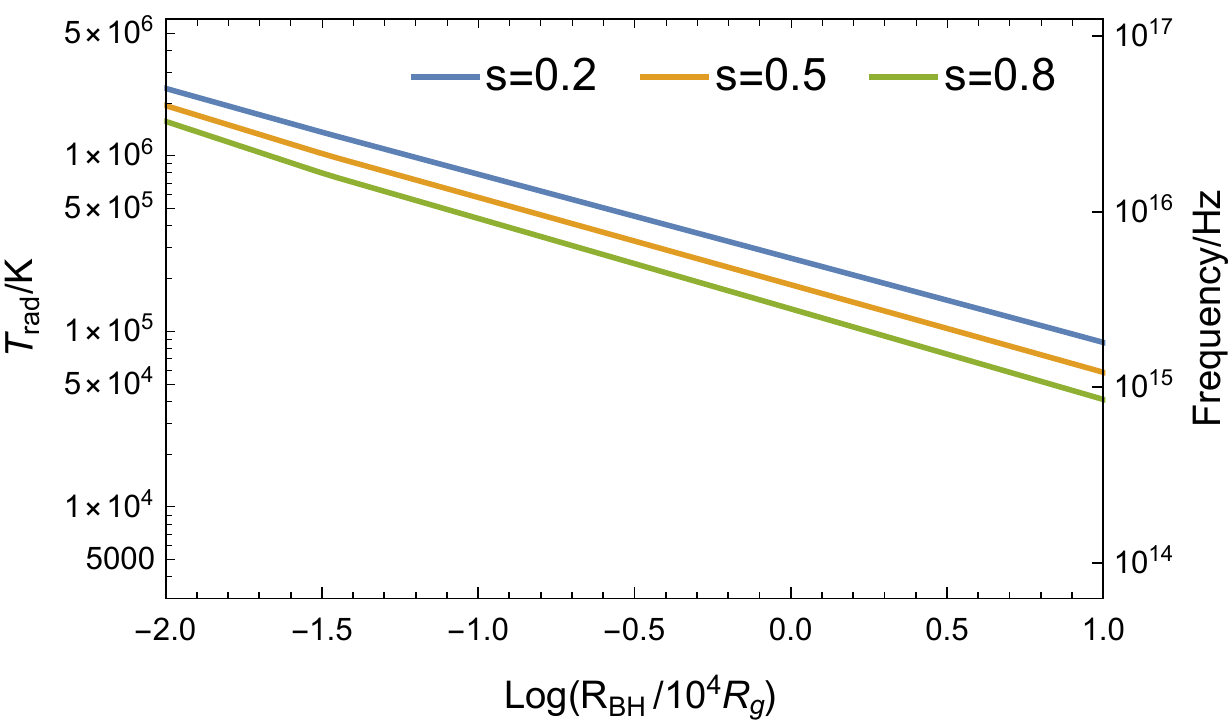}
		\includegraphics[width=0.325\textwidth]{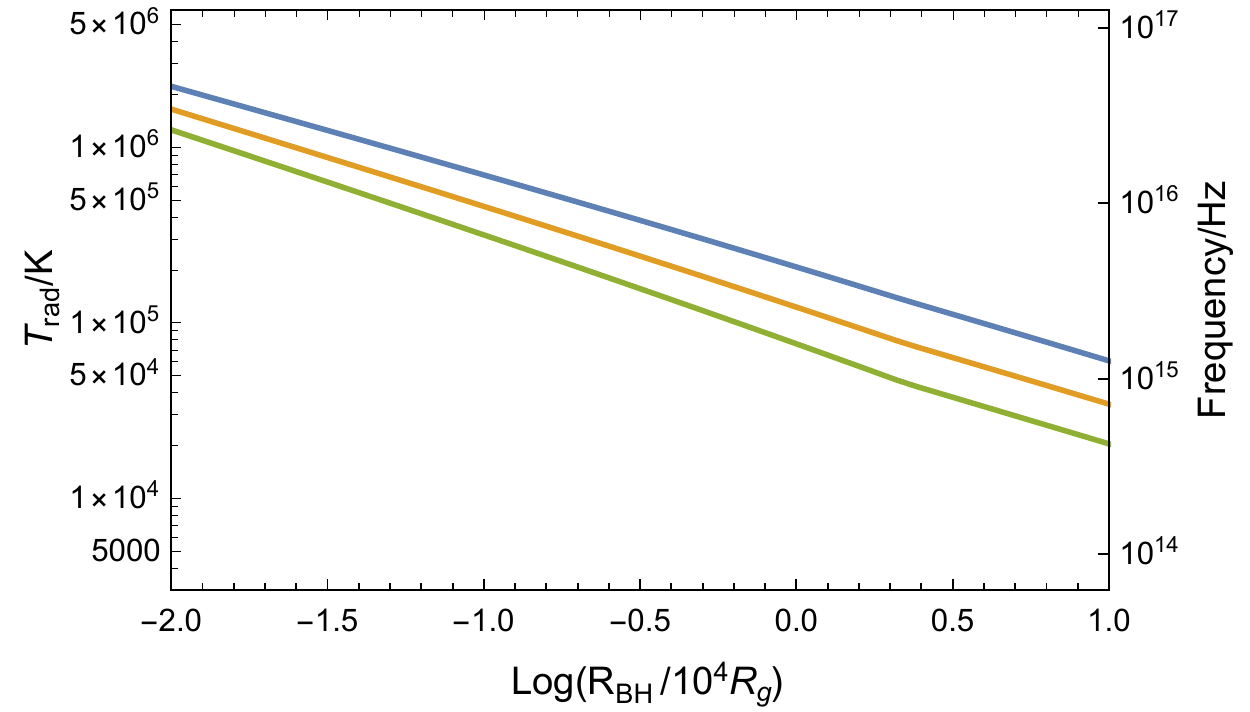}
		\includegraphics[width=0.325\textwidth]{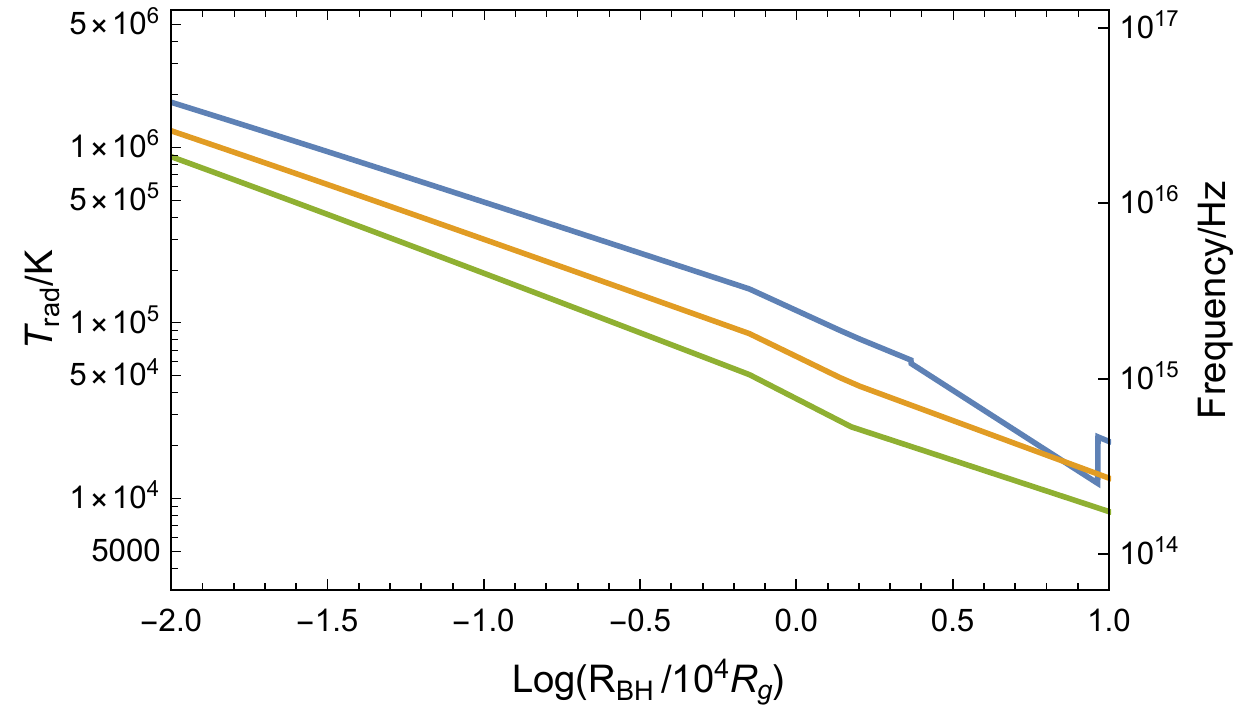}
	\end{center}
	\caption{The radiation features of the breakout outflow-driven shocked 
	shell. Top panels represent the estimated luminosity $L_{\rm{bre}}$
	of the breakout 
	photon for different cases of $s=0.2$, $s=0.5$, and $s=0.8$, 
	with the bolometric luminosity of AGN shown by the orange lines; 
	bottom panels represent the blackbody temperature $T_{\rm{rad}}$ of 
	the breakout photon. Each column represents 
	case with different SMBH mass of 
	$M=10^6 M_\odot$, $M=10^7 M_\odot$, and $M=10^8 M_\odot$, respectively. 
    The discontinuity and the steep drop derive from 
    the disk truncation by outflow.}
	\label{Fig:Lbre}
\end{figure*}

\begin{figure*}
	\begin{center}
		\includegraphics[width=0.325\textwidth]{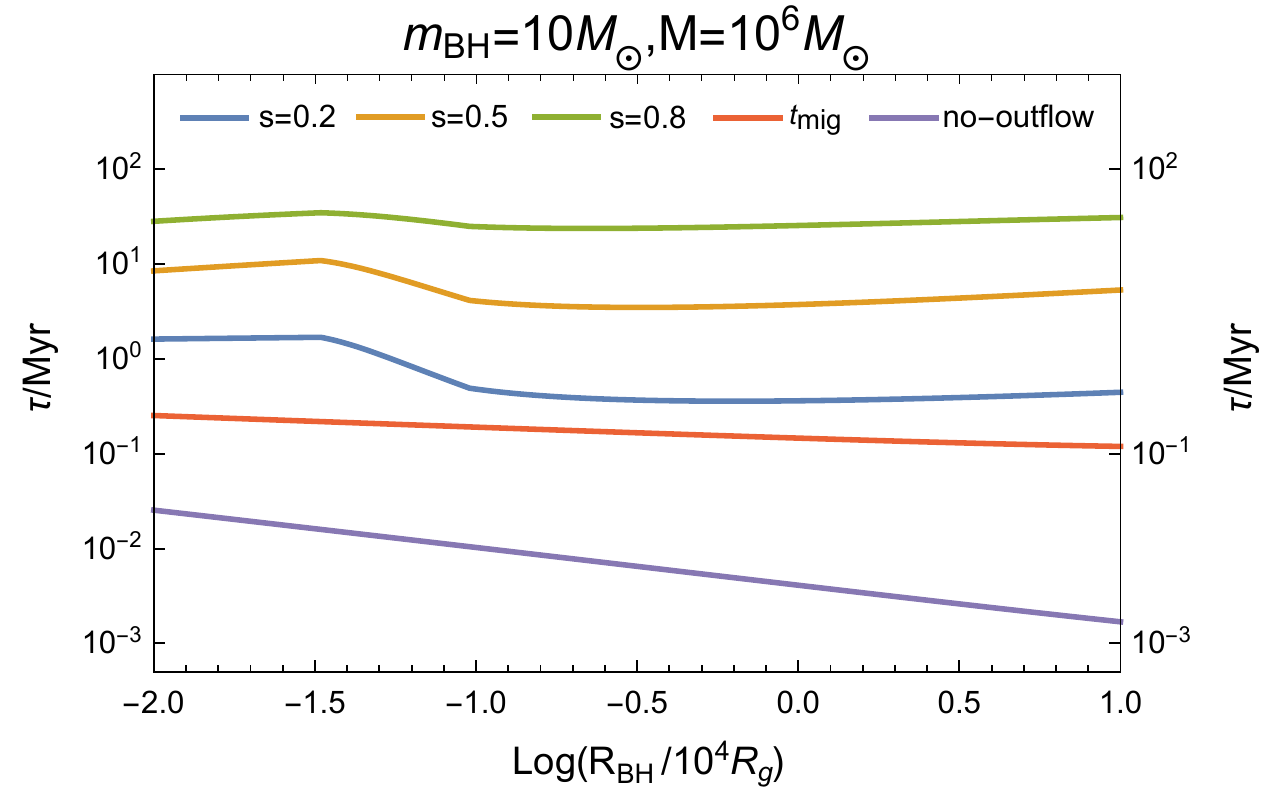}
		\includegraphics[width=0.325\textwidth]{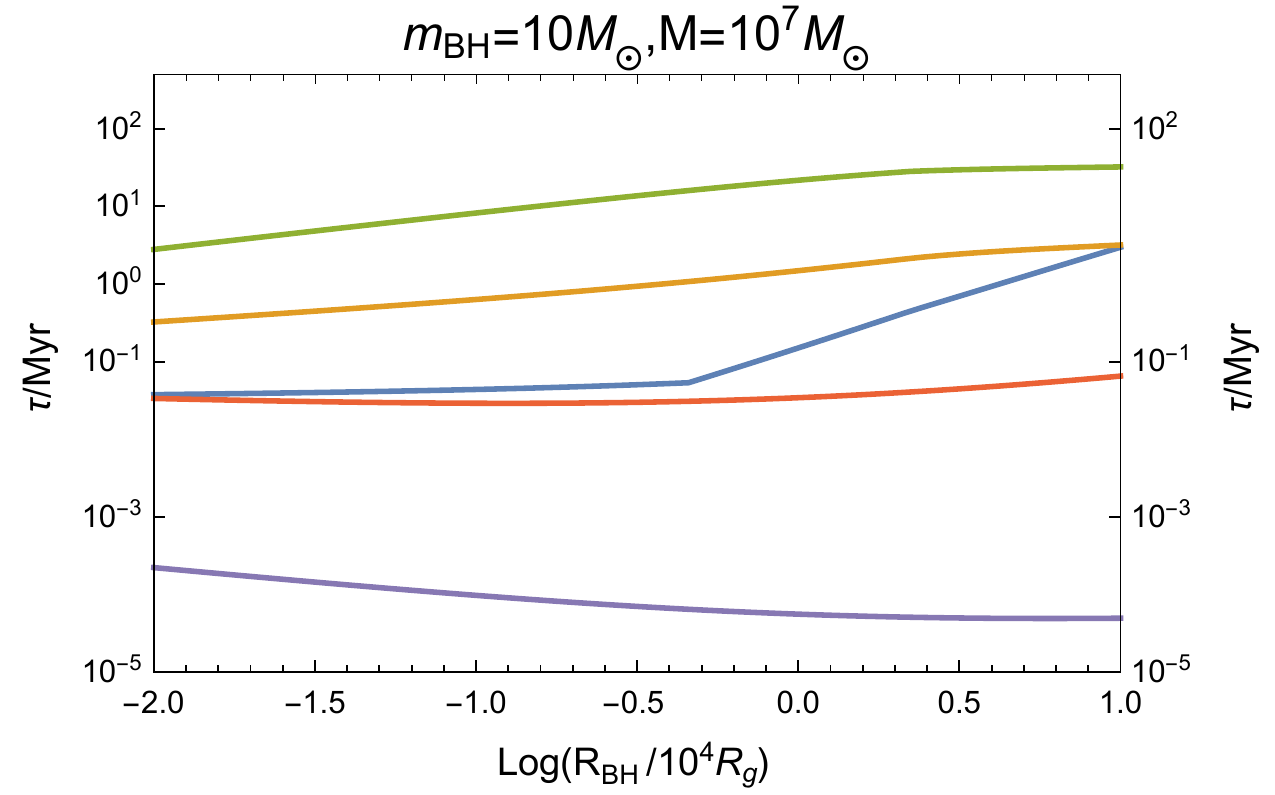}
		\includegraphics[width=0.325\textwidth]{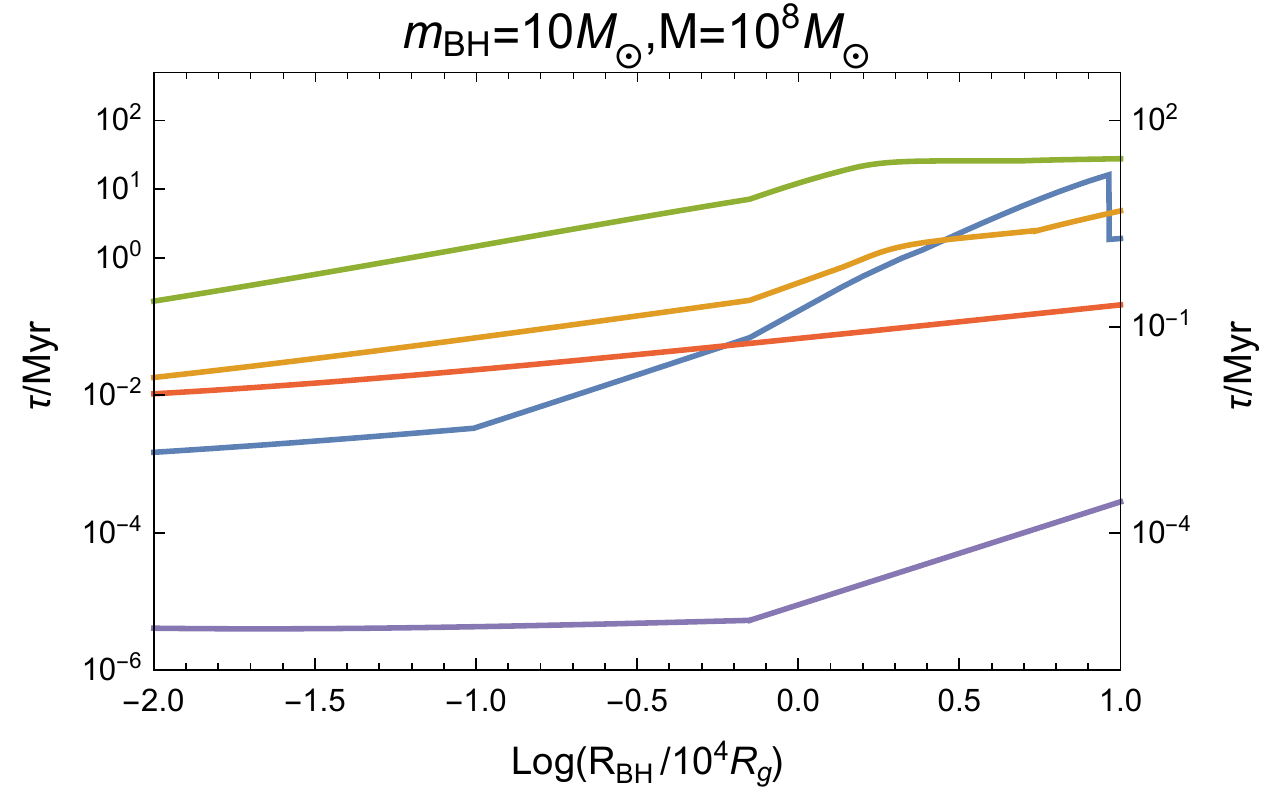}
	\end{center}
	\caption{The characteristic timescales linked to the mass 
	growth and the migration of CO embedded in the AGN disk, e.g., 
    a $m_{\rm{BH}}=10 M_{\odot}$ BH. The lines of $s=0.2$, 
    $s=0.5$, $s=0.8$ represent $t_{\rm{grow}}$ for the cases of 
    varying outflow strength, respectively. The orange and purple
    lines represent the migration timescale $t_{\rm{mig}}$, 
    of which the AGN disk structure is specifically expressed as 
    Equation (\ref{Eq:disk}), 
    and the mass growth timescale $t_{\rm{nowind}}$ leaving out 
    the outflow formation, respectively. Basically,
    $t_{\rm{nowind}} \ll t_{\rm{mig}} < t_{\rm{grow}}$ is valid.}
	\label{Fig:timescales}
\end{figure*}

\subsection{Jet Effect and Outflow EM Radiation} \label{subs44}
An additional process extracting the spin energy of an 
accreting BH via the formation of a jet would come into being 
when the BH is rotating, i.e. the Blandford-Znajek 
mechanism \citep{Blandford77}, the power of which 
$P_{\rm{BZ}} \propto B_{\perp}^2 a_{\rm{BH}}^2$, where $B_{\perp}$ is 
the magnetic field crossing the BH horizon
and $a_{\rm{BH}}$ is the BH’s dimensionless spin parameter. By two 
valid methods, we  can estimate the magnetic field strength around 
the BH: one is the winds produced by hyper-accreting circum-BH 
disk stretch the disk local magnetic fields to make large-scale 
poloidal components, or the AGN disk environment directly provides the
ordered poloidal magnetic fields, both of which are continuously advected 
inwards and accumulate at the vicinity of BH,
ultimately leading to a magnetically arrested disk \citep{Kimura21a, Kaaz22}, 
the maximum jet power is $L_{\rm{MAD}} \sim 2.2 a_{\rm{BH}}^2 
\dot{M}_{\rm{BH}} c^2$ for $h=0.5$ \citep[e.g.][]{Davis20, Curd23}; 
another is via 
building pressure balance $P_{\rm{mag}}=\alpha_{\rm{CO}}P_{\rm{disk}}$ at the inner 
boundary of the circum-BH disk with the assumption that 
magnetic fields are local, the jet power is 
$L_{\rm{BZ}} \sim 1/64 \ a_{\rm{BH}}^2 \dot{M}_{\rm{BH}} c^2$ 
\citep[e.g.][]{Wang21b}. The relative strength between the wind, 
Equation (\ref{Lw}), and the jet is
\begin{align}
L_{\rm{w}}/&L_{\rm{MAD(BZ)}}\sim 0.45(64) f_w a_{\rm{BH}}^{-2}
\frac{r_g}{r_{\rm{in}}} \nonumber \\
&\times\left[\frac{1-(r_{\rm{out}}/r_{\rm{in}} )^{s-1}}{1-s}-
\frac{1-(r_{\rm{out}}/r_{\rm{in}} )^{-3/2}}{3/2} \right] , 
\end{align}
the ratio of which is nearly independent of AGN disk 
parameters $M$ and $R_{\rm{CO}}$, and mainly dependent on index
$s$; for instance, $L_{\rm{w}}/L_{\rm{MAD}}\sim 0.03,\ 0.06,\ 0.18$ 
or $L_{\rm{w}}/L_{\rm{BZ}}\sim 4.3,\ 8.5,\ 25.6$
for $a_{\rm{BH}}=0.7$ and $s=0.2,\ 0.5,\ 0.8$ cases. 
Two estimations of the jet power cause inversely relative strength 
to the wind, so a more physical description of the magnetic 
field around BH is important to determine the jet power and the relevant 
feedback strength, and needs to be detailedly studied.

The jet-cocoon evolution can also generate mechanical feedback to 
suppress the accretion of BH in the AGN disk \citep{Kaaz21,Tagawa22}.
Assuredly, $L_{\rm{MAD}}>L_{\rm{w}}$ may cause jet other than wind 
dominants the feedback. However, the jet is highly anisotropic with 
narrow opening angle, and thus the size of cocoon cavity is smaller 
than $r_{\rm{Hill}}$ (Figure 1 in \citealt{Tagawa22}); as a contrast, we confirm 
that the wind outflow is nearly isotropic, and open a cavity much 
larger than $r_{\rm{Hill}}$, as shown in Figure \ref{Fig:cav}. Meanwhile, 
for the accretors being a BH with low spin $a_{\rm{BH}}$ or a NS and when the
magnetic field around BH is weak, the jet is feeble 
or even absent, leading to the feedback dominated by wind outflow. 
Therefore, we suggest that the jet feedback would play a secondary but
non-negligible role, and for the high spin BH case, a more 
complete feedback mechanism containing both jet and wind needs to be built.

We also estimate the radiation features of the wind-outflow-driven 
shocked shell during breakout, which may provide an opportunity to 
observe the cavity evolution. Following \cite{Kimura21b}, the luminosity during
shell breakout can be achieved from the photon energy released within 
the photon diffusion timescale, and the blackbody
temperature of the breakout photon can be estimated from the shock jump 
condition, i.e.,
\begin{equation}
	L_{\rm{bre}}\approx \frac{9 \pi H^2 \rho_{\rm{d}} 
	v_{\rm{shell}}(t_{\rm{bre}} )^3}{4} ,
\end{equation}
and
\begin{equation}
	T_{\rm{rad}}\approx \left[\frac{9 \rho_{\rm{d}} 
		v_{\rm{shell}}( t_{\rm{bre}} )^2}{4a} \right]^{1/4} ,
\end{equation}
where $a$ is the radiation constant; 
the variations of $L_{\rm{bre}}$ and $T_{\rm{rad}}$ under 
various parameters are shown in Figure \ref{Fig:Lbre}. As can be seen, the 
breakout radiation emerges as a soft X-ray/UV/optical transient depending on 
the CO location; meanwhile, the AGN emission is also powerful in 
X-ray/UV/optical bands \citep{Shang11}. By comparing $L_{\rm{bre}}$ with 
the AGN bolometric luminosity $L_{\rm{AGN}} \simeq 0.1 \dot{M}_{\rm{AGN}} c^2$, 
we loosely infer that the breakout events are 
general dimmer than the background emission, even though the extreme 
$s=0.2$ case, so the AGN disk would not be outshone. 
We confirm that the outflow cavity breakout events 
taking place in the AGN disk are generally unobservable.
As in \cite{Wang21a}, the subsequent interaction between 
the outflow bubble and the broad-line region gas may 
generate $\rm{TeV}$ flare, which can be detected in non-blazar AGNs.

\subsection{Potential role of Outflow Cavity Surrounding CO} \label{subs45}
There are two noteworthy features of the outflow cavity surrounding 
CO: CO is commonly harbored in the cavity environment because of 
the always satisfied $t_{\rm{acc}} \ll (t_{\rm{cav}}+t_{\rm{ref}} )$, 
and the gas density of cavity is much lower than the otherwise 
unperturbed AGN disk. These features would significantly affect the 
CO evolution and the CO-related events in AGN disk.

When embedded in the AGN disk, CO can 
not only capture the gas within its 
gravity sphere but also gravitational 
interact with the large-scale disk gas; 
correspondingly, the whole AGN disk gas exerts gravitational torques 
onto CO from the Lindblad and corotation resonances, 
leading to the CO migration \citep[e.g.][]{Ward97}, of which the 
timescale is expressed as \citep[e.g.][]{Kanagawa18}: 
\begin{align}
	t_{\rm{mig}}= &
	0.25\left(1+0.04K\right) \nonumber\\
	&\times\left(\frac{m_{\rm{CO}}}{M}\right)^{-1}\left(\frac{M}{\Sigma_{\rm{d}}
		R_{\rm{CO}}^2}\right)
	\left(\frac{H}{R_{\rm{CO}}}\right)^{2}\tilde{\Omega}_{\rm K}^{-1} ,
\end{align}
where $K=(m_{\rm{CO}}/M)^2(H/R_{\rm{CO}})^{-5}\alpha^{-1}$. 
Though the cavity with large size and long duration must modify
migration, the Lindblad resonance primarily originates from the 
whole background gas \citep[e.g.][]{Armitage07}, most of which 
are unaffected to provide torque, and during the cavity refilling 
process $t_{\rm{ref}}$, AGN disk gas can enter the cavity 
to gravitationally interact with the CO, again producing torque; so we 
ignore the effect of outflow cavity on migration for simplicity. Because 
CO mainly stays in the underdense cavity, the averaged mass accretion rate 
of CO reduces significantly, and thereby 
the mass growth rate of CO reduces as well. We estimate the CO mass 
growth timescale containing outflow feedback as 
$t_{\rm{grow}}=m_{\rm{CO}}/\dot{M}_{\rm{CO,red}}$ and the  
timescale omitting outflow as 
$t_{\rm{nowind}}=m_{\rm{CO}}/\dot{M}_{\rm{obd}}$. By comparing relevant 
timescales as shown in Figure \ref{Fig:timescales}, we find that the 
outflow feedback markedly alters the nature of CO evolution. 
If not bringing in outflow, 
the extremely large mass accretion rate results in a 
violent mass growth, $t_{\rm{nowind}}\ll t_{\rm{mig}}$, thus the CO 
evolves in situ without migration. Conversely, the formation of 
outflow and the concomitant feedback lead to a moderate mass growth, 
basically $t_{\rm{grow}} > t_{\rm{mig}}$, so the CO can observably migrate 
before significantly becoming heavier. In a word, the accretion-induced 
outflow strangles the mass growth of CO.

As have already been mentioned, a variety of EM events involving CO 
are likely to take place in the AGN disk. Since the striking difference 
between the cavity and AGN disk environment (e.g. density, 
temperature, opacity), EM events would possess distinguishable 
features when they separately occur in the two environments. 
Propagating within 
the unperturbed dense AGN disk, the jet, ejecta, and outflow generated by 
the EM events undergo rapid deceleration, which can produce 
characteristic radiations. %notably different from the same events occurring 
%in the classical interstellar medium environment. 
On the other hand, as the cavity is much more tenuous 
than the AGN disk but still denser than the interstellar medium, these 
jet, ejecta, and outflow would not be significantly decelerated 
but still interact with the cavity medium \citep[e.g.][]{Yuan22}, the produced
EM emissions would propagate within the cavity environment as well
\citep[e.g.][]{Kimura21b}, which may thus lead to characteristic 
signals differing from the ones in an unperturbed AGN disk or in a classical 
interstellar medium. Also, as the underdense cavity has a limited 
size, matter penetrating along the AGN disk 
will firstly collide with cavity and then AGN disk gas, causing a latter 
effective loss of the kinetic energy and thereby latter brightening, 
which may apply to the events with wide-angle ejecta and outflow, 
as a star tidally disrupted by a stellar mass BH  
\citep[e.g.][]{Perets16, Yang22}, 
or BH-NS and NS-NS mergers \citep[e.g.][]{Zhu21a, Ren22}. 
A more precise description of the cavity structure 
and the properties of various EM events occurring 
in the cavity-AGN disk environment are left in a follow-up work.

\cite{Tagawa20a} showed that single-single gas-capture channel dominates 
the binary formation in the AGN disk, where the gas dynamical friction 
acting on CO, which is approximately proportional to the 
gas density surrounding CO, i.e.,
$a_{\rm{GDF}} \propto \rho_{\rm{CO}}$ \citep{Ostriker99}, effectively
removes the binary binding energy during the two bodies encountering in their mutual 
Hill radius, and thereby a CO binary forms. However, as  
COs are mainly harbored in the cavity environment, of which the gas density 
is much lower than the otherwise unperturbed AGN disk
and the size is much larger than the Hill radius, 
as shown in Figure \ref{Fig:cav}, the gas dynamical friction can be 
significantly weakened, $a_{\rm{GDF,cav}} \sim O(10^{-4}-10^{-2} ) \ a_{\rm{GDF}}$, 
hence the efficiency of single-single gas capture channel 
may be markedly reduced. In brief, the outflow 
feedback plays an important role 
and should not be neglected when studying the 
binary formation in the AGN disk.

\subsection{Accretion Feedback of CO with Large Bulk Velocity} \label{subs46}
CO can possess a large bulk velocity relative to the AGN disk gas, 
i.e. $v_{\rm{bulk}} \gg \tilde{c}_s \ and \ v_{\rm{shear}}$, 
e.g., the orbiting CO in the galactic nucleus holding 
nonzero inclination w.r.t. the disk plane will periodically cross 
the AGN disk with considerable perpendicular velocity 
\citep[e.g.][]{Fabj20, Nasim22}, or the remnant of a binary CO merger will 
be ejected at large recoil velocity \citep[e.g.][]{Campanelli07}, 
or the CO can leave triple 
system with large escape velocity after the chaotic interaction 
\citep{Valtonen06}, and so on. In the case of supersonic motion, 
CO triggers the formation of a bow shock, producing a downstream overdense  
shocked gas wake, which drags the CO and is accreted 
onto it \citep[e.g.][]{Antoni19}. Meanwhile, the BHL mass inflow rate is still 
hyper-Eddington, e.g. $\dot{M}_{\rm{BHL}}=1.6 \times 10^3 \dot{M}_{\rm{Edd}}
(m_{\rm{CO}}/M_\odot)(\rho_{\rm{d}}/10^{-10}\,{\rm g\,cm^{-3}} )(v_{\rm{bulk}}/
10^{8}\,{\cm \s^{-1}} )^{-3}$, so the radiation-driven outflow should emerge, 
the feedback of which results in the redistribution of  
bow shock separating the outflow and the surrounding gas, and an underdense wake 
region may be built, depending on the outflow geometry \citep{Li20}. 
The accretion and dynamical friction on CO, 
which are important for the evolution of CO supersonically 
moving within AGN disk and the capture of CO by AGN 
disk \citep{Fabj20, Nasim22}, would be significantly affected 
by the potential outflow \citep{Gruzinov20, Li20, Bosch-Ramon22, Kaaz22}.

However, whether an outflow really forms and the 
concrete mechanism of its ejection from
the supersonically moving CO have not been 
well studied, much less in the particular AGN disk 
environment. For BHL accretion with large 
$v_{\rm{bulk}}$, the inflow would be highly quasi-spherical 
\citep[e.g.][]{Mellah15}, so no region is left to release outflow, 
and photons are trapped and advected inward with 
inflow at hyper-Eddington rate \citep[e.g.][]{Begelman79, Blondin86}. 
Vast mass and energy ultimately flow into the
horizon for the case of BH, or eventually hit the 
hard surface and potentially release huge energy 
for the case of NS \citep[e.g.][]{Shakura15}.
%or are blocked at the 
%magnetosphere boundary and enter the magnetosphere 
%via Rayleigh-Taylor instability for the case of NS 
%with strong magnetic field \citep{Shakura15}; or
%would directly interact with the hard 
%surface and potentially release huge energy  
%for the case of NS with weak magnetic field. 
On the other hand, the initial infalling 
gas may carry a small but unnegligible amount of 
angular momentum depending on the AGN disk properties, 
and thus an accretion disk may 
form around CO to launch winds via the 
mechanism discussed in Section \ref{2-2}; 
a self-consistent evolution process should come into being 
(probably alike \citealt{Wang21a}).

Since the accretion feedback of CO with a large bulk velocity 
is vital but poorly understood, we would prepare another separate
work to study the launch mechanisms and the properties of 
outflow; the interaction between the outflow and the AGN 
gas medium, the effects of outflow on CO accretion process 
and dynamical drag with the induced 
acceleration/deceleration of CO, the ability of 
AGN disk on capturing the crossing CO when 
considering accretion feedback, and the relevant EM 
radiations would be studied as well.

\section{Summary} \label{sec:Summary}

In this work, we have explored the role of outflow feedback 
on accretion of the CO embedded in an AGN disk. 
We have revealed that winds launched from the 
circum-CO hyper-Eddington accretion disk are 
asymptotically isotropic and would truncate the disk to halt 
CO accretion. Interaction between the outflow and the
AGN disk gas gives rise to a cyclic process of formation 
and refilling of a long-running outflow cavity, 
of which the size is larger 
than the CO gravity sphere (BHL and Hill radius) 
and the density is lower than that of the unperturbed environment. 
Efficient CO accretion takes place in the AGN disk 
rather the cavity environment, and therefore the outflow feedback 
leads to an accretion duty-circle of $O(10^{-4})-O(10^{-1})$.
We have found that if taking the influences of outflow and its 
feedback into account, then the mass accretion rate onto a BH, 
which now ranges $O(10)-O(10^{5})\ \dot{M}_{\rm{Edd}}$ depending on 
the AGN disk parameters, the BH's location, 
and the outflow strength, is extremely 
reduced in comparison to the initial gas captured rate.
Although the outflow feedback itself is hard to be 
observed, it remarkably changes the CO evolution character,
causing considerable CO migration before sizable mass 
growth via the reduction of time-averaged CO mass 
growth rate; and it prevents the AGN disk from being depleted by 
the otherwise violent CO accretion.

We have proved that both cavity formation and mass accretion 
reduction are the universal roles of outflow 
feedback, independent of the cooling efficiency of  
shocked gas and whether the gravity gap opens or not. Moreover, 
the underdense cavity environment can impact 
the features of CO-related EM events and the CO binary 
formation efficiency, which are significant roles of 
outflow feedback as well.
We suggest that jet feedback would also play a 
role for the cases of BH with large spin and strong 
magnetic field around its horizon.

In addition, we have investigated the role of 
outflow feedback on NS accretion due to the existence
of NS magnetic field and hard surface,  
of which the general properties are essentially 
unchanged. But for a NS case, the accretion process 
is relatively complex, because the more powerful outflow 
feedback derived from the energy released near NS surface weakens 
the accretion, reversely the magnetospheric 
disk truncation makes mass rate accreted onto NS larger. 
So we suggest that the NS evolution and NS-related 
events need to be specifically studied.

\begin{acknowledgements}
	
We would like to thank the referee for helpful 
and valuable comments and suggestions.
This work was supported by the National Key Research 
and Development Program of China (grant No. 2017YFA0402600),
the National SKA Program of China (grant No. 2020SKA0120300), 
and the National Natural Science Foundation of China 
(grant No. 11833003)

\end{acknowledgements}

\begin{appendix}
	
\section{Self-similar Solutions of Hyper-Eddington Accretion Disk}
\label{Appendix-disk}
Although the hyper-Eddington accretion process around CO is complicated 
and should be investigated via numerical
simulation methods \citep[e.g.][et al.]{Ohsuga05, Yang14, Jiang14, 
Sadowski151, Kitaki18}, 
analytic solutions are valuable because they can roughly 
match the simulation results \citep{Jiao15}, and it is convenient 
to use these solutions to study the accretion systems under wide 
parameter space. When analytically solving the disk structure, 
self-similar assumption has been widely adopted 
\citep[et al.]{Blandford99, Blandford04, Begelman12, Gu12,Gu15, 
Wu22, Ghoreyshi20, Zeraatgari20}; 
though these solutions would fail near the inner and the outer 
boundaries, most regions of the disk can be well described. 
In this section, we build the self-similar disk with 
wind outflow taken into account, and estimate the 
contribution of outflow to carrying away the viscous heat of 
the hyper-Eddington accretion disk.

We consider a steady state axisymmetric accretion flow. For simplicity, 
we use the Newtonian point-mass potential $ \Phi=-G m_{\mathrm{CO}} /r $,  
and we ignore the vertical structure of the inflow and 
assume that all flow variables depend only on radial distance $r$. 
The vertical scale height of the disk is 
$H_{\rm{CCOD}} = c_{\rm{s}}/\Omega_{\rm{K}}$, where 
$c_{\rm{s}} = (P/\rho)^{1/2}$ denotes the sound 
speed, among which $P$ and $\rho$ are pressure and density of the 
inflow, respectively. We only consider $r\phi$-component of the 
shear stress tensor and the viscosity is 
$\nu = \alpha_{\rm{CO}} c_{\rm{s}} H_{\rm{CCOD}}$.

We assume that the mass inflow rate $\dot{M}$ varies with radius $r$ 
due to the mass loss via outflow \citep{Blandford99, Wu22} and is
described by
\begin{equation}
\dot{M} =-2\pi r\Sigma v_{\rm{r}}=\dot{M}_{\rm{out}}\left(\frac{r}
{r_{\rm{out}}} \right)^{s} \ , \label{Macc}
\end{equation} 
the outflow mass rate $\dot{M}_{\rm{w}}(r)\sim \dot{M}(r)$, 
accordingly the wind loss rate per unit area 
$\dot{m}_{\rm w}(r)$ can be calculated by:
\begin{equation}
\dot{M}_{\rm w}(r) = \int_{r_{\rm{in}}}^{r} 4\pi r^\prime\dot{m}_{\rm w}
(r^\prime)dr^\prime \ , \label{}
\end{equation}
and so
\begin{equation}
\dot{m}_{\rm w}=\frac{s\dot{M}}{4\pi r^2} \ .
\end{equation}

Including the influence of outflow \citep[e.g.][]{Knigge99}, 
the continuity equation, the vertically integrated radial momentum and 
azimuthal momentum equations are
\begin{equation}
\frac{1}{r}\frac{d}{d r}(r\Sigma v_{\rm{r}} )+\frac{1}{2\pi r}\frac
{d \dot{M}_{\rm w}}{d r} = 0 \ , \label{eq4}
\end{equation}
\begin{equation}
v_{\rm r}\frac{dv_{\rm r}}{dr} + (\Omega_{\rm K}^2-\Omega^2)r + 
\frac{1}{\rho}\frac{dP}{dr} = 0 \ , \label{eq5}
\end{equation}
\begin{equation}
-\frac{1}{r}\frac{d}{d r}\left(r^3\Sigma v_{\rm r}\Omega \right)+\frac{1}{r}
\frac{d}{d r}\left(r^3\nu \Sigma\frac{d \Omega}{d r} \right)-
\frac{(lr)^2\Omega}{2\pi r}\frac{d \dot{M}_{\rm {w}}}{d r} = 0 \ ,
\label{eq6}
\end{equation}
where Equation~(\ref{eq6}) contains the loss of angular momentum via 
outflow; $l\geqslant1$ reflects the amount of specific angular momentum carried 
away by outflow, where $l=1$ represents the wind carrying away the local 
angular momentum of inflow at the radius where wind 
originates \citep{Knigge99}. 
Combining Equations~(\ref{eq4}) and~(\ref{eq6}) with the adoption of a 
torque-free inner boundary condition, i.e. 
$T_{r\phi}(r_{\rm in} )=r\nu\Sigma d\Omega/dr=0$, 
and $\Omega\varpropto r^{-3/2}$, we can get
\begin{equation}
\nu\Sigma=\frac{1}{3\pi}\left(1-\frac{l^2s}{s+\frac{1}{2}} \right)\dot{M}\left[1-\frac{\Omega_{\rm in}}{\Omega}\left(\frac{r_{\rm in}}{r} \right)^{s+2}\right]\ , \label{Qvis}
\end{equation} 
where $r_{\rm in}$ is set to be the inner boundary of the disk. 
The energy equation is
\begin{equation}
Q_{\rm{vis}} = Q_{\rm adv} + Q_{\rm w} \ , \label{eq8}
\end{equation}
where $Q_{\rm{vis}}$, $Q_{\rm{adv}}$ and $Q_{\rm{w}}$ are the viscous 
heating rate, the advective cooling rate and the energy rate taken away by 
outflow per unit area, respectively. 
We approximatively ignore the convection cooling term $Q_{\rm con}$ 
and the radiation cooling term $Q_{\rm rad}$ because for hyper-Eddington
accretion flow, the convection is relatively less important than 
the advection \citep{Jiao15}, and the radiation is effectively 
trapped in the flow \citep{Begelman12}, leading to a convectively 
and radiatively inefficient flow. The expressions of the energy 
terms are as follows \citep[e.g.][]{Abramowicz95}:
\begin{equation}
Q_{\rm{vis}} = \nu \Sigma\left(r\frac{d\Omega}{dr} \right)^2 \ , \label{}
\end{equation}
\begin{equation}
Q_{\rm adv} = \Sigma v_{\rm r} T\frac{dS}{dr} = \Sigma v_{\rm r}\left(\frac{1}
{\gamma-1}\frac{dc_{\rm s}^2}{dr} - \frac{c_{\rm s}^2}{\rho}\frac{d\rho}{dr}
\right) \ , \label{}
\end{equation}
where $S$ is the specific entropy and $\gamma$ is the ratio of 
specific heat. The energy taken away by outflow is simply assumed to be
 comparable with the local gas kinetic energy:
\begin{equation}
Q_{\rm w} = \eta \dot{m}_{\rm w} v_{\rm K}^2
\left[1-\frac{\Omega_{\rm in}}{\Omega}
\left(\frac{r_{\rm in}}{r} \right)^{s+2}\right] \ ,
\end{equation}
where $\eta$ reflects the specific energy of outflow, and the square 
brackets part is added to ensure the self-similarity of Equation~(\ref{eq8}).

Therefore, the specific expressions of the self-similar variables, 
$H_{\rm{CCOD}}\varpropto r$, $\Omega\varpropto r^{-3/2}$, 
$v_{\rm r}\varpropto r^{-1/2}$, $\rho\varpropto r^{s-3/2}$, 
and $P\varpropto r^{s-5/2}$, can be solved using the 
Equations~(\ref{eq4}),~(\ref{eq5}),~(\ref{eq6}),~(\ref{eq8})
and the expression of $c_s$. 
In this work, we primarily focus on the total heat energy 
and the outflow feedback energy of the hyper-Eddington accretion disk, 
the ratio between these two is:
\begin{equation}
\frac{Q_{\rm{w}}}{Q_{\rm{vis}}}=\frac{\eta \dot{m}_{\rm w} v_{\rm K}^2\left[1-\frac{\Omega_{\rm in}}{\Omega}(\frac{r_{\rm in}}{r})^{s+2}\right]}{\nu \Sigma\left(r\frac{d\Omega}{dr}\right)^2}=\frac{\eta s(1+2s)\Omega^2_{K}}{3(1+2s-2l^2s)\Omega^2} > \frac{1}{3}\eta s(1+2s)\ ,
\end{equation}
where the inequality holds for the criteria $l\geqslant1$, $0<s\leqslant1$ and 
$\Omega < \Omega_{K}$, which are satisfied in hyper-Eddington 
accretion disk. The outflow cooling rate needs to be less than the 
total viscous heating rate, yet the outflow should be fast enough 
to propagate far away (but see the breeze solutions in \citealt{Begelman12}), 
thereby restricts $1\leqslant\eta<3/s(1+2s)$. 
We take $(Q_{\rm w}+Q_{\rm rad})/Q_{\rm vis}
\sim Q_{\rm w}/Q_{\rm vis}\sim O(10^{-1})$ as a reasonable ratio. 
The exact values of $\eta$ and $s$ in the large-scale hyper-Eddington 
accretion system should be determined by numerical simulations.
 
\section{Gas Density and Height of Hyper-Eddington Accretion Disk}
\label{appendix-disk-density}
Circum-CO disk with diverse values of the initial mass inflow 
rate $\dot{M}_{\rm {obd}}$ would manifest piecewise structure, 
as discussed in Section \ref{2-2}. Inside $r_{\rm{tr}}$, the 
disk is advection-cooling dominated, gas density of the disk
region is 
\setcounter{equation}{0}
\begin{equation}
	\rho_{\rm{CCOD,adv}}=\frac{\Sigma}{2H_{\rm{CCOD}}}\simeq 
	\frac{\dot{M}_{\rm {in}}}{4 \pi r^2 h v_r}=
	\frac{\dot{M}_{\rm {in}}}{4 \pi \alpha_{\rm{CO}} r^2 h^3 v_{\rm{K}}}, 
	\label{B1}
\end{equation}
where $r$ is the actual value of disk radius. 
Outside $r_{\rm{tr}}$, the disk is 
radiation-cooling dominated, thereby the standard disk structure can 
probably describe the region \citep[e.g.][]{Kato08}.
To be consistent with the expressions of
AGN disk structure, i.e. Equation (\ref{Eq:disk}), we consider pure hydrogen 
medium; the pressure is the sum of radiation and 
gas components, $P=P_{\rm{rad}}+P_{\rm{gas}}=a T^4 /3+ 2\rho k T/m_p$;
the energy equation is $Q_{\rm{vis}}=Q_{\rm{rad}}$, where 
$Q_{\rm{rad}}=16 a c T^4/3 \kappa \Sigma$; and the main opacity sources 
include electron scattering and free-free absorption, 
$\kappa=\kappa_{\rm{es}}+\kappa_{\rm{ff}}$, where 
$\kappa_{\rm{es}}=0.40 \cm^2\g^{-1}$ and 
$\kappa_{\rm{ff}}=0.64 \times 10^{23}\cm^2\g^{-1}\rho T^{-7/2}$. 
The disk structure can be solved by combining the energy equation 
and Equation (\ref{Macc}). For the inner region where $P=P_{\rm{rad}}$ and 
$\kappa=\kappa_{\rm{es}}$, the disk density and height are 
\begin{align}
\rho_{\rm{CCOD,rad}}=5.5 \times 10^{-12} \g \cm^{-3}
\alpha_{\rm{CO}}^{-1} m^{-5/2} \dot{m}^{-2} r^{3/2},
\end{align}
\begin{align}
H_{\rm{CCOD,rad}}=6.6 \times 10^{4} \cm \ m \ \dot{m},
\end{align}
where $m=m_{\rm{CO}}/M_\odot$ and 
$\dot{m} \equiv \dot{M}_{\rm {obd}}/\dot{M}_{Edd}$; 
for the middle region where 
$P=P_{\rm{gas}}$ and $\kappa=\kappa_{\rm{es}}$, the disk density
and height are
\begin{align}
\rho_{\rm{CCOD,gas}}=8.8 \times 10^{9} \g \cm^{-3}
\alpha_{\rm{CO}}^{-7/10} m^{19/20} \dot{m}^{2/5} r^{-33/20},
\end{align}
\begin{align}
H_{\rm{CCOD,gas}}=4.8 \times 10^{-3} \cm \ 
\alpha_{\rm{CO}}^{-1/10} m^{-3/20} \dot{m}^{1/5} r^{21/20},
\end{align}
and for the outer region where $P=P_{\rm{gas}}$ and 
$\kappa=\kappa_{\rm{ff}}$, the disk density and height are
\begin{align}
\rho_{\rm{CCOD,ff}}=8.8 \times 10^{11} \g \cm^{-3}
\alpha_{\rm{CO}}^{-7/10} m^{47/40} \dot{m}^{11/20} r^{-15/8},
\end{align}
\begin{align}
H_{\rm{CCOD,ff}}=1.0 \times 10^{-3} \cm \ 
\alpha_{\rm{CO}}^{-1/10} m^{-9/40} \dot{m}^{3/20} r^{9/8}.
\end{align}
The boundary between these three regions are radii where 
$P_{\rm{rad}}=P_{\rm{gas}}$ and $\kappa_{\rm{es}}=\kappa_{\rm{ff}}$, i.e.,
\begin{equation}
	r_{\rm{rad-gas}}=36 r_{g} \alpha_{\rm{CO}}^{2/21} m^{2/21} \dot{m}^{16/21}
	=5.4\times 10^6 \cm \ \alpha_{\rm{CO}}^{2/21} m^{23/21} \dot{m}^{16/21},
\end{equation}
and
\begin{equation}
r_{\rm{es-ff}}=4.7\times 10^3 r_{g} \dot{m}^{2/3}
=7.1\times 10^8 \cm \ m \ \dot{m}^{2/3}.
\end{equation}

We tentatively use the standard disk model, in spite of the potential 
thermal instabilities at the radiation-cooling inner 
region \citep[e.g.][]{Shen14}, because the instability is still poorly 
understood in the theory of accretion disks, with some simulations show the 
thermally stable disk \citep{Hirose09}, and the general viscosity 
$\nu \propto \alpha_{\rm{CO}}P^{1-\mu}P_{\rm{gas}}^\mu$ can relieve the 
instability. Also, we note that when the initial mass inflow rate is 
extremely large, there can be $r_{\rm{tr}}>r_{\rm{rad-gas}}$, 
namely the disk skips over 
the radiation-cooling and radiation pressure dominated state to directly 
become advection-dominated, which is unphysical and leads to the jump of 
relevant features, e.g. the discontinuity in Figure \ref{Fig:Lbre} and 
Figure \ref{Fig:timescales}. The inaccuracy derives from the 
simplified expression of the boundary $r_{\rm{tr}}$  
and the directly connection between the outer 
standard disk and the inner advective disk.
The accurate construction of the hyper-Eddington accretion 
disk should simultaneously 
contain advection and radiation term all the radius, 
and should deal with the formation of disk wind 
more specifically \citep[e.g.][]{Blandford04}. In this work we  
build the circum-CO disk roughly,  which we think is enough to reveal the disk 
features, to focus on the outflow feedback;
we leave the specific structure of
hyper-Eddington accretion disk in a future work.

\section{Evolution of Shell in Momentum Conservation or Adiabatic Case}
\label{appendix-extreme-case}
For the momentum conservation case, the shocked outflow and the swept AGN disk gas 
are assumed to lose all thermal energy immediately, the expansion of the
shell is completely driven by the outflow momentum, 
i.e. Equation (\ref{pwind}). The radius and velocity of the expanding shell 
evolve with time as \citep{Dyson97}:
\setcounter{equation}{0}
\begin{equation}
	r_{\rm{shell,m}}=0.83\left(\frac{\dot{p}_{\rm{w}} t^{2}}{\rho_{\rm{d}}}\right)^{1/4} ,  \label{eqrshellm}
\end{equation}
and
\begin{equation}
	v_{\rm{shell,m}}=0.42\left(\frac{\dot{p}_{\rm{w}}}{\rho_{\rm{d}}t^{2}}\right)^{1/4} . \label{eqvshellm}
\end{equation}
After the stop of CO accretion, the evolution of shell follows:
\begin{equation}
	r_{\rm{sh}}^{2} \dot{r}_{\rm{sh}} \simeq r_{\rm{shell,m}}^{2}(t_{\rm{acc}}) 
	v_{\rm{shell,m}}(t_{\rm{acc}}) . \label{eqmm}
\end{equation}
Also, if the efficient accretion stops before breakout, 
the swept shell expands roughly
driven by the total momentum injected by the outflow, 
$P_{\rm{w}}=\dot{p}_{\rm{w}} t_{\rm{acc}}$, 
the radius and velocity are approximately given by \citep{Dyson97} 
\begin{equation}
	r_{\rm{shellP}}\simeq \left(\frac{P_{\rm{w}} t}{\rho_{\rm{d}}}\right)^{1/4} , \label{eqrshellP}
\end{equation}
and
\begin{equation}
	v_{\rm{shellP}}\simeq 0.25\left(\frac{P_{\rm{w}}}{\rho_{\rm{d}}t^{3}}\right)^{1/4} . \label{eqvshellP}
\end{equation}
After the breakout, the evolution of shell follows:
\begin{equation}
	r_{\rm{shP}}^{2} \dot{r}_{\rm{shP}} \simeq r_{\rm{shellP}}^{2}(t_{\rm{breP}}) 
	v_{\rm{shellP}}(t_{\rm{breP}}) . \label{eqmP}
\end{equation} 
The half-width, the formation and refilling timescales of cavity can be 
achieved analogously as Equations (\ref{eqrcav}), (\ref{eqtcav}) 
and (\ref{eqref}). Or if the shell fails to punch the AGN disk, we use 
$ v_{\rm{shellP}}=\tilde{c}_{\rm{s}}$ as a criterion to study the 
evolution of shell.

For the adiabatic case, we roughly assume the whole accretion-released energy 
$E_{\rm{w}}=L_{\rm{w}}t_{\rm{acc}}$ drives the shell expansion all the time, 
thereby the radius and velocity are again approximately given as 
\begin{equation}
	r_{\rm{shellE}}=\left(\frac{E_{\rm{w}} t^{2}}{\rho_{\rm{d}}}\right)^{1/5} , 
\end{equation}
and
\begin{equation}
	v_{\rm{shellE}}=0.4\left(\frac{E_{\rm{w}}}{\rho_{\rm{d}}t^{3}}\right)^{1/5} . 
\end{equation}
The half-width, the formation and refilling timescales of cavity
can be calculated via the criterion where
\begin{equation}
	v_{\rm{shellE}}=\tilde{c}_{\rm{s}} . 
\end{equation}
\end{appendix}


\begin{thebibliography}{63}

\bibitem[Abbott et al.(2020)]{Abbott20} Abbott, R., Abbott, T.~D., Abraham, S., et al.\ 2020, \prl, 125, 101102. 10.1103/PhysRevLett.125.101102

\bibitem[Abramowicz et al.(1995)]{Abramowicz95} Abramowicz, M.~A., Chen, X., Kato, S., et al.\ 1995, \apjl, 438, L37. doi:10.1086/187709

\bibitem[Antoni et al.(2019)]{Antoni19} Antoni, A., MacLeod, M., \& Ramirez-Ruiz, E.\ 2019, \apj, 884, 22. doi:10.3847/1538-4357/ab3466

\bibitem[Armitage(2007)]{Armitage07} Armitage, P.~J.\ 2007, astro-ph/0701485

\bibitem[Ayliffe \& Bate(2009)]{Ayliffe09} Ayliffe, B.~A. \& Bate, M.~R.\ 2009, \mnras, 397, 657. doi:10.1111/j.1365-2966.2009.15002.x

\bibitem[Bartos et al.(2017)]{Bartos17} Bartos, I., Kocsis, B., Haiman, Z., et al.\ 2017, \apj, 835, 165. doi:10.3847/1538-4357/835/2/165

\bibitem[Baruteau et al.(2014)]{Baruteau14} Baruteau, C., Crida, A., Paardekooper, S.-J., et al.\ 2014, Protostars and Planets VI, 667. doi:10.2458/azu\_uapress\_9780816531240-ch029

\bibitem[Begelman(1979)]{Begelman79} Begelman, M.~C.\ 1979, \mnras, 187, 237. doi:10.1093/mnras/187.2.237

\bibitem[Begelman(2012)]{Begelman12} Begelman, M.~C.\ 2012, \mnras, 420, 2912. doi:10.1111/j.1365-2966.2011.20071.x

\bibitem[Bellovary et al.(2016)]{Bellovary16} Bellovary, J.~M., Mac Low, M.-M., McKernan, B., et al.\ 2016, \apjl, 819, L17. doi:10.3847/2041-8205/819/2/L17

\bibitem[Blandford \& Znajek(1977)]{Blandford77} Blandford, R.~D. \& Znajek, R.~L.\ 1977, \mnras, 179, 433. doi:10.1093/mnras/179.3.433

\bibitem[Blandford \& Begelman(1999)]{Blandford99} Blandford, R.~D. \& Begelman, M.~C.\ 1999, \mnras, 303, L1. doi:10.1046/j.1365-8711.1999.02358.x

\bibitem[Blandford \& Begelman(2004)]{Blandford04} Blandford, R.~D. \& Begelman, M.~C.\ 2004, \mnras, 349, 68. doi:10.1111/j.1365-2966.2004.07425.x

\bibitem[Blondin(1986)]{Blondin86} Blondin, J.~M.\ 1986, \apj, 308, 755. doi:10.1086/164548

\bibitem[Bosch-Ramon(2022)]{Bosch-Ramon22} Bosch-Ramon, V.\ 2022, \aap, 660, A5. doi:10.1051/0004-6361/202142821

\bibitem[Campanelli et al.(2007)]{Campanelli07} Campanelli, M., Lousto, C.~O., Zlochower, Y., et al.\ 2007, \prl, 98, 231102. doi:10.1103/PhysRevLett.98.231102

\bibitem[Chashkina et al.(2019)]{Chashkina19} Chashkina, A., Lipunova, G., Abolmasov, P., et al.\ 2019, \aap, 626, A18. doi:10.1051/0004-6361/201834414

\bibitem[Cheng \& Wang(1999)]{Cheng99} Cheng, K.~S. \& Wang, J.-M.\ 1999, \apj, 521, 502. doi:10.1086/307572

\bibitem[Coughlin \& Begelman(2014)]{Coughlin14} Coughlin, E.~R. \& Begelman, M.~C.\ 2014, \apj, 781, 82. doi:10.1088/0004-637X/781/2/82

\bibitem[Curd \& Narayan(2023)]{Curd23} Curd, B. \& Narayan, R.\ 2023, \mnras, 518, 3441. doi:10.1093/mnras/stac3330

\bibitem[Davis \& Tchekhovskoy(2020)]{Davis20} Davis, S.~W. \& Tchekhovskoy, A.\ 2020, \araa, 58, 407. doi:10.1146/annurev-astro-081817-051905

\bibitem[Dittmann et al.(2021)]{Dittmann21} Dittmann, A.~J., Cantiello, M., \& Jermyn, A.~S.\ 2021, \apj, 916, 48. doi:10.3847/1538-4357/ac042c

\bibitem[Dyson \& Williams(1997)]{Dyson97} Dyson, J.~E. \& Williams, D.~A.\ 1997, The physics of the interstellar medium.  Edition: 2nd ed. Publisher: Bristol: Institute of Physics Publishing, 1997. Edited by J. E. Dyson and D. A. Williams. Series: The graduate series in astronomy. ISBN: 0750303069. doi:10.1201/9780585368115

\bibitem[Edgar(2004)]{Edgar04} Edgar, R.\ 2004, \nar, 48, 843. doi:10.1016/j.newar.2004.06.001

\bibitem[El Mellah \& Casse(2015)]{Mellah15} El Mellah, I. \& Casse, F.\ 2015, \mnras, 454, 2657. doi:10.1093/mnras/stv2184

\bibitem[Fabj et al.(2020)]{Fabj20} Fabj, G., Nasim, S.~S., Caban, F., et al.\ 2020, \mnras, 499, 2608. doi:10.1093/mnras/staa3004

\bibitem[Ghoreyshi \& Shadmehri(2020)]{Ghoreyshi20} Ghoreyshi, S.~M. \& Shadmehri, M.\ 2020, \mnras, 493, 5107. doi:10.1093/mnras/staa599

\bibitem[Ghosh \& Lamb(1979)]{Ghosh79} Ghosh, P. \& Lamb, F.~K.\ 1979, \apj, 234, 296. doi:10.1086/157498

\bibitem[Gilbaum \& Stone(2022)]{Gilbaum22} Gilbaum, S. \& Stone, N.~C.\ 2022, \apj, 928, 191. doi:10.3847/1538-4357/ac4ded

\bibitem[Goodman(2003)]{Goodman03} Goodman, J.\ 2003, \mnras, 339, 937. doi:10.1046/j.1365-8711.2003.06241.x

\bibitem[Graham et al.(2020)]{Graham20} Graham, M.~J., Ford, K.~E.~S., McKernan, B., et al.\ 2020, \prl, 124, 251102. doi:10.1103/PhysRevLett.124.251102

\bibitem[Grishin et al.(2021)]{Grishin21} Grishin, E., Bobrick, A., Hirai, R., et al.\ 2021, \mnras, 507, 156. doi:10.1093/mnras/stab1957

\bibitem[Gruzinov et al.(2020)]{Gruzinov20} Gruzinov, A., Levin, Y., \& Matzner, C.~D.\ 2020, \mnras, 492, 2755. doi:10.1093/mnras/staa013

\bibitem[Gu(2012)]{Gu12} Gu, W.-M.\ 2012, \apj, 753, 118. doi:10.1088/0004-637X/753/2/118

\bibitem[Gu(2015)]{Gu15} Gu, W.-M.\ 2015, \apj, 799, 71. doi:10.1088/0004-637X/799/1/71

\bibitem[Hashizume et al.(2015)]{Hashizume15} Hashizume, K., Ohsuga, K., Kawashima, T., et al.\ 2015, \pasj, 67, 58. doi:10.1093/pasj/psu132

\bibitem[Hirose et al.(2009)]{Hirose09} Hirose, S., Krolik, J.~H., \& Blaes, O.\ 2009, \apj, 691, 16. doi:10.1088/0004-637X/691/1/16

\bibitem[Hu et al.(2022)]{Hu22} Hu, H., Inayoshi, K., Haiman, Z., et al.\ 2022, \apj, 934, 132. doi:10.3847/1538-4357/ac75d8

\bibitem[Jermyn et al.(2021)]{Jermyn21} Jermyn, A.~S., Dittmann, A.~J., Cantiello, M., et al.\ 2021, \apj, 914, 105. doi:10.3847/1538-4357/abfb67

\bibitem[Jiang et al.(2014)]{Jiang14} Jiang, Y.-F., Stone, J.~M., \& Davis, S.~W.\ 2014, \apj, 796, 106. doi:10.1088/0004-637X/796/2/106

\bibitem[Jiao et al.(2015)]{Jiao15} Jiao, C.-L., Mineshige, S., Takeuchi, S., et al.\ 2015, \apj, 806, 93. doi:10.1088/0004-637X/806/1/93

\bibitem[Kaaz et al.(2021)]{Kaaz21} Kaaz, N., Schr{\o}der, S.~L., Andrews, J.~J., et al.\ 2021, arXiv:2103.12088

\bibitem[Kaaz et al.(2022)]{Kaaz22} Kaaz, N., Murguia-Berthier, A., Chatterjee, K., et al.\ 2022, arXiv:2201.11753

\bibitem[Kanagawa et al.(2015)]{Kanagawa15} Kanagawa, K.~D., Muto, T., Tanaka, H., et al.\ 2015, \apjl, 806, L15. doi:10.1088/2041-8205/806/1/L15

\bibitem[Kanagawa et al.(2016)]{Kanagawa16} Kanagawa, K.~D., Muto, T., Tanaka, H., et al.\ 2016, \pasj, 68, 43. doi:10.1093/pasj/psw037

\bibitem[Kanagawa et al.(2017)]{Kanagawa17} Kanagawa, K.~D., Tanaka, H., Muto, T., et al.\ 2017, \pasj, 69, 97. doi:10.1093/pasj/psx114

\bibitem[Kanagawa et al.(2018)]{Kanagawa18} Kanagawa, K.~D., Tanaka, H., \& Szuszkiewicz, E.\ 2018, \apj, 861, 140. doi:10.3847/1538-4357/aac8d9

\bibitem[Kato et al.(2008)]{Kato08} Kato, S., Fukue, J., \& Mineshige, S.\ 2008, Black-Hole Accretion Disks --- Towards a New Paradigm ---, 549 pages, including 12 Chapters, 9 Appendices,  ISBN 978-4-87698-740-5, Kyoto University Press (Kyoto, Japan), 2008.

\bibitem[Kimura et al.(2021a)]{Kimura21a} Kimura, S.~S., Sudoh, T., Kashiyama, K., et al.\ 2021a, \apj, 915, 31. doi:10.3847/1538-4357/abff58

\bibitem[Kimura et al.(2021b)]{Kimura21b} Kimura, S.~S., Murase, K., \& Bartos, I.\ 2021b, \apj, 916, 111. doi:10.3847/1538-4357/ac0535

\bibitem[Kitaki et al.(2018)]{Kitaki18} Kitaki, T., Mineshige, S., Ohsuga, K., et al.\ 2018, \pasj, 70, 108. doi:10.1093/pasj/psy110

\bibitem[Kitaki et al.(2021)]{Kitaki21} Kitaki, T., Mineshige, S., Ohsuga, K., et al.\ 2021, \pasj, 73, 450. doi:10.1093/pasj/psab011

\bibitem[Knigge(1999)]{Knigge99} Knigge, C.\ 1999, \mnras, 309, 409. doi:10.1046/j.1365-8711.1999.02839.x

\bibitem[Kocsis et al.(2011)]{Kocsis11} Kocsis, B., Yunes, N., \& Loeb, A.\ 2011, \prd, 84, 024032. doi:10.1103/PhysRevD.84.024032

\bibitem[Kohri et al.(2005)]{Kohri05} Kohri, K., Narayan, R., \& Piran, T.\ 2005, \apj, 629, 341. doi:10.1086/431354

\bibitem[Kremer et al.(2019)]{Kremer19} Kremer, K., Lu, W., Rodriguez, C.~L., et al.\ 2019, \apj, 881, 75. doi:10.3847/1538-4357/ab2e0c

\bibitem[Kumar et al.(2008)]{Kumar08} Kumar, P., Narayan, R., \& Johnson, J.~L.\ 2008, \mnras, 388, 1729. doi:10.1111/j.1365-2966.2008.13493.x

\bibitem[Lai(2014)]{Lai14} Lai, D.\ 2014, European Physical Journal Web of Conferences, 64, 01001. doi:10.1051/epjconf/20136401001

\bibitem[Li et al.(2020)]{Li20} Li, X., Chang, P., Levin, Y., et al.\ 2020, \mnras, 494, 2327. doi:10.1093/mnras/staa900

\bibitem[Mac Low \& McCray(1988)]{Mac Low88} Mac Low, M.-M. \& McCray, R.\ 1988, \apj, 324, 776. doi:10.1086/165936

\bibitem[McKernan et al.(2012)]{McKernan12} McKernan, B., Ford, K.~E.~S., Lyra, W., et al.\ 2012, \mnras, 425, 460. doi:10.1111/j.1365-2966.2012.21486.x

\bibitem[McKernan et al.(2020a)]{McKernan20a} McKernan, B., Ford, K.~E.~S., O'Shaugnessy, R., et al.\ 2020, \mnras, 494, 1203. doi:10.1093/mnras/staa740

\bibitem[McKernan et al.(2020b)]{McKernan20b} McKernan, B., Ford, K.~E.~S., \& O'Shaughnessy, R.\ 2020, \mnras, 498, 4088. doi:10.1093/mnras/staa2681

\bibitem[Metzger(2012)]{Metzger12} Metzger, B.~D.\ 2012, \mnras, 419, 827. doi:10.1111/j.1365-2966.2011.19747.x

\bibitem[Mishra et al.(2020)]{Mishra20} Mishra, B., Begelman, M.~C., Armitage, P.~J., et al.\ 2020, \mnras, 492, 1855. doi:10.1093/mnras/stz3572

\bibitem[Moranchel-Basurto et al.(2021)]{Moranchel21} Moranchel-Basurto, A., S{\'a}nchez-Salcedo, F.~J., Chametla, R.~O., et al.\ 2021, \apj, 906, 15. doi:10.3847/1538-4357/abca88

\bibitem[Narayan \& Yi(1994)]{Narayan94} Narayan, R. \& Yi, I.\ 1994, \apjl, 428, L13. doi:10.1086/187381

\bibitem[Narayan \& Yi(1995)]{Narayan95} Narayan, R. \& Yi, I.\ 1995, \apj, 444, 231. doi:10.1086/175599

\bibitem[Nasim et al.(2022)]{Nasim22} Nasim, S.~S., Fabj, G., Caban, F., et al.\ 2022, arXiv:2207.09540

\bibitem[Olano(2009)]{Olano09} Olano, C.~A.\ 2009, \aap, 506, 1215. doi:10.1051/0004-6361/200912602

\bibitem[Ohsuga et al.(2005)]{Ohsuga05} Ohsuga, K., Mori, M., Nakamoto, T., et al.\ 2005, \apj, 628, 368. doi:10.1086/430728

\bibitem[Ostriker(1983)]{Ostriker83} Ostriker, J.~P.\ 1983, \apj, 273, 99. doi:10.1086/161351

\bibitem[Ostriker \& McKee(1988)]{Ostriker88} Ostriker, J.~P. \& McKee, C.~F.\ 1988, Reviews of Modern Physics, 60, 1. doi:10.1103/RevModPhys.60.1

\bibitem[Ostriker(1999)]{Ostriker99} Ostriker, E.~C.\ 1999, \apj, 513, 252. doi:10.1086/306858

\bibitem[Pan \& Yang(2021a)]{Pan21a} Pan, Z. \& Yang, H.\ 2021a, \prd, 103, 103018. doi:10.1103/PhysRevD.103.103018

\bibitem[Pan \& Yang(2021b)]{Pan21b} Pan, Z. \& Yang, H.\ 2021b, \apj, 923, 173. doi:10.3847/1538-4357/ac249c

\bibitem[Pan et al.(2022)]{Pan22} Pan, Z., Lyu, Z., \& Yang, H.\ 2022, \prd, 105, 083005. doi:10.1103/PhysRevD.105.083005

\bibitem[Perets et al.(2016)]{Perets16} Perets, H.~B., Li, Z., Lombardi, J.~C., et al.\ 2016, \apj, 823, 113. doi:10.3847/0004-637X/823/2/113

\bibitem[Perna et al.(2021a)]{Perna21a} Perna, R., Lazzati, D., \& Cantiello, M.\ 2021a, \apjl, 906, L7. doi:10.3847/2041-8213/abd319

\bibitem[Perna et al.(2021b)]{Perna21b} Perna, R., Tagawa, H., Haiman, Z., et al.\ 2021b, \apj, 915, 10. doi:10.3847/1538-4357/abfdb4

\bibitem[Piro \& Lu(2020)]{Piro20} Piro, A.~L. \& Lu, W.\ 2020, \apj, 894, 2. doi:10.3847/1538-4357/ab83f6

\bibitem[Ren et al.(2022)]{Ren22} Ren, J., Chen, K., Wang, Y., et al.\ 2022, \apjl, 940, L44. doi:10.3847/2041-8213/aca025

\bibitem[Romanova \& Owocki(2015)]{Romanova15} Romanova, M.~M. \& Owocki, S.~P.\ 2015, \ssr, 191, 339. doi:10.1007/s11214-015-0200-9

\bibitem[Rozyczka et al.(1995)]{Rozyczka95} Rozyczka, M., Bodenheimer, P., \& Lin, D.~N.~C.\ 1995, \mnras, 276, 597. doi:10.1093/mnras/276.2.597

\bibitem[S{\c a}dowski et al.(2015)]{Sadowski151} S{\c a}dowski, A., Narayan, R., Tchekhovskoy, A., et al.\ 2015, \mnras, 447, 49. doi:10.1093/mnras/stu2387

\bibitem[S{\c a}dowski \& Narayan(2015)]{Sadowski152} S{\c a}dowski, A. \& Narayan, R.\ 2015, \mnras, 453, 3213. doi:10.1093/mnras/stv1802

\bibitem[S{\c a}dowski et al.(2016)]{Sadowski161} S{\c a}dowski, A., Lasota, J.-P., Abramowicz, M.~A., et al.\ 2016, \mnras, 456, 3915. doi:10.1093/mnras/stv2854

\bibitem[S{\c a}dowski \& Narayan(2016)]{Sadowski162} S{\c a}dowski, A. \& Narayan, R.\ 2016, \mnras, 456, 3929. doi:10.1093/mnras/stv2941

\bibitem[Samsing et al.(2022)]{Samsing22} Samsing, J., Bartos, I., D'Orazio, D.~J., et al.\ 2022, \nat, 603, 237. doi:10.1038/s41586-021-04333-1

\bibitem[Schiano(1985)]{Schiano85} Schiano, A.~V.~R.\ 1985, \apj, 299, 24. doi:10.1086/163680

\bibitem[Secunda et al.(2019)]{Secunda19} Secunda, A., Bellovary, J., Mac Low, M.-M., et al.\ 2019, \apj, 878, 85. doi:10.3847/1538-4357/ab20ca

\bibitem[Shakura \& Sunyaev(1973)]{Shakura73} Shakura, N.~I. \& Sunyaev, R.~A.\ 1973, \aap, 24, 337

\bibitem[Shakura et al.(2015)]{Shakura15} Shakura, N.~I., Postnov, K.~A., Kochetkova, A.~Y., et al.\ 2015, Astronomy Reports, 59, 645. doi:10.1134/S1063772915070112

\bibitem[Shang et al.(2011)]{Shang11} Shang, Z., Brotherton, M.~S., Wills, B.~J., et al.\ 2011, \apjs, 196, 2. doi:10.1088/0067-0049/196/1/2

\bibitem[Shen \& Matzner(2014)]{Shen14} Shen, R.-F. \& Matzner, C.~D.\ 2014, \apj, 784, 87. doi:10.1088/0004-637X/784/2/87

\bibitem[Sirko \& Goodman(2003)]{SG03} Sirko, E. \& Goodman, J.\ 2003, \mnras, 341, 501. doi:10.1046/j.1365-8711.2003.06431.x

\bibitem[Syer et al.(1991)]{Syer91} Syer, D., Clarke, C.~J., \& Rees, M.~J.\ 1991, \mnras, 250, 505. doi:10.1093/mnras/250.3.505


%\bibitem[S\k{a}dowski \& Narayan(2015)]{Sadowski15} S\k{a}dowski, A. \& Narayan, R.\ 2015, \mnras, 453, 3213. doi:10.1093/mnras/stv1802

%\bibitem[S\k{a}dowski et al.(2016)]{Sadowski16} S\k{a}dowski, A., Lasota, J.-P., Abramowicz, M.~A., et al.\ 2016, \mnras, 456, 3915. doi:10.1093/mnras/stv2854

\bibitem[Tagawa et al.(2020a)]{Tagawa20a} Tagawa, H., Haiman, Z., \& Kocsis, B.\ 2020a, \apj, 898, 25. doi:10.3847/1538-4357/ab9b8c

\bibitem[Tagawa et al.(2020b)]{Tagawa20b} Tagawa, H., Haiman, Z., Bartos, I., et al.\ 2020b, \apj, 899, 26. doi:10.3847/1538-4357/aba2cc

\bibitem[Tagawa et al.(2021)]{Tagawa21} Tagawa, H., Kocsis, B., Haiman, Z., et al.\ 2021, \apj, 908, 194. doi:10.3847/1538-4357/abd555

\bibitem[Tagawa et al.(2022)]{Tagawa22} Tagawa, H., Kimura, S.~S., Haiman, Z., et al.\ 2022, \apj, 927, 41. doi:10.3847/1538-4357/ac45f8

\bibitem[Takahashi et al.(2018)]{Takahashi18} Takahashi, H.~R., Mineshige, S., \& Ohsuga, K.\ 2018, \apj, 853, 45. doi:10.3847/1538-4357/aaa082

\bibitem[Takeo et al.(2020)]{Takeo20} Takeo, E., Inayoshi, K., \& Mineshige, S.\ 2020, \mnras, 497, 302. doi:10.1093/mnras/staa1906

\bibitem[Tanigawa et al.(2012)]{Tanigawa12} Tanigawa, T., Ohtsuki, K., \& Machida, M.~N.\ 2012, \apj, 747, 47. doi:10.1088/0004-637X/747/1/47

\bibitem[Tanigawa \& Tanaka(2016)]{Tanigawa16} Tanigawa, T. \& Tanaka, H.\ 2016, \apj, 823, 48. doi:10.3847/0004-637X/823/1/48

\bibitem[Thompson et al.(2005)]{TQM05} Thompson, T.~A., Quataert, E., \& Murray, N.\ 2005, \apj, 630, 167. doi:10.1086/431923

\bibitem[Toomre(1964)]{Toomre64} Toomre, A.\ 1964, \apj, 139, 1217. doi:10.1086/147861

\bibitem[Valtonen \& Karttunen(2006)]{Valtonen06} Valtonen, M. \& Karttunen, H.\ 2006, The Three-Body Problem, by Mauri Valtonen and Hannu Karttunen, pp. . ISBN 0521852242. Cambridge, UK: Cambridge University Press,  2006.

\bibitem[Wang et al.(2021a)]{Wang21a} Wang, J.-M., Liu, J.-R., Ho, L.~C., et al.\ 2021a, \apjl, 911, L14. doi:10.3847/2041-8213/abee81

\bibitem[Wang et al.(2021b)]{Wang21b} Wang, J.-M., Liu, J.-R., Ho, L.~C., et al.\ 2021b, \apjl, 916, L17. doi:10.3847/2041-8213/ac0b46

\bibitem[Wang et al.(2022)]{Wang22} Wang, Y.-H., Lazzati, D., \& Perna, R.\ 2022, \mnras. doi:10.1093/mnras/stac1968

\bibitem[Ward(1997)]{Ward97} Ward, W.~R.\ 1997, \icarus, 126, 261. doi:10.1006/icar.1996.5647

\bibitem[Weaver et al.(1977)]{Weaver77} Weaver, R., McCray, R., Castor, J., et al.\ 1977, \apj, 218, 377. doi:10.1086/155692

\bibitem[Wu et al.(2022)]{Wu22} Wu, W.-B., Gu, W.-M., \& Sun, M.\ 2022, \apj, 930, 108. doi:10.3847/1538-4357/ac6588

\bibitem[Yang et al.(2014)]{Yang14} Yang, X.-H., Yuan, F., Ohsuga, K., et al.\ 2014, \apj, 780, 79. doi:10.1088/0004-637X/780/1/79

\bibitem[Yang et al.(2022)]{Yang22} Yang, Y., Bartos, I., Fragione, G., et al.\ 2022, \apjl, 933, L28. doi:10.3847/2041-8213/ac7c0b

\bibitem[Yuan et al.(2022)]{Yuan22} Yuan, C., Murase, K., Guetta, D., et al.\ 2022, \apj, 932, 80. doi:10.3847/1538-4357/ac6ddf

\bibitem[Yuan \& Narayan(2014)]{Yuan14} Yuan, F. \& Narayan, R.\ 2014, \araa, 52, 529. doi:10.1146/annurev-astro-082812-141003

\bibitem[Zahra Zeraatgari et al.(2020)]{Zeraatgari20} Zahra Zeraatgari, F., Mosallanezhad, A., Yuan, Y.-F., et al.\ 2020, \apj, 888, 86. doi:10.3847/1538-4357/ab594f

\bibitem[Zhang \& Dai(2009)]{Zhang09} Zhang, D. \& Dai, Z.~G.\ 2009, \apj, 703, 461. doi:10.1088/0004-637X/703/1/461

\bibitem[Zhu et al.(2021a)]{Zhu21a} Zhu, J.-P., Zhang, B., Yu, Y.-W., et al.\ 2021a, \apjl, 906, L11. doi:10.3847/2041-8213/abd412

\bibitem[Zhu et al.(2021b)]{Zhu21b} Zhu, J.-P., Yang, Y.-P., Zhang, B., et al.\ 2021b, \apjl, 914, L19. doi:10.3847/2041-8213/abff5a


\end{thebibliography}
\end{document}